\documentclass[12pt,a4]{article}
\usepackage{amsthm,amsmath,latexsym,multibox,amssymb,amsfonts,amsbsy}
\usepackage{graphicx,lscape,fancyhdr,array,stmaryrd,euscript,wrapfig}
\usepackage{cite,stmaryrd}
\usepackage{amsthm,amsmath,latexsym,multibox,amssymb,amsfonts}
\usepackage{graphicx,lscape,fancyhdr,array,stmaryrd}

\newcommand{\bep}{\begin{picture}}
\newcommand{\eep}{\end{picture}}

\newcommand{\SlNProduct}{{\bep(160,170)(0,-20){%
\put(0,-20){\RectT{1}{2}{\TextRight{$ \alpha_{N+1}$}}}%
\put(0,0){\RectT{5}{4}{\TextTopLeftIn{$ s_N$}{$ p_N$}}}%
\put(50,20){\RectT{1}{2}{\TextRight{$ \alpha_N$}}}%
\put(0,40){\RectT{9}{5}{\TextTopLeftIn{$ s_2$}{$ p_2$}}}%
\put(90,60){\RectT{1}{3}{\TextRight{$ \alpha_2$}}}%
\put(0,90){\RectT{14}{6}{\TextTopLeftIn{$ s_1$}{$ p_1$}}}%
\put(140,120){\RectT{1}{3}{\TextRight{$ \alpha_1$}}}%
}\eep}}

\newcommand{\TracedTensorProduct}{{\bep(160,170){%
\put(0,0){\RectT{1}{2}{\TextRight{$\scriptstyle \alpha_{N+1}$}}}
\put(0,20){\ABRectT{5}{4}{1}{1}{N}}%
\put(0,60){\ABRectT{9}{5}{2}{2}{2}}%
\put(0,110){\ABRectT{14}{6}{2}{2}{1}}%
}\eep}}

\newcommand{\TracedTensorProductABG}{{\bep(200,200){%
\put(0,0){\RectT{1}{2}{\TextRight{$\scriptstyle \alpha_{N+1}$}}}
\put(0,20){\ABGRectMT{6}{5}{1}{1}{3}{0}{N}}%
\put(0,70){\ABGRectMT{12}{6}{2}{1}{3}{6}{2}}%
\put(0,130){\ABGRectMT{18}{7}{2}{2}{3}{12}{1}}%
}\eep}}

\newcommand{\SpecialCaseA}{\bep(40,40){%
\put(0,10){\RectRaggedTop{4}{2}}
\put(5,0){$\gamma_N=0$}
}\eep}

\newcommand{\SpecialCaseB}{\bep(80,60){%
\put(0,0){\line(1,0){20}}\put(0,10){\line(1,0){20}}\put(20,0){\line(0,1){40}}%
\put(0,30){\line(1,0){20}}\put(20,40){\line(1,0){50}}\put(70,40){\line(0,1){15}}%
\put(22,0){$\gamma_i=\Delta_i$}\put(22,15){$\epsilon_i$}\put(22,30){$\gamma_{i-1}=0$}%
}\eep}

\newcommand{\SpecialCaseC}{\bep(80,75){%
\put(0,0){\line(1,0){20}}\put(20,0){\line(0,1){20}}\put(20,20){\line(1,0){10}}%
\put(30,20){\line(0,1){30}}\put(0,40){\line(1,0){30}}\put(30,50){\line(1,0){50}}\put(80,50){\line(0,1){15}}%
\put(22,5){$\beta_i>0$}\put(32,40){$\gamma_{i-1}=0$}\put(32,25){$\epsilon_i$}%
}\eep}

\newcommand{\SpecialCaseD}{\bep(80,75){%
\put(10,0){\line(0,1){50}}\put(10,50){\line(1,0){70}}\put(0,40){\line(1,0){50}}\put(50,40){\line(0,1){10}}\put(80,50){\line(0,1){15}}%
\put(10,20){\line(1,0){10}}\put(20,20){\line(0,1){20}}
\put(22,25){$\alpha_{i+1}$}\put(12,42){$\scriptstyle 0<\gamma_{i}<\Delta_i$}%
}\eep}

\newcommand{\SpecialCaseF}{\bep(60,55){%
\put(0,0){\line(1,0){20}}\put(0,10){\line(1,0){20}}\put(20,0){\line(0,1){50}}%
\put(20,30){\line(1,0){10}}\put(30,30){\line(0,1){20}}\put(20,50){\line(1,0){20}}%
\put(22,0){$\gamma_i=\Delta_i$}\put(22,15){$\epsilon_i$}\put(32,35){$\alpha_{i}$}%
}\eep}

\newcommand{\SpecialCaseG}{\bep(60,65){%
\put(0,0){\line(1,0){10}}\put(10,0){\line(0,1){20}}\put(10,20){\line(1,0){10}}%
\put(20,20){\line(0,1){20}}\put(20,40){\line(1,0){10}}\put(30,40){\line(0,1){20}}\put(30,60){\line(1,0){20}}%
\put(12,5){$\beta_i$}\put(22,25){$\epsilon_i$}\put(32,45){$\alpha_i$}%
}\eep}

\newcounter{YoungHeight}\newcounter{YoungWidth}

\newcounter{Mul1}\newcounter{Mul2}\newcounter{Mul3}\newcounter{Mul4}
\newcounter{A0}\newcounter{A1}\newcounter{A2}\newcounter{A3}\newcounter{A4}\newcounter{A5}\newcounter{A6}
\newcounter{B0}\newcounter{B1}\newcounter{B2}\newcounter{B3}
\newcounter{C1}\newcounter{C2}\newcounter{C3}\newcounter{C4}\newcounter{C6}\newcounter{C7}
\newcounter{D1}\newcounter{D2}\newcounter{D3}\newcounter{D4}\newcounter{D5}
\newcounter{T0}\newcounter{T1}

\newcounter{TGR0}
\newcounter{R0}\newcounter{R1}\newcounter{R2}\newcounter{R3}
\newcounter{AR0}\newcounter{AR1}\newcounter{AR2}\newcounter{AR3}\newcounter{AR4}\newcounter{AR5}\newcounter{AR6}\newcounter{AR7}
\newcounter{Dotted0}\newcounter{Dotted1}\newcounter{Dotted2}\newcounter{Dotted3}
\newcounter{CircX}\newcounter{CircY}
\newcounter{reptA}

\newlength{\txtHShift}

\newlength{\txtWidth}

\newcommand{\HalfLength}[2]{\setcounter{Mul1}{#1}\setcounter{Mul2}{#1}\addtocounter{Mul1}{\value{Mul2}}\addtocounter{Mul1}{\value{Mul2}}%
\addtocounter{Mul1}{\value{Mul2}}\addtocounter{Mul1}{\value{Mul2}}\setcounter{#2}{\value{Mul1}}}

\newcommand{\Add}[3]{\setcounter{#1}{#2}\addtocounter{#1}{#3}}

\newcommand{\Length}[1]{#10}
\newcommand{\YoungScale}{}

\newcommand{\shiftedText}[2]{{\hspace{#1}#2}}
\newcommand{\calcHShift}[1]{\settowidth{\txtWidth}{#1}\setlength{\txtHShift}{-0.5\txtWidth}}

\newcommand{\TextCenter}[3]{{\HalfLength{#2}{T0}%
\HalfLength{#3}{T1}\addtocounter{T1}{-3}\calcHShift{#1}%
\put(\value{T0},\value{T1}){\shiftedText{\txtHShift}{#1}}}}

\newcommand{\TextCenterA}[3]{{\HalfLength{#2}{T0}%
\HalfLength{#3}{T1}\addtocounter{T1}{-3}\addtocounter{T0}{10}\calcHShift{#1}%
\put(\value{T0},\value{T1}){\shiftedText{\txtHShift}{#1}}}}
\newcommand{\TextCenterB}[3]{{\calcHShift{#1}\HalfLength{#2}{T0}\Add{T1}{\Length{#3}}{-7}\put(\value{T0},\value{T1}){\shiftedText{\txtHShift}{#1}}}}

\newcommand{\TextTop}[3]{{\calcHShift{#1}\HalfLength{#2}{T0}\Add{T1}{\Length{#3}}{-9}\put(\value{T0},\value{T1}){\shiftedText{\txtHShift}{#1}}}}

\newcommand{\TextLeftIn}[3]{{\HalfLength{#3}{T1}\addtocounter{T1}{-5}\put(2,\value{T1}){#1}}}

\newcommand{\TextRight}[3]{{\HalfLength{#3}{T1}\addtocounter{T1}{-5}\Add{T0}{\Length{#2}}{2}\put(\value{T0},\value{T1}){#1}}}

\newcommand{\TextTopLeftIn}[4]{{\TextTop{#1}{#3}{#4}\TextLeftIn{#2}{#3}{#4}}}

\newcommand{\dottedline}[2]{{\HalfLength{#1}{Dotted0}\HalfLength{#2}{Dotted1}\Add{Dotted2}{#1}{#2}\HalfLength{\value{Dotted2}}{Dotted3}%
\qbezier[\value{Dotted3}](0,0)(\value{Dotted0},\value{Dotted1})(\Length{#1},\Length{#2})}}

\newcommand{\BlockA}[2]{{\YoungScale\bep(\Length{#1},\Length{#2}){\Add{A1}{#1}{1}\Add{A2}{#2}{1}}%
\multiput(0,0)(10,0){\value{A1}}{\line(0,1){\Length{#2}}}\multiput(0,0)(0,10){\value{A2}}{\line(1,0){\Length{#1}}}%
\setcounter{YoungHeight}{\Length{#2}}\setcounter{YoungWidth}{\Length{#1}}\eep}}

\newcommand{\BlockADotted}[2]{{\YoungScale\bep(\Length{#1},\Length{#2}){\Add{A1}{#1}{1}\Add{A2}{#2}{1}}%
\multiput(0,0)(10,0){\value{A1}}{\dottedline{0}{#2}}\multiput(0,0)(0,10){\value{A2}}{\dottedline{#1}{0}}%
\setcounter{YoungHeight}{\Length{#2}}\setcounter{YoungWidth}{\Length{#1}}\eep}}

\newcommand{\ColumnShadded}[1]{\bep(10,\Length{#1})\put(0,0){\Rect{1}{#1}}%
\multiput(0,0)(0,10){#1}{\line(1,1){10}}%
\eep}

\newcommand{\Rect}[2]{\bep(\Length{#1},\Length{#2})\put(0,0){\line(1,0){\Length{#1}}}\put(0,0){\line(0,1){\Length{#2}}}%
\put(\Length{#1},\Length{#2}){\line(-1,0){\Length{#1}}}\put(\Length{#1},\Length{#2}){\line(0,-1){\Length{#2}}}\eep}

\newcommand{\RectRaggedTop}[2]{\bep(\Length{#1},\Length{#2})%
\put(0,0){\line(1,0){\Length{#1}}}\put(0,0){\line(0,1){\Length{#2}}}%
\put(\Length{#1},0){\line(0,1){\Length{#2}}}%
\HalfLength{#1}{R0}\HalfLength{#2}{R1}\Add{R2}{#1}{#1}\Add{R3}{\value{R0}}{#1}\addtocounter{R3}{#1}%
\put(0,\Length{#2}){\qbezier(0,0)(\value{R2},5)(\value{R0},0)\qbezier(\value{R0},0)(\value{R3},-5)(\Length{#1},0)}%
\eep}

\newcommand{\RectT}[3]{\bep(\Length{#1},\Length{#2})\put(0,0){\line(1,0){\Length{#1}}}\put(0,0){\line(0,1){\Length{#2}}}%
\put(\Length{#1},\Length{#2}){\line(-1,0){\Length{#1}}}\put(\Length{#1},\Length{#2}){\line(0,-1){\Length{#2}}}#3{#1}{#2}\eep}

\newcommand{\RectTDotted}[3]{\bep(\Length{#1},\Length{#2})\put(0,0){\dottedline{#1}{0}}\put(0,0){\dottedline{0}{#2}}%
\put(0,\Length{#2}){\dottedline{#1}{0}}\put(\Length{#1},0){\dottedline{0}{#2}}#3{#1}{#2}\eep}

\newcommand{\ColumnTDotted}[3]{\RectTDotted{#1}{#2}{\TextCenter{$\scriptstyle \ #3$}}}

\newcommand{\RectARow}[2]{{\bep(\Length{#1},10)\put(0,0){\RectT{#1}{1}{\TextTop{#2}}}\eep}}

\newcommand{\RectBRow}[4]{{\bep(\Length{#1},20)\put(0,0){\RectT{#2}{1}{\TextTop{#4}}}%
\put(0,10){\RectT{#1}{1}{\TextTop{#3}}}\eep}}

\newcommand{\RectCRow}[6]{{\bep(\Length{#1},30)\put(0,0){\RectT{#3}{1}{\TextTop{#6}}}%
\put(0,10){\RectT{#2}{1}{\TextTop{#5}}}\put(0,20){\RectT{#1}{1}{\TextTop{#4}}}\eep}}

\newcommand{\RectARowUp}[2]{{\bep(\Length{#1},10)\put(0,0){\RectT{#1}{1}{\TextCenterB{#2}}}\eep}}

\newcommand{\RectBRowUp}[4]{{\bep(\Length{#1},20)\put(0,0){\RectT{#2}{1}{\TextCenterB{#4}}}%
\put(0,10){\RectT{#1}{1}{\TextCenterB{#3}}}\eep}}

\newcommand{\RectCRowUp}[6]{{\bep(\Length{#1},30)\put(0,0){\RectT{#3}{1}{\TextCenterB{#6}}}%
\put(0,10){\RectT{#2}{1}{\TextCenterB{#5}}}\put(0,20){\RectT{#1}{1}{\TextCenterB{#4}}}\eep}}

\newcommand{\RectAYoung}[3]{{\bep(0,0)#3\eep\Add{A0}{\value{YoungWidth}}{\Length{#1}}%
\Add{A1}{\value{YoungHeight}}{-10}\bep(\value{A0},\value{YoungHeight})%
\put(\value{YoungWidth},\value{A1}){\RectT{#1}{1}{\TextTop{#2}}}\eep}}

\newcommand{\RectBYoung}[3]{{\bep(0,0)\put(0,0){#3}\eep\Add{A0}{\value{YoungHeight}}{10}%
\bep(\Length{#1},\value{A0})%
\put(0,\value{YoungHeight}){\RectT{#1}{1}{\TextTop{#2}}}\eep}}

\newcommand{\RectCYoung}[5]{{\bep(0,0)\put(0,0){#5}\eep\Add{A0}{\value{YoungHeight}}{20}%
\bep(\Length{#1},\value{A0})%
\put(0,\value{YoungHeight}){\RectBRow{#1}{#2}{#3}{#4}}\eep}}

\newcommand{\RectDYoung}[7]{{\bep(0,0)\put(0,0){#7}\eep\Add{A0}{\value{YoungHeight}}{30}%
\bep(\Length{#1},\value{A0})%
\put(0,\value{YoungHeight}){\RectCRow{#1}{#2}{#3}{#4}{#5}{#6}}\eep}}

\newcommand{\RectDotted}[2]{\bep(\Length{#1},\Length{#2})\put(0,0){\dottedline{#1}{0}}\put(0,0){\dottedline{0}{#2}}%
\put(\Length{#1},0){\dottedline{0}{#2}}\put(0,\Length{#2}){\dottedline{#1}{0}}\eep}

\newcommand{\BRectT}[5]{{\Add{AR0}{\Length{#1}}{0}\Add{AR1}{\Length{#2}}{-\Length{#3}}\bep(\value{AR0},\Length{#2}){%
\Add{AR2}{\Length{#1}}{-10}\put(0,0){\line(1,0){\value{AR2}}}\put(\value{AR2},0){\line(0,1){\Length{#4}}}%
\put(\value{AR2},\Length{#4}){\line(1,0){10}}\put(\value{AR2},0){\dottedline{1}{0}}%
\put(\Length{#1},0){\dottedline{0}{#4}}%
\Add{AR3}{\Length{#2}}{-\Length{#3}}\addtocounter{AR3}{-\Length{#4}}%
\put(0,0){\line(0,1){\Length{#2}}}\put(0,\Length{#2}){\line(1,0){\value{AR0}}}%
\put(\Length{#1},\Length{#4}){\line(0,1){\value{AR3}}}%
\put(\value{AR0},\value{AR1}){\line(0,1){\Length{#3}}}%
\put(\value{AR2},0){\TextCenter{$ \ t_{#5}$}{1}{#4}}%
\put(0,0){\TextLeftIn{$ p_{#5}$}{#1}{#2}}%
\put(0,0){\TextTop{$ s_{#5}$}{#1}{#2}}%
}\eep}}

\newcommand{\BRectTM}[5]{{\Add{AR0}{\Length{#1}}{0}\Add{AR1}{\Length{#2}}{-\Length{#3}}\bep(\value{AR0},\Length{#2}){%
\Add{AR2}{\Length{#1}}{-10}\put(0,0){\line(1,0){\value{AR2}}}\put(\value{AR2},0){\line(0,1){\Length{#4}}}%
\put(\value{AR2},\Length{#4}){\line(1,0){10}}\put(\value{AR2},0){\dottedline{1}{0}}%
\put(\Length{#1},0){\dottedline{0}{#4}}%
\Add{AR3}{\Length{#2}}{-\Length{#3}}\addtocounter{AR3}{-\Length{#4}}%
\put(0,0){\line(0,1){\Length{#2}}}\put(0,\Length{#2}){\line(1,0){\value{AR0}}}%
\put(\Length{#1},\Length{#4}){\line(0,1){\value{AR3}}}%
\put(\value{AR0},\value{AR1}){\line(0,1){\Length{#3}}}%
\put(0,0){\TextLeftIn{$ p_{#5}$}{#1}{#2}}%
\put(0,0){\TextTop{$ s_{#5}$}{#1}{#2}}%
}\eep}}

\newcommand{\BRectTA}[5]{{\Add{AR0}{\Length{#1}}{0}\Add{AR1}{\Length{#2}}{-\Length{#3}}\bep(\value{AR0},\Length{#2}){%
\Add{AR2}{\Length{#1}}{-10}\put(0,0){\line(1,0){\value{AR2}}}\put(\value{AR2},0){\line(0,1){\Length{#4}}}%
\put(\value{AR2},\Length{#4}){\line(1,0){10}}\put(\value{AR2},0){\dottedline{1}{0}}%
\put(\Length{#1},0){\dottedline{0}{#4}}%
\Add{AR3}{\Length{#2}}{-\Length{#3}}\addtocounter{AR3}{-\Length{#4}}%
\put(0,0){\line(0,1){\Length{#2}}}\put(0,\Length{#2}){\line(1,0){\value{AR0}}}%
\put(\Length{#1},\Length{#4}){\line(0,1){\value{AR3}}}%
\put(\value{AR0},\value{AR1}){\line(0,1){\Length{#3}}}%
\put(\value{AR2},0){\TextCenter{$ \ t_{#5}$}{1}{#4}}%
\put(0,0){\TextLeftIn{$ p_{#5}-1$}{#1}{#2}}%
\put(0,0){\TextTop{$ s_{#5}$}{#1}{#2}}%
}\eep}}

\newcommand{\BRectTMA}[5]{{\Add{AR0}{\Length{#1}}{0}\Add{AR1}{\Length{#2}}{-\Length{#3}}\bep(\value{AR0},\Length{#2}){%
\Add{AR2}{\Length{#1}}{-10}\put(0,0){\line(1,0){\value{AR2}}}\put(\value{AR2},0){\line(0,1){\Length{#4}}}%
\put(\value{AR2},\Length{#4}){\line(1,0){10}}\put(\value{AR2},0){\dottedline{1}{0}}%
\put(\Length{#1},0){\dottedline{0}{#4}}%
\Add{AR3}{\Length{#2}}{-\Length{#3}}\addtocounter{AR3}{-\Length{#4}}%
\put(0,0){\line(0,1){\Length{#2}}}\put(0,\Length{#2}){\line(1,0){\value{AR0}}}%
\put(\Length{#1},\Length{#4}){\line(0,1){\value{AR3}}}%
\put(\value{AR0},\value{AR1}){\line(0,1){\Length{#3}}}%
\put(0,0){\TextLeftIn{$ p_{#5}-1$}{#1}{#2}}%
\put(0,0){\TextTop{$ s_{#5}$}{#1}{#2}}%
}\eep}}

\newcommand{\ABRectT}[5]{{\Add{AR0}{\Length{#1}}{10}\Add{AR1}{\Length{#2}}{-\Length{#3}}\bep(\value{AR0},\Length{#2}){%
\Add{AR2}{\Length{#1}}{-10}\put(0,0){\line(1,0){\value{AR2}}}\put(\value{AR2},0){\line(0,1){\Length{#4}}}%
\put(\value{AR2},\Length{#4}){\line(1,0){10}}\put(\value{AR2},0){\dottedline{1}{0}}%
\put(\Length{#1},0){\dottedline{0}{#4}}%
\Add{AR3}{\Length{#2}}{-\Length{#3}}\addtocounter{AR3}{-\Length{#4}}%
\put(0,0){\line(0,1){\Length{#2}}}\put(0,\Length{#2}){\line(1,0){\value{AR0}}}%
\put(\Length{#1},\Length{#4}){\line(0,1){\value{AR3}}}\put(\Length{#1},\value{AR1}){\line(1,0){10}}%
\put(\value{AR0},\value{AR1}){\line(0,1){\Length{#3}}}\put(\Length{#1},\value{AR1}){\dottedline{0}{#3}}%
\put(\value{AR2},0){\TextCenter{$\scriptstyle \ \beta_{#5}$}{1}{#4}}%
\put(\Length{#1},\value{AR1}){\TextCenter{$\scriptstyle \alpha_{#5}$}{1}{#3}}%
\Add{AR5}{#2}{-#3}\addtocounter{AR5}{-#4}\put(\Length{#1},\Length{#4}){\TextLeftIn{$\scriptstyle \epsilon_{#5}$}{2}{\value{AR5}}}%
\put(0,0){\TextLeftIn{$ p_{#5}$}{#1}{#2}}%
\put(0,0){\TextCenter{$ s_{#5}$}{#1}{#2}}%
}\eep}}

\newcommand{\ABGRectMT}[7]{{\Add{AR0}{\Length{#1}}{10}\Add{AR1}{\Length{#2}}{-\Length{#3}}\bep(\value{AR0},\Length{#2}){%
\Add{AR2}{\Length{#1}}{-10}\put(0,0){\line(1,0){\Length{#6}}}\put(0,10){\line(1,0){\Length{#6}}}%
\put(\Length{#6},0){\RectT{#5}{1}{\TextCenter{\ensuremath{\scriptstyle \gamma_{#7}}}}}%
\Add{AR4}{\value{AR2}}{-\Length{#5}}\addtocounter{AR4}{-\Length{#6}}\Add{AR6}{\Length{#5}}{\Length{#6}}%
\put(\value{AR6},10){\line(1,0){\value{AR4}}}\put(\value{AR2},10){\line(0,1){\Length{#4}}}%
\Add{AR6}{\Length{#4}}{10}%
\put(\value{AR2},\value{AR6}){\line(1,0){10}}\put(\value{AR2},10){\dottedline{1}{0}}%
\put(\Length{#1},10){\dottedline{0}{#4}}%
\Add{AR3}{\Length{#2}}{-\Length{#3}}\addtocounter{AR3}{-\Length{#4}}\addtocounter{AR3}{-10}%
\put(0,0){\line(0,1){\Length{#2}}}\put(0,\Length{#2}){\line(1,0){\value{AR0}}}%
\put(\Length{#1},\value{AR6}){\line(0,1){\value{AR3}}}\put(\Length{#1},\value{AR1}){\line(1,0){10}}%
\put(\value{AR0},\value{AR1}){\line(0,1){\Length{#3}}}\put(\Length{#1},\value{AR1}){\dottedline{0}{#3}}%
\put(\value{AR2},10){\TextCenter{$\scriptstyle\ \beta_{#7}$}{1}{#4}}%
\put(\Length{#1},\value{AR1}){\TextCenter{$\scriptstyle \alpha_{#7}$}{1}{#3}}%
\Add{AR5}{#2}{-#3}\addtocounter{AR5}{-#4}\addtocounter{AR5}{-1}%
\put(\Length{#1},\value{AR6}){\TextLeftIn{$\scriptstyle \,\,\epsilon_{#7}$}{2}{\value{AR5}}}%
\Add{AR7}{#2}{-1}
\put(0,10){\TextLeftIn{$ p_{#7}\!-\!1$}{#1}{\value{AR7}}}%
\put(0,10){\TextCenterA{$ s_{#7}$}{#1}{\value{AR7}}}%
}\eep}}

\newcommand{\YoungA}{\BlockA{1}{1}}
\newcommand{\YoungB}{\BlockA{2}{1}}
\newcommand{\YoungC}{\BlockA{3}{1}}

\newcommand{\BlockCcA}[2]{{\YoungScale\bep(\Length{#1},\Length{#2}){\Add{A1}{#1}{1}\Add{A2}{#2}{1}}%
\multiput(0,0)(10,0){\value{A1}}{\line(0,1){\Length{#2}}}\multiput(0,0)(0,10){\value{A2}}{\line(1,0){\Length{#1}}}\setcounter{YoungHeight}{\Length{#2}}\setcounter{YoungWidth}{\Length{#1}}%
\multiput(0,0)(10,0){#1}{\multiput(0,0)(0,10){#2}{\qbezier(2,2)(5,5)(8,8)\qbezier(2,8)(5,5)(8,2)}}\eep}}

\newcommand{\BlockCcB}[4]{{\YoungScale\Add{B3}{\Length{#2}}{\Length{#4}}\bep(\Length{#1},\value{B3})\put(0,\Length{#4}){\BlockCcA{#1}{#2}}%
\put(0,0){\BlockCcA{#3}{#4}}\setcounter{YoungHeight}{\value{B3}}\setcounter{YoungWidth}{\Length{#1}}\eep}}

\newcommand{\YoungCcA}{\BlockCcA{1}{1}}

\newcommand{\YoungCcC}{\BlockCcA{3}{1}}

\newcommand{\YoungCcAA}{\BlockCcA{1}{2}}

\newcommand{\YoungCcAAA}{\BlockCcA{1}{3}}
\newcommand{\YoungCcBAA}{\BlockCcB{2}{1}{1}{2}}

\renewcommand{\theequation}{\arabic{section}.\arabic{equation}}

\newcolumntype{x}[1]{%
>{\centering\hspace{0pt}}m{#1}}%
\newcolumntype{w}[1]{%
>{\raggedright\hspace{0pt}}m{#1}}%
\newcolumntype{z}[1]{%
>{\raggedleft\hspace{0pt}}m{#1}}%

\newcommand{\circledDigit}[1]{{\bep(12,12)\put(4,4){\circle{11}}\put(1.1,0){$#1$}\eep}}

\newcommand{\BYoung}[4]{{{\smallpic{\RectBRow{#1}{#2}{${\scriptstyle #3}$}{${\scriptstyle #4}$}}}}}

\newcommand{\CYoung}[6]{{{\smallpic{\RectCRow{#1}{#2}{#3}{${\scriptstyle #4}$}{${\scriptstyle #5}$}{${\scriptstyle #6}$}}}}}

\newcommand{\BYoungM}[4]{{{\minorpic{\RectBRow{#1}{#2}{${\scriptstyle #3}$}{${\scriptstyle #4}$}}}}}
\newcommand{\AYoungM}[2]{{{\minorpic{\RectARow{#1}{${\scriptstyle #2}$}}}}}
\newcommand{\CYoungM}[6]{{{\minorpic{\RectCRow{#1}{#2}{#3}{${\scriptstyle #4}$}{${\scriptstyle #5}$}{${\scriptstyle #6}$}}}}}

\newcommand{\emptypic}[1]{{\bep(0,#1)\eep}}

\newcommand{\CYYoung}[7]{{{\smallpic{\RectDYoung{#1}{#2}{#3}{${\scriptstyle #4}$}{${\scriptstyle #5}$}{${\scriptstyle #6}$}{#7}}}}}

\newcommand{\stackB}[2]{{\begin{tabular}{c} #1\\ #2\\ \end{tabular}}}

\newcommand{\msp}{{\mbox{msp}}}
\newcommand{\mspr}{{\mbox{mspr}}}
\newcommand{\hwp}{{\mbox{hwp}}}
\newcommand{\res}{{\mbox{Res}}}

\newcommand{\Rham}{{\mathcal{R}}}
\newcommand{\hRham}{{\mathsf{h}\mathcal{R}}}

\newcommand{\emptypar}[1]{{\parbox[c][#1][c]{0pt}{}}}

\newcommand{\piLambda}{{\ensuremath{\pi_{\boldsymbol{\Lambda}}}}}
\newcommand{\piDs}{{\ensuremath{\pi_\Ds}}}
\newcommand{\piCross}{{\ensuremath{\pi_{\boldsymbol{cr}}}}}
\newcommand{\piTot}{{\ensuremath{\pi_{\boldsymbol{tot}}}}}

\newcommand{\Y}[1]{{\ensuremath{\mathbb{Y}(#1)}}}
\newcommand{\Ya}[1]{{\ensuremath{\mathbb{Y}[#1]}}}
\newcommand{\Yb}[1]{{\ensuremath{\mathbb{Y}\{#1\}}}}
\newcommand{\Yy}{\ensuremath{\mathbf{Y}}}
\newcommand{\Xx}{\ensuremath{\mathbf{X}}}
\newcommand{\Zz}{\ensuremath{\mathbf{Z}}}
\newcommand{\Ss}{\ensuremath{\mathbf{S}}}
\newcommand{\Ll}{\ensuremath{\mathbf{L}}}
\newcommand{\Ds}{\ensuremath{\mathbf{A}}}

\newcommand{\Comp}{{\ensuremath{\mathcal{C}}}}

\newcommand{\DL}{{D}}
\newcommand{\DD}{{{\mathcal{D}}}}
\newcommand{\DO}{{D_{\Omega}}}

\newcommand{\coh}{{\mathsf{H}}}

\newcommand{\partition}[1]{{\ensuremath{\mathtt{P}(#1)}}}

\newcommand{\WW}{{\ensuremath{\mathcal{W}}}}

\theoremstyle{definition}

\theoremstyle{plain}
\newtheorem*{Theorem}{Theorem}

\newtheorem*{Corollary}{Corollary}

\theoremstyle{remark}

\newcommand{\Sigp}{{\ensuremath{{\boldsymbol \sigma}_+}}}
\newcommand{\Sigm}{{\ensuremath{{\boldsymbol \sigma}_-}}}
\newcommand{\Sigpm}{{\ensuremath{{\boldsymbol \sigma}_\pm}}}

\newcommand{\ComplexC}[3]{\ensuremath{0 \longrightarrow #1\longrightarrow #2 \longrightarrow #3 \longrightarrow 0}}

\newcommand{\AlgebraFont}[1]{\mathfrak{#1}}

\newcommand{\DSADS}{{\ensuremath{\boldsymbol{(A)dS_d}}}}
\newcommand{\dSAdS}{{\ensuremath{(A)dS_d}}}
\newcommand{\AdS}{{\ensuremath{AdS_d}}}
\newcommand{\dS}{{\ensuremath{dS_d}}}

\newcommand{\ads}{{\ensuremath{\AlgebraFont{so}(d-1,2)}}}
\newcommand{\ds}{{\ensuremath{\AlgebraFont{so}(d,1)}}}

\newcommand{\lorentz}{{\ensuremath{\AlgebraFont{so}(d-1,1)}}}

\newcommand{\msv}{{\ensuremath{\AlgebraFont{so}(d-1)}}}

\newcommand{\sod}{{\ensuremath{so(d)}}}
\newcommand{\sodd}{{\ensuremath{so(d+1)}}}
\newcommand{\soddd}{{\ensuremath{so(d+2)}}}
\newcommand{\son}{{\ensuremath{so(n)}}}

\newcommand{\sld}{{\ensuremath{sl(d)}}}

\newcommand{\sldd}{{\ensuremath{sl(d+1)}}}
\newcommand{\slddd}{{\ensuremath{sl(d+2)}}}

\newcommand{\sods}{{\ensuremath{\scriptstyle{so(d)}}}}

\newcommand{\slds}{{\ensuremath{\scriptstyle{sl(d)}}}}
\newcommand{\sldds}{{\ensuremath{\scriptstyle{sl(d+1)}}}}

\newcommand{\Verma}[2]{\ensuremath{\EuScript{D}\left({\textstyle{#1}};#2\right)}}

\newcommand{\Irrep}[2]{\ensuremath{\EuScript{H}\left({\textstyle{#1}};#2\right)}}

\newcommand{\pl}{\partial}

\newcommand{\fm}[1]{_{\mathbf{{#1}}}}

\newcommand{\be}{\begin{equation}}
\newcommand{\ee}{\end{equation}}

\newcommand{\Sa}{{\ensuremath{\mathbf{a}}}}

\newcommand{\aA}{{\ensuremath{\mathcal{A}}}}
\newcommand{\aB}{{\ensuremath{\mathcal{B}}}}
\newcommand{\aC}{{\ensuremath{\mathcal{C}}}}
\newcommand{\aD}{{\ensuremath{\mathcal{D}}}}
\newcommand{\aO}{{\ensuremath{\mathcal{O}}}}

\newcommand{\smallpic}[1]{{\unitlength=0.2mm#1}}
\newcommand{\minorpic}[1]{{\unitlength=0.25mm#1}}
\newcommand{\boldpic}[1]{{\linethickness{0.4mm}#1}}

\begin{document}
\renewcommand{\thefootnote}{\fnsymbol{footnote}}
{\begin{titlepage}
\begin{flushright}
\vspace{1mm}
FIAN/TD/23-09\\
\end{flushright}

\vspace{1cm}

\begin{center}
{\bf \Large Gauge fields in $\DSADS$ within the unfolded approach: algebraic aspects} \vspace{1cm}

\textsc{E.D. Skvortsov\footnote{skvortsov@lpi.ru}}

\vspace{.7cm}

{ I.E.Tamm Department of Theoretical Physics, P.N.Lebedev Physical Institute,\\Leninsky prospect 53, 119991, Moscow, Russia}
\end{center}

\vspace{0.5cm}

\begin{abstract}
It has recently been shown that generalized connections of the $\dSAdS$ symmetry algebra provide an
effective geometric and algebraic framework for all types of gauge fields in $\dSAdS$, both for
massless and partially-massless. The equations of motion are equipped with a nilpotent operator
called $\sigma_-$ whose cohomology groups correspond to the dynamically relevant quantities like
differential gauge parameters, dynamical fields, gauge invariant field equations, Bianchi
identities etc. In the paper the $\sigma_-$-cohomology is computed for all gauge theories of this
type and the field-theoretical interpretation is discussed. In the simplest cases the
$\sigma_-$-cohomology is equivalent to the ordinary Lie algebra cohomology.
\end{abstract}
\end{titlepage}

\renewcommand{\thefootnote}{\arabic{footnote}}
\setcounter{footnote}{0}
\section*{Introduction}\setcounter{equation}{0}
This paper is devoted to studying gauge fields in the (anti)-de Sitter background within the
framework of the unfolded approach. The background (anti)-de Sitter space \dSAdS{} can have any
dimension $d\geq4$. The gauge fields under consideration are the fields of the most general spin
type, so-called mixed-symmetry fields \cite{Curtright:1980yk, Labastida:1987kw, Zinoviev:2002ye,
Bekaert:2002dt, Zinoviev:2003dd, Zinoviev:2003ix, Alkalaev:2003hc, Alkalaev:2006hq, Bekaert:2006ix,
Moshin:2007jt, Buchbinder:2007ix, Skvortsov:2008sh, Reshetnyak:2008gp, Skvortsov:2008vs,
Campoleoni:2008jq, Zinoviev:2008ve,
Alkalaev:2008gi,Zinoviev:2009vy,Skvortsov:2009zu,Zinoviev:2009gh,Alkalaev:2009vm}, whose spin
degrees of freedom are described by tensors having the symmetry of arbitrary Young diagrams.

The gauge fields in Minkowski space are presented only by massless fields with arbitrary spin. By
contrast, for given mixed-symmetry field in \dSAdS{} there are different types of gauge modes, each type appearing at certain critical value of the mass parameter
\cite{Metsaev:1995re, Metsaev:1997nj, Brink:2000ag}, so one may talk about different types of massless fields with the same spin. Only one member of the family of massless fields with the same spin is unitary in $\AdS$ \cite{Metsaev:1995re, Metsaev:1997nj, Brink:2000ag}. In addition there are partially-massless
fields \cite{Deser:1983mm, Deser:1983tm, Higuchi:1986wu, Deser:2001pe, Deser:2001xr, Deser:2001us,
Deser:2003gw, Zinoviev:2001dt, Zinoviev:2008ve, Boulanger:2008up, Boulanger:2008kw,
Zinoviev:2009vy}, which, due to higher derivative gauge transformation law, have more degrees of
freedom than massless fields but less than massive ones and have no counterparts in Minkowski case.

In this paper we study all types of gauge fields in $\dSAdS$, both unitary and nonunitary, in order to better understand
the peculiarity of the former within the framework of the unfolded approach. The goal will be to construct the nonlinear theory with fields of mixed-symmetry type in the spectrum, which is still lacking.

A new object, which can be referred to as generalized Yang-Mills field, shows up naturally in the
unfolded approach. Generalized Yang-Mills field (or generalized connection) of the (anti)-de Sitter
algebra is a degree-$q$ differential form over (anti)-de Sitter space with values in arbitrary
representation of its symmetry algebra, which is $\ds$ for de Sitter and $\ads$ for anti-de Sitter.
Since the space-time symmetry algebra is just an orthogonal algebra, which can be realized as a Lie
algebra of antisymmetric matrices, the ordinary Yang-Mills field emerges when $q=1$ and the
representation is an irreducible one on rank-two antisymmetric tensors. The ordinary Yang-Mills
field of the (anti)-de Sitter algebra is known \cite{MacDowell:1977jt,Stelle:1979aj} to describe
the (anti)-de Sitter gravity, i.e. the theory of massless spin-two field.

The main statement of \cite{Skvortsov:2009zu}, extending the results of \cite{Vasiliev:2001wa,
Alkalaev:2003qv, Skvortsov:2006at}, is that each gauge field in $\dSAdS$ can be described by
certain generalized Yang-Mills field (connection) of the (anti)-de Sitter algebra.

The unfolded approach provides an effective framework for field theories. The unfolded approach
\cite{Vasiliev:1988xc, Vasiliev:1988sa} is a reformulation of differential equations in first
order form by making use of the de Rham differential and exterior product of differential forms.
The underlying algebraic structure is the Free Differential Algebra \cite{Sullivan77,
D'Auria:1982nx, D'Auria:1982pm, Nieuwenhuizen:1982zf} whereto Lie algebras, their modules and
Chevalley-Eilenberg cohomology belong. The main achievements of the unfolded approach are the full
classical nonlinear theory of totally-symmetric massless fields of spins $s=0,1,...,\infty$ \cite{Vasiliev:1989yr,
Vasiliev:1990en, Vasiliev:2003ev} and the coordinate-free description of black-holes
\cite{Didenko:2008va, Didenko:2009tc}.

Every linearized unfolded system comes equipped with a nilpotent operator called \Sigm{},
representing the algebraic part of the generalized covariant derivative acting on the fields. The
\Sigm-cohomology group $\coh(\Sigm)$ contains all information about the dynamically relevant
independent quantities of a given unfolded system \cite{Lopatin:1987hz, Shaynkman:2000ts,
Vasiliev:2007yc, Skvortsov:2008vs}. Differential gauge parameters, dynamical fields, gauge
invariant equations of motion and Bianchi identities are the representatives of $\coh(\Sigm)$. The
\Sigm-cohomology is a nice tool which allows to avoid solving differential equations. This paper is
written to present the results on the \Sigm-cohomology for the unfolded equations describing
arbitrary-spin massless and partially-massless fields in (anti)-de Sitter space.

A typical linearized unfolded system consists of two parts coupled together via an appropriate
Chevalley-Eilenberg cocycle. The one containing the forms of degree higher than zero is referred to
as the gauge module and describes the gauge sector, another one containing zero-degree forms is
referred to as the Weyl module and describes the physical degrees of freedom.

The generalized Yang-Mills fields provide the explicit construction for the gauge module of every
gauge field in $\dSAdS$. We compute $\coh(\Sigm)$ for the gauge module, the result is presented in
section \ref{SbSGCSigmaResult}. The same technique, developed in Appendix B, can be applied to the
Weyl module, showing that the cohomology matching condition between the gauge module and the Weyl
module is fulfilled. Certain special cases were considered in \cite{Shaynkman:2000ts,
Alkalaev:2003qv, Skvortsov:2006at, Boulanger:2008kw}.

That massless fields in Minkowski space are the systems with the first class constraints and the
massive fields both in Minkowski and $\dSAdS$ are the systems with the second class constraints is
mirrored in certain dualities on $\coh(\Sigm)$. The gauge fields in $\dSAdS$ are the systems with
both the first and the second class constraints present. Therefore, the duality on $\coh(\Sigm)$ is
found to be more complicated.

A generic element of the \Sigm-complex $\Comp(\Sigm)$ is a differential form with values in some
finite-dimensional tensor representation of the Lorentz algebra, the latter results from the
restriction of the irreducible module of the (anti)-de Sitter algebra, in which a generalized
Yang-Mills field takes values, to its Lorentz subalgebra. The Lorentz algebra commutes with the
action of \Sigm. Therefore, the representatives of $\coh(\Sigm)$ are labelled by Young diagrams of
the Lorentz algebra. They correspond to the fields, gauge parameters, etc. of the minimal
formulation in terms of metric-like fields, which turns out to be very complicated in contrast to
the formulation in terms of generalized Yang-Mills fields.

Quite surprisingly, the symmetry types of the representatives of $\coh(\Sigm)$ turn out to be
given by what may be called 'the maximally symmetric part' of the tensor product, i.e. the
corresponding Young diagrams tend to be as symmetric as it is possible, the rest of diagrams that
are less symmetric label acyclic subcomplexes of $\Comp(\Sigm)$. Making essential use of Young
diagrams allows us to present the results in a simple form.

The paper is organized as follows. We begin in section \ref{SFieldsConnections} by presenting the
classification of the gauge fields in $\dSAdS$, including all types of massless and
partially-massless fields. Essential facts of the unfolded approach are quickly summarized in
section \ref{SUnfolding}, where a linear gauge theory with generalized Yang-Mills fields of the
(anti)-de Sitter algebra is defined. The \Sigm{} operator is discussed in section
\ref{SSigmaMinus}, where the results on the \Sigm-cohomology are stated, the proof is in Appendix
B. The field-theoretical interpretation of the \Sigm-cohomology is in section
\ref{SInterpretation}. In section \ref{SExampleTwoRow} the example of a two-row massless field is
given to illustrate the general formalism. The conclusions are in section \ref{SConclusions}.

\section*{Preliminaries}
Whereas only tensor representations are considered in the paper, we do not make any distinction
between irreducible finite-dimensional highest weight modules of \sld{} and \sod{} with highest
weight $(s_1,...,s_n,0_{n+1},...,0_\nu)$, $\nu=d-1$ for \sld{}, $\nu=[d/2]$ for \sod{}, and Young
diagrams of shape $\Y{s_1,...,s_n}$. For simplicity, we ignore the Young diagrams of height close
to the maximal admissible height $\nu$, so that it is assumed $n<\nu-1$, thus for \sod{} we will
not consider (anti)-self dual representations.

A Young diagram $\Xx$ can be defined in several ways: (1) by specifying the lengths of its rows
$\Xx=\Y{s_1,...,s_n}$ (row notation), $n$ being the number of rows and $s_i$ being the length of
the $i$-th row, $s_i\geq s_{i+1}$; (2) by specifying the widths and heights of its subblocks
\be\nonumber\Yb{(s_1,p_1),...,(s_N,p_N)}\equiv\Y{\overbrace{s_1,...,s_1}^{p_1},...,\overbrace{s_N,...,s_N}^{p_N}},\ee
where $s_i$ and $p_i$ are the width and the height of the $i$-th subblock, $N$ is the number of
subblocks; (3) by specifying the heights of columns $\Ya{h_1,...,h_n}$, $h_i$ being the height of
the $i$-th column, $h_i\geq h_{i+1}$.

Let $\mathfrak{f}$ be some orthogonal algebra $\mathfrak{so}(p,q)$, the orthogonal algebras of
interest being \msv, \lorentz, \ds{} and \ads. A tensor $C^\Xx$ of $\mathfrak{f}$ or \sld{} is said
to have the symmetry of some Young diagram $\Xx$ if its indices realize the irreducible
representation of the permutation group labelled by $\Xx$, which for \sld{} guarantees the
irreducibility of the tensor.

If a tensor $C^\Xx$ having the symmetry of $\Xx=\Y{s_1,...,s_n}$ needs to be written explicitly, it
is always taken in the symmetric basis, meaning that \circledDigit{1} it has $n$ groups of indices,
the $k$-th group containing $s_k$ indices; \circledDigit{2} it is manifestly symmetric with respect
to permutations of indices within any group; \circledDigit{3} it satisfies the Young condition \be
C^{a(s_1),...,b(s_k),...,bc(s_i-1),...,u(s_n)}\equiv0,\qquad 1\leq k<i\leq n,\nonumber\ee where a
group of symmetric indices is denoted by one letter with the number of indices indicated in
brackets, e.g. $a(s_1)\equiv a_1a_2...a_{s_1}$, and the (normalized) sum over all permutations of
two or more (groups of) indices denoted by the same letter is implied, e.g.
$b(s_k),...,bc(s_i-1)\equiv\frac1{(s_k+1)!}\sum_\sigma
b_{\sigma(1)}...b_{\sigma(s_k)},...,b_{\sigma(s_{k}+1)}c_1...c_{s_i-1}$.

A tensor of the orthogonal algebra is said to be an irreducible tensor having the symmetry of $\Xx$
iff in addition to having the symmetry of $\Xx$ it is completely traceless, i.e. contraction of the
invariant metric tensor with any two indices vanishes identically.

\section{Gauge Fields in (anti)-de Sitter space}\setcounter{equation}{0}\label{SFieldsConnections}
According to \cite{Skvortsov:2009zu}, which is a generalization of numerous results
\cite{Deser:1983mm, Higuchi:1986wu, Metsaev:1995re, Metsaev:1997nj, Brink:2000ag} concerned with
gauge fields in $\dSAdS$, any gauge field in $\dSAdS$ is completely determined by a triple
$(\Ss,q,t)$, where $\Ss$ is a finite-dimensional irreducible representation of the (anti)-de Sitter
'Wigner little algebra' \msv, specified by some Young diagram $\Ss=\Y{s_1,...,s_n}$; $q$ is an
integer in the range $1$,...,$n$ such that $(s_q-s_{q+1})>0$; $t$ is an integer in the range
$1$,...,$(s_q-s_{q+1})$, which is equal to the order of derivative in the gauge transformation law.

For a given triple $(\Ss,q,t)$ the irreducible module $\Irrep{E_0}{\Ss_0\equiv\Ss}$ of the
(anti)-de Sitter algebra that is referred to as a massless ($t=1$) or partially-massless ($t>1$)
field\footnote{In the de Sitter case the construction of \ds{} modules is different because the
corresponding representations of \ds{} are not of the lowest weight type. Nevertheless, the notion
of the lowest energy $E_0$ can be introduced \cite{Higuchi:1986wu,Deser:2003gw}. As for field
equations, the situation is more simple inasmuch as the change $\lambda^2\longrightarrow-\lambda^2$
makes the transition from \AdS{} to \dS.} is defined by the following exact sequence
\cite{Shaynkman:2004vu, Skvortsov:2009zu}
\be0\rightarrow\Verma{E_q}{\Ss_q}\longrightarrow...\longrightarrow\Verma{E_1}{\Ss_1}\longrightarrow\Verma{E_0}{\Ss_0}\longrightarrow\Irrep{E_0}{\Ss_0}\rightarrow0\label{ExactSequence},\ee
where for the anti-de Sitter case $\Verma{E'}{\Ss'}$ is a Verma module freely generated by
the positive grade operators of \ads{} from the vacuum $|E',\Ss'\rangle$ annihilated by the
negative grade generators of \ads{}, which is an irreducible representation of the maximal compact
subgroup $\AlgebraFont{so}(2)\oplus\msv$ of \ads{} defined by the lowest weights $E'$ and $\Ss'$ of
$\AlgebraFont{so}(2)$ and $\msv$, respectively.

The lowest weights $E_i$ and $\Ss_i$ of $\AlgebraFont{so}(2)\oplus\msv$ are defined as \begin{align} E_i&=\begin{cases} d+s_q-t-q-1,& i=0,\\
d+s_{q-i+1}-(q-i+1)-1, & i=1,...,q,\end{cases}\label{AllLevelsE}\\ \Ss_i&=\begin{cases} \Y{s_1,...,s_n}\equiv\Ss,& i=0,\\
\Y{s_1,s_2,...,s_{q-1},s_q-t,s_{q+1},...,s_n}, & i=1,\\
\Y{s_1,...,s_{q-i},s_{q-i+2}-1,...,s_q-1,s_q-t,s_{q+1},...,s_n}, & i=2,...,q-1, \\
\Y{s_2-1,s_3-1,...,s_q-1,s_q-t,s_{q+1},...,s_n}, & i=q.\label{AllLevelsS}\end{cases}\end{align}

Given $E'$ and $\Ss'=\Y{s'_1,...,s'_n}$, $\Verma{E'}{\Ss'}$ can be realized on the solutions of
\begin{align}
(\Box+{m'}^2)&C^{a(s'_1),...,u(s'_n)}=0, &&\label{FullSystemA}\\
\DL_m &C^{a(s'_1),...,mc(s'_i-1),...,u(s'_n)}=0,&&i=1,...,n,\label{FullSystemB}
\end{align}
where $C^{a(s'_1),...,u(s'_n)}(x)$ is an irreducible Lorentz tensor field having the symmetry of $\Ss'$, $\Box\equiv\DL^m \DL_m$
and the mass-like parameter ${m'}^2$ is related to $E'$ and $\Ss'$ as \be
{m'}^2=\lambda^2\left(E'(E'-d+1)-s'_1-...-s'_n\right).\label{WEMassFormula}\ee

Therefore, a field-theoretical model for $\Irrep{E_0}{\Ss_0}$ is given by the irreducible Lorentz field
$\phi^{\Ss}\equiv\phi^{a(s_1),...,u(s_n)}(x)$ having the symmetry of $\Ss$ and satisfying equations
(\ref{FullSystemA})-(\ref{FullSystemB}) with the mass-like parameter determined by $E_0$ and $\Ss_0\equiv\Ss$. The exactness of
$\Verma{E_1}{\Ss_1}\longrightarrow\Verma{E_0}{\Ss_0}\longrightarrow\Irrep{E_0}{\Ss_0}$ implies that at $E_0$ and $\Ss_0$
equations (\ref{FullSystemA})-(\ref{FullSystemB}) become invariant under gauge transformations of the form
\be\label{GaugeTransformations} \delta\phi^{a(s_1),...,u(s_n)}=\overbrace{D^c...D^c}^t
\xi^{a(s_1),...,b(s_{q-1}),c(s_q-t),d(s_{q+1}),...,u(s_n)}+...,\ee where '$...$' stands for certain lower derivative terms and
for the terms that project onto the Young symmetry $\Ss$. The gauge parameter $\xi^{\Ss_1}$ is an irreducible Lorentz tensor
field having the symmetry of $\Ss_1$ and satisfying equations analogous to (\ref{FullSystemA})-(\ref{FullSystemB}). The rest of
$\Verma{E_i}{\Ss_i}$ with $i=2,..,q$ corresponds to reducible gauge symmetries.

The exact sequence (\ref{ExactSequence}) can be extended to the right with
$\Verma{E_{-1}}{\Ss_{-1}}\longrightarrow\Verma{E_{-2}}{\Ss_{-2}}\longrightarrow...$, implying that for a field $\phi^{\Ss_0}$
with gauge transformations (\ref{GaugeTransformations}) one can construct the generalized Weyl tensor $C^{\Ss_{-1}}$ having the
symmetry of $\Ss_{-1}$, \cite{Boulanger:2008up}. $C^{\Ss_{-1}}$ is obtained by applying $(s_q-s_{q+1}-t+1)$ derivatives to $\phi^{\Ss_0}$. By definition,
the generalized Weyl tensor $C^{\Ss_{-1}}$ is the lowest order nontrivial on-mass-shell gauge invariant under
$\xi^{\Ss_1}$-transformations (\ref{GaugeTransformations}) tensor built from $\phi^{\Ss_0}$. That the Weyl tensor does exist, its
symmetry type and order of derivative follows from the analysis of $\coh(\Sigm)$ and, of course, from the structure of singular
vectors in $\Verma{E'}{\Ss'}$.

Consequently, the space of gauge invariant differential expressions constructed from $\phi^{\Ss_0}$ is generated by
(\ref{FullSystemA})-(\ref{FullSystemB}) and by the generalized Weyl tensor.

To illustrate, the Young diagrams $\Ss_1$ of the gauge parameter, $\Ss_0$ of the spin and $\Ss_{-1}$ of the generalized Weyl
tensor have the form
\begin{align}
&\Ss_1=\parbox{3.6cm}{{\bep(80,90)\unitlength=0.38mm%
\put(0,0){\RectCRowUp{4}{3}{2}{$s_{q+2}$}{$...$}{$s_n$}}%
\put(0,30){\RectCRowUp{8}{6}{5}{$s_{q-1}$}{$s_q-t$}{$s_{q+1}$}}
\put(0,60){\RectBRowUp{10}{9}{$s_1$}{$...$}}\eep}}%
&& \Ss_0=\parbox{3.6cm}{{\bep(80,90)\unitlength=0.38mm%
\put(0,0){\RectCRowUp{4}{3}{2}{$s_{q+2}$}{$...$}{$s_n$}}%
\put(0,30){\RectCRowUp{8}{7}{5}{$s_{q-1}$}{$s_q$}{$s_{q+1}$}}%
\put(0,60){\RectBRowUp{10}{9}{$s_1$}{$...$}}\eep}} &&
\Ss_{-1}=\parbox{3.6cm}{{\bep(80,90)\unitlength=0.38mm%
\put(0,0){\RectCRowUp{4}{3}{2}{$s_{q+2}$}{$...$}{$s_n$}}%
\put(0,30){\RectCRowUp{8}{7}{6}{$s_{q-1}$}{$s_q$}{$s_{q}-t+1$}}%
\put(0,60){\RectBRowUp{10}{9}{$s_1$}{$...$}}\eep}}\nonumber\end{align}

\section{Unfolded approach}\setcounter{equation}{0}\label{SUnfolding}
\paragraph{General definition.} A set of differential equations is said to have the unfolded form
\cite{Vasiliev:1988xc, Vasiliev:1988sa, Vasiliev:1992gr} if it can be written as the zero curvature
condition \be\label{UnfldEquations} R^\aA\equiv d W^\aA + F^\aA(W)=0,\ee where $W^\aA$ is a set of
differential forms on some manifold $\mathcal{M}_d$ with values in vector spaces labelled by $\aA$,
so that indices $\aA,\aB,...$ indicate the vector space rather than components in a particular
basis. $|\aA|$ is the form-degree of $W^\aA$, $d$ is the exterior differential on $\mathcal{M}_d$
and $F^\aA(W)$ is a degree-$(|\aA|+1)$ function of $W$ assumed to be expandable in terms of
exterior (wedge) products only\footnote{The wedge symbol $\wedge$ will be systematically omitted
henceforth.} \be F^\aA(W)=\sum_{n=1}^{\infty}\sum_{|\aB_1|+...+|\aB_n|=|\aA|+1} f^\aA_{\phantom{\aA}
\aB_1 ... \aB_n}W^{\aB_1}\wedge...\wedge W^{\aB_n},\ee where $f^\aA_{\phantom{\aA} \aB_1 ...
\aB_n}$ are some $x$-independent elements of $\mbox{Hom}(\aB_1\otimes...\otimes\aB_n,\aA)$. In
order to guarantee the formal consistency of (\ref{UnfldEquations}) $F^\aA(W)$ must satisfy the
integrability condition (referred to as either the generalized Jacobi identity or the Bianchi
identity) obtained by applying $d$ to (\ref{UnfldEquations}) \be\label{UnfldBianchi}
F^\aB\frac{\delta F^\aA}{\delta W^\aB}\equiv0.\ee Any solution of (\ref{UnfldBianchi}) defines a
free differential algebra (FDA) \cite{Sullivan77, D'Auria:1982nx, D'Auria:1982pm,
Nieuwenhuizen:1982zf}. If the Jacobi identity (\ref{UnfldBianchi}) is satisfied regardless of
$\mathcal{M}_d$ dimension, the free differential algebra is referred to as universal
\cite{Bekaert:2005vh, Vasiliev:2007yc}. It is the universal algebras only that will be considered
henceforth. Equations (\ref{UnfldEquations}) are invariant under the gauge transformations
\begin{align}\label{UnfldGauge}\delta W^\aA&=d\epsilon^\aA-\epsilon^\aB\frac{\delta F^\aA}{\delta W^\aB},&& \mbox{if }|\aA|>0,\\
\label{UnfldGaugeZeroForm}\delta W^\aA&=-\epsilon^{\aB'}\frac{\delta F^\aA}{\delta W^{\aB'}},&& \aB': |\aB'|=1,\qquad \mbox{if
}|\aA|=0,\end{align} where $\epsilon^\aA$ is a degree-$(|\aA|-1)$ form taking values in the same space as $W^\aA$.

\paragraph{Linearization.} In what follows we consider the linearized unfolded systems. Let the base manifold $\mathcal{M}_d$
be a homogeneous space\footnote{The homogeneous spaces of interest are given by Minkowski space
$\mathfrak{G}=ISO(d-1,1)$, $\mathfrak{H}=SO(d-1,1)$, anti-de Sitter space $\mathfrak{G}=SO(d-1,2)$,
$\mathfrak{H}=SO(d-1,1)$, de Sitter space $\mathfrak{G}=SO(d,1)$, $\mathfrak{H}=SO(d-1,1)$, and by
the space with $\mathfrak{G}=SP(8)$ and $\mathfrak{H}$ being the maximal parabolic subgroup of
$\mathfrak{G}$, in which the symmetries of $4$d higher-spin fields gets realized geometrically
\cite{Fronsdal:1985pd, Vasiliev:2001dc}.} $\mathcal{M}_d=\mathfrak{G}/\mathfrak{H}$, with
$\mathfrak{g}$ and $\mathfrak{h}$ being the Lie algebras of $\mathfrak{G}$ and $\mathfrak{H}$.
Typically, $\mathfrak{h}$ is the Lorentz algebra, \lorentz. The most general unfolded equations
linearized over $\mathcal{M}_d$ have the form \cite{Barnich:2004cr, Barnich:2006pc, Vasiliev:2007yc}
\begin{align} R^I&=d\Omega^I+f^I_{JK}\Omega^J\Omega^K=0,\label{UnfldFlatness}\\ R^\aA&=d
W^\aA+{f}^\aA_{\phantom{\aA}\aB}W^\aB+\Omega^I{f_I}^\aA_{\phantom{\aA}\aB}W^\aB+...
+\Omega^{I_1}...\Omega^{I_k}{f_{I_1...I_k}}^\aA_{\phantom{\aA}\aB}W^\aB=0\label{UnfldLinearized},
\end{align} where $\Omega$ is a one-form connection of $\mathfrak{g}$ with $\Omega^I$ being the components of $\Omega$ in some base,
$f^I_{JK}$ are the structure constants of $\mathfrak{g}$; ${f_{I_1...I_k}}^\aA_{\phantom{\aA}\aB}\equiv0$ unless
$|\aA|+1=|\aB|+k$. The set of fields consists of two subsets: degree-one forms $\Omega^I$ that are assumed to have the zeroth
order and the matter fields $W^\aA$, which may have various degrees, are assumed to have the first order so that the unfolded
equations are linear in $W^\aA$. $\Omega^I$ can be recognized as the Cartan connection on $\mathcal{M}_d$, it describes the
background geometry by virtue of the flatness condition (\ref{UnfldFlatness}).

The Jacobi identity (\ref{UnfldBianchi}) implies that ${f_{I_1...I_k}}^\aA_{\phantom{\aA}\aB}$ are closely related to the Lie
algebra $\mathfrak{g}$ of the space-time symmetry group $\mathfrak{G}$. Namely, ${f_{I_1...I_k}}^\aA_{\phantom{\aA}\aB}$ is a
Chevalley-Eilenberg $k$-cocycle of $\mathfrak{g}$ with coefficients in $\mbox{Hom}(\aB,\aA)$ \cite{Vasiliev:2007yc}. If $k=1$ and
$\aA=\aB$, ${f_I}^\aA_{\phantom{\aA}\aA}$ is just a representation of $\mathfrak{g}$ in the vector space $\aA$. Coboundary
cocycles can be removed from the equations (\ref{UnfldLinearized}) by a nonsingular field redefinition, thus,
${f_{I_1...I_k}}^\aA_{\phantom{\aA}\aB}$ can be assumed to be nontrivial representatives of Chevalley-Eilenberg cohomology
groups.

Having fixed the background connection $\Omega^I$, it is useful to introduce \cite{Barnich:2004cr, Skvortsov:2008vs} the
generalized\footnote{If $\DD$ consists only of the expression in brackets, it reduces to the ordinary covariant derivative.}
covariant derivative $\DD$ \be\label{UnfldGeneralizedDerivative}\DD={f}^\aA_{\phantom{\aA}\aB}+(\delta^{\aA}_{\aB}d
+\Omega^I{f_I}^\aA_{\phantom{\aA}\aB})+... +\Omega^{I_1}...\Omega^{I_k}{f_{I_1...I_k}}^\aA_{\phantom{\aA}\aB},\ee which acts on
the whole space $\WW\fm{q}$ of matter fields \be\WW\fm{q}=\{W^\aB,W^\aC,...,W^\aD\},\qquad\quad q=\max_\aA|\aA|\ee We define
$\WW\fm{q\pm i}$ to be the spaces of differential forms with values in the same vector spaces as $\WW\fm{q}$ but with the form
degrees shifted by $\pm i$. If for some $\aB$ and $i$ we have $|\aB|-i<0$ the corresponding element of $\WW\fm{q-i}$ becomes
trivial.

In this special case, the nilpotency of $\DD$, $\DD^2=0$, is equivalent to the generalized Jacobi
identity (\ref{UnfldBianchi}). Then, the gauge transformations for a matter field $W\fm{q}$,
$W\fm{q}\in\WW\fm{q}$ read $\delta W\fm{q}=\DD\xi\fm{q-1}$ with $\xi\fm{q-1}\in\WW\fm{q-1}$. The
gauge invariant field curvature $R\fm{q+1}=\DD W\fm{q}$ belongs to $\WW\fm{q+1}$, the rest of
$\WW\fm{q\pm i}$ corresponds either to reducible gauge symmetries or to reducible Bianchi
identities. Therefore, we have the unfolded complex $\Comp(\WW,\DD)$,
\be\label{UnfldComplex}\WW\fm{0}\xrightarrow{\phantom{a}\DD\phantom{a}}\WW\fm{1}\xrightarrow{\phantom{a}\DD\phantom{a}}...
\xrightarrow{\phantom{a}\DD\phantom{a}}\WW\fm{q-1}\xrightarrow{\phantom{a}\DD\phantom{a}}
\WW\fm{q}\xrightarrow{\phantom{a}\DD\phantom{a}}...\quad.\ee We can always split $\Comp(\WW,\DD)$
as \be\Comp(\WW,\DD)=\Comp(\WW^{\mbox{contr}},\DD)\oplus\Comp(\WW^{\mbox{gauge}},\DD)
\supsetplus\Comp(\WW^{\mbox{Weyl}},\DD),\ee where $\WW^{\mbox{contr}}$ is a contractible part
\cite{Sullivan77, Boulanger:2008up}, which can be consistently set to zero. The equations
$\DD\WW^{\mbox{contr}}$ are of the form $d W^1+W^2+...=0$ so that $W^1$ can be gauged away. In what
follows contractible parts will never appear. $\WW^{\mbox{gauge}}$ is referred to as the gauge
module, it contains the forms of degree greater than zero, which are necessarily gauge fields by
virtue of (\ref{UnfldGauge}). The zero degree forms constitute the Weyl module $\WW^{\mbox{Weyl}}$,
which carries physical degrees of freedom since the field equations (\ref{UnfldEquations}) can be
treated as a cocycle condition, having pure gauge solutions in the sector of $k$-forms with $k>0$
by virtue of the Poincare lemma, hence, only zero degree forms parameterize a solution to
(\ref{UnfldEquations}). The semidirect sum sign $\supsetplus$ is due to the Chevalley-Eilenberg cocycle
that glues the Weyl module to the gauge module.

\paragraph{Generalized Yang-Mills connections of the (anti)-de Sitter algebra.}
Below we define a family of gauge modules which is natural to consider in (anti)-de Sitter space.

Let $\Omega$ be a flat connection of the (anti)-de Sitter algebra $\mathfrak{g}$,
$\mathfrak{g}=\ds$ (de Sitter) and $\mathfrak{g}=\ads$ (anti-de Sitter). Given an arbitrary
irreducible representation of the (anti)-de Sitter algebra $\Ds$ we define the complex
$\Comp(\Ds,\DO)$
\begin{align}
& \WW^\Ds\fm{0}&&\xrightarrow{\DO}&&...&&\xrightarrow{\DO}&&\WW^\Ds\fm{q-1}&&\xrightarrow{\DO}&&\WW^\Ds\fm{q}
&&\xrightarrow{\DO}&&\WW^\Ds\fm{q+1}&&\xrightarrow{\DO}&&\WW^\Ds\fm{q+2}&&\xrightarrow{\DO}&&...,\nonumber
\end{align}
where $\WW^\Ds\fm{i}$ is a $\mathfrak{g}$-module of degree-$i$ differential forms with values in
$\Ds$. The flatness condition implies the nilpotency of $\DO$, $\DO^2=0$. When the dimension of
$\dSAdS$ is reached, $\WW^\Ds\fm{i}$ becomes trivial, i.e. $\WW^\Ds\fm{i}=\emptyset$ if $i>d$.

Given a distinguished degree $q>0$, for the gauge field $W^\Ds\fm{q}\in\WW\fm{q}^\Ds$ one can
easily define \cite{Vasiliev:2001wa, Alkalaev:2003qv, Skvortsov:2006at} the field curvature
$R^\Ds\fm{q+1}=\DO W^\Ds\fm{q}$ that is invariant under the gauge transformations $\delta
W^\Ds\fm{q}=\DO \xi^\Ds\fm{q-1}$ and satisfies the Bianchi identity $\DO R^\Ds\fm{q+1}\equiv0$. The
lower degree elements $\xi^\Ds\fm{q-i}$ of $\WW^\Ds\fm{q-i}$, $i=2,...,q$ of the complex
$\Comp(\Ds,\DO)$ correspond to the reducible gauge transformations $\delta \xi^\Ds\fm{q-i+1}=\DO
\xi^\Ds\fm{q-i}$. The higher degree elements $\WW^\Ds\fm{q+i}$, $i=2,...$ correspond to the
reducibility of Bianchi identities. Thus, $\Comp(\Ds,\DO)$ is a particular realization of
$\Comp(\WW^{\mbox{gauge}},\DD)$ with $\DD=\DO$.

\paragraph{\DSADS-background geometry.} The connection $\Omega$ can be presented in components by a one-form
$\Omega^{A,B}\equiv\Omega^{A,B}_\mu dx^\mu$ antisymmetric in its fiber indices of $\mathfrak{g}$,
$\Omega^{A,B}=-\Omega^{B,A}$, $A,B,...=0,...,d$, with the flatness condition having the form \be
d\Omega^{A,B}+\Omega^{A,}_{\phantom{A,}C}\wedge\Omega^{C,B}=0\label{AntideSitterFlatness}.\ee

To interpret the fields in terms of the Lorentz algebra the manifest local (anti)-de Sitter
symmetry must be lost. The local Lorentz algebra \lorentz{} is identified with the subalgebra of
the local (anti)-de Sitter algebra that annihilates some vector field $V^A(x)$, called compensator
\cite{Stelle:1979aj, Vasiliev:2001wa}, which is convenient to normalize as\footnote{Upper/lower
sign corresponds to the de Sitter/anti-de Sitter case hereinafter.} $V^BV_B=\mp 1$. The generalized
vielbein field $E^A_\mu dx^\mu$ \be\lambda E^A=\DO
V^A=dV^A+\Omega^{A,}_{\phantom{A,}B}V^B\label{UnfldGeneralizedE}\ee is required to have the maximal
rank, thus giving rise to a nonsingular vielbein field $h^a_\mu$. $E^BV_B=0$ by virtue of
(\ref{UnfldGeneralizedE}). The connection of the Lorentz algebra \be\Omega_L^{A,B}=\Omega^{A,B}\mp
\lambda(V^AE^B-E^AV^B)\label{BGLorentzConnection}\ee allows to define the Lorentz covariant
derivative $\DL=d+\Omega_L$. Both the compensator $V^A$ and the generalized vielbein $E^A$ are
Lorentz-covariantly constant \be\DL V^A=0, \qquad \DL E^A=0.\label{BGCovariantConstancy}\ee For the
further convenience we introduce
\be\Omega^{A,B}=\Omega_L^{A,B}+\nabla_-^{A,B}-\nabla_+^{A,B},\qquad \nabla_-^{A,B}=\pm \lambda
V^AE^B, \quad \nabla_+^{A,B}=\pm\lambda E^AV^B.\label{ConnectionDecomposition}\ee

One can always choose the 'standard gauge' for the compensator field $V_A=\delta^A_{\bullet}$, then $\lambda
E^A=\Omega^{A,}_{\phantom{A,}\bullet}$, $E^\bullet_\mu=0$ and $\Omega_L^{a,b}=\Omega^{a,b}$, so that the vielbein field $h^a_\mu$
and the Lorentz spin-connection $\varpi^{a,b}$ are defined as
\begin{align} \lambda h^a=\Omega^{a,}_{\phantom{a,}\bullet}, && \varpi^{a,b}=\Omega^{a,b}.\label{UnfldSplitting}\end{align}
In terms of $h^a$ and $\varpi^{a,b}$ the flatness condition (\ref{AntideSitterFlatness}) reads
\begin{align}
        &&&dh^a+\varpi^{a,}_{\phantom{a,}b}\wedge h^b=0,&&&&\label{UnfldIsoFlatnessA}\\
        &&&d\varpi^{a,b}+\varpi^{a,}_{\phantom{a,}c}\wedge\varpi^{c,b}\pm\lambda^2h^a\wedge h^b=0.&&&&\label{UnfldIsoFlatnessB}
\end{align}

Having identified the Lorentz algebra together with the Lorentz covariant derivative $\DL$, the
complex $\Comp(\Ds,\DO)$ can be interpreted in terms of (generalized) connections of the Lorentz
algebra. Passing to connections of the Lorentz algebra, the manifest (anti)-de Sitter symmetry gets
lost. After converting the connections of the Lorentz algebra into fully world or fully fiber
(metric-like) tensors with the help of the background vielbein $h^a_\mu$, in terms of metric-like
fields the gauge theory acquires a very complicated form due to Young symmetrizers and because of a
large number of component metric-like fields, most of which are auxiliary or pure gauge. The
\Sigm-technique allows us to find out which fields are dynamical and hence to give an
interpretation of $\Comp(\Ds,\DO)$ in terms of the metric-like fields of section
\ref{SFieldsConnections}.

\paragraph{Unfolding gauge fields in (anti)-de Sitter.} We assume $\mathfrak{g}$ is \ds{} or \ads, the Cartan connection $\Omega^I$
is presented by a one-form $\Omega^{A,B}\equiv\Omega^{A,B}_\mu dx^\mu$ with the flatness condition
(\ref{UnfldFlatness}) having the form (\ref{AntideSitterFlatness}). The unfolded equations for a
gauge field $(\Ss,q,t)$ are expected to have the following form, which is a special case of the
unfolded complex (\ref{UnfldComplex}),
\begin{align}
\DO W^{\Ds}\fm{q}&=f(E,...,E)(C\fm{0}), \label{UnfoldedGaugeModule}\\
\widetilde{\DO} C\fm{0}&=0, \label{UnfoldedWeylModule}
\end{align}
where $W^\Ds\fm{q}$ is a $\mathfrak{g}$-connection (\ref{GaugeFieldConnection}) associated with $(\Ss,q,t)$-field, $C\fm{0}$ is a certain infinite-dimensional $\mathfrak{g}$-module, the Weyl module, whose restriction to the Lorentz algebra yields a direct sum of an infinite number of irreducible Lorentz tensors. $f(E,...,E)$ is a Chevalley-Eilenberg cocycle gluing the gauge module to
the Weyl module. $\widetilde{\DO}$ is a $\mathfrak{g}$-covariant derivative in the Weyl module. Note that only finitely many
Lorentz modules constituting the Weyl module are glued with the gauge module. These are given by the Weyl tensor together with
certain of its descendants.

The explicit constructions known up-to-date are given by a massless spin-$s$ field
\cite{Vasiliev:1988xc, Vasiliev:1990en, Vasiliev:1992gr, Vasiliev:2001wa, Vasiliev:2003ev};
partially-massless spin-$s$ fields can be rewritten in the same way since a simple change of variables gives all the coefficients of $\widetilde{\DO}$ from those of $\DO$, \cite{Skvortsov:2006at}; for the series $(\Ss,q_{min},1)$, where $q_{min}$
is the height of the shortest column of $\Ss$, the free unfolded equations were obtained in
\cite{Boulanger:2008up, Boulanger:2008kw}. However, it is still lacking for arbitrary-spin massless
and partially-massless gauge fields in $\dSAdS$. We expect it may be extracted analogously to the
series $(\Ss,q_{min},1)$ from the unfolded equations for massive arbitrary-spin field in $\dSAdS$
of \cite{Boulanger:2008up, Boulanger:2008kw}, which are obtained as the constrained radial reduction of unfolded
equations for massless fields in Minkowski space \cite{Skvortsov:2008vs}. However, it is not obvious at the moment how the equations can be cast into the form (\ref{UnfoldedGaugeModule}). The equations for the Weyl module (\ref{UnfoldedWeylModule}) were given in \cite{Boulanger:2008up, Boulanger:2008kw} for arbitrary case together with the constraints that single out the different cases, i.e. massive, massless or partially-massless.
\section{The Sigma-minus operator}\setcounter{equation}{0}\label{SSigmaMinus}
The unfolded form of any field-theoretical system has many advantages in that it is formulated in terms of connections of the
space-time symmetry algebra. As compared to the minimal formulation in terms of metric-like fields the unfolded form requires
more component fields with many of them playing auxiliary role. Therefore, given some unfolded equations whose field-theoretical
interpretation is not clear or while unfolding some known field system there comes the question of what fields are the true
dynamical ones and what gauge parameters are the true differential ones, etc.

A natural gauge sector of an unfolded complex in (anti)-de Sitter space is presented by the complex
$\Comp(\Ds,\DO)$ of gauge connections of the (anti)-de Sitter algebra. Actually, every
finite-dimensional irreducible gauge module is given by some $W^\Ds\fm{q}$. With the help of the
\Sigm-cohomology technique \cite{Lopatin:1987hz, Shaynkman:2000ts, Vasiliev:2007yc,
Skvortsov:2008vs} we classify all dynamically relevant independent quantities in $\Comp(\Ds,\DO)$,
which gives the full list of dynamical fields contained in $W^\Ds\fm{q}$, differential gauge
parameters in $\xi^\Ds\fm{q-1}$ and the gauge invariant equations that can be imposed on
$W^\Ds\fm{q}$ in terms of $R^\Ds\fm{q+1}$.

The initial data for \Sigm{} are a flat connection $\Omega$ of the (anti)-de Sitter algebra
$\mathfrak{g}$ together with an irreducible $\mathfrak{g}$-module $\Ds$. The starting point is
that, according to (\ref{ConnectionDecomposition}), the (anti)-de Sitter covariant derivative $\DO$
splits as \be \DO=\DL+\nabla_--\nabla_+,\label{ConnectionSplitting}\ee where $(\nabla_\pm)$ are
nilpotent algebraic operators, $(\nabla_\pm)^2=0$. The operators $\nabla_\pm$ preserve Young
symmetry properties.

First, in \ref{SbSInterpretation} we introduce the operator \Sigm{} in abstract terms and give an
overview of the well-known facts on the interpretation of the \Sigm-cohomology
\cite{Lopatin:1987hz, Shaynkman:2000ts, Vasiliev:2007yc, Skvortsov:2008vs}. Then, to strictly
define \Sigm{} for the complex $\Comp(\Ds,\DO)$ of gauge connections we need some details about the
restriction of irreducible modules , which are in section \ref{SbSRestriction}. Before giving a
formal definition for \Sigm{} in \ref{SbGCSigmaMinusDef}, two simple examples are considered in
\ref{SbSTwoSimpleCases}. To present the result on \Sigm{}-cohomology in a simple form, which is
done in \ref{SbSGCSigmaResult} with the proof being in Appendix B, we define the highest weight
part and the maximally symmetric part of tensor products in \ref{SbSHWPMSP} and introduce in
\ref{SbSRestrictionTensorProduct} a special structure joining restriction of modules with maximally
symmetric part of tensor products. A number of useful examples on \Sigm-cohomology is given in
\ref{SbSExamplesSigmaCohomology}.

\subsection{Interpretation of cohomology}\label{SbSInterpretation}
Suppose that the field content of some unfolded system of equation is given by a graded collection
of degree-$q$ differential forms\footnote{The analysis is not affected if the form degree varies
with the grade, $\omega^g\fm{q_g}$, as it occurs for massless fields in Minkowski space
\cite{Skvortsov:2008vs}.} $\omega^g\fm{q}$, $g=0,1,...$. The corresponding gauge parameters are the
forms of degree-$(q-1)$ with the values in the same spaces as $\omega^g\fm{q}$. If $q>1$ there are
reducible gauge symmetries with parameters $\xi^g\fm{q-i}$, $i=2,...,q$. Suppose also that the
gauge transformations, the field curvatures and the Bianchi identities have the form, which is a
special case of the unfolded complex (\ref{UnfldComplex}),
\begin{align} &\hspace{3cm}&\delta \xi^g\fm{1}&=&&\DL\xi^k\fm{0}\phantom{\scriptstyle -2|}&&+&&\Sigm\left(\xi^{g+1}\fm{0}\right), && \hspace{3cm}\\
&&...&=&&...,&&&&&&\\
&&\delta \xi^g\fm{q-1}&=&&\DL\xi^g\fm{q-2}&&+&&\Sigm\left(\xi^{g+1}\fm{q-2}\right),&&\\
&&\delta \omega^g\fm{q}&=&&\DL\xi^g\fm{q-1}&&+&&\Sigm\left(\xi^{g+1}\fm{q-1}\right),&&\\
&&R^g\fm{q+1}&=&&\DL\omega^g\fm{q}&&+&&\Sigm\left(\omega^{g+1}\fm{q}\right),&&\\
&&0&=&&\DL R^g\fm{q+1}&&+&&\Sigm\left(R^{g+1}\fm{q+1}\right), && \end{align} where \Sigm{} is an
algebraic operator that decreases the grade by one and increases the form degree by one. The only
differential part is in the Lorentz covariant derivative $\DL$. The formal consistency of the
system requires (1) $\{\DL,\Sigm\}=0$, which trivially holds in the systems of interest since
\Sigm{} is built of the background vielbein and $\DL h^a=0$ is equivalent to
(\ref{UnfldIsoFlatnessA}); (2) \Sigm{} is a nilpotent operator, $\Sigm^2=0$.

Having a nilpotent operator suggests the cohomology problem. The \Sigm-cohomo\-lo\-gy turns out to
have a very clear field-theoretical meaning, classifying all dynamically relevant independent
quantities. Indeed, the degree-$k$, $k=1,...,q-1$, gauge parameters $\xi^{g}\fm{k}$ that are
\Sigm-exact can be gauged away with the help of the reducible algebraic gauge symmetry with
$\xi^{g+1}\fm{k-1}$. The leftover gauge symmetry $0=\delta \xi^{g}\fm{k}=\DL \xi^{g}\fm{k-1}
+\Sigm\left(\xi^{g+1}\fm{k-1}\right)$ just expresses $\xi^{g+1}\fm{k-1}$ modulo \Sigm-closed part
in terms of $\xi^{g}\fm{k-1}$. Therefore, the true differential gauge parameters are \Sigm-closed
and are not \Sigm-exact thus being representatives of the \Sigm-cohomology groups
$\coh^{k}(\Sigm)$. Quite analogously, the dynamical fields, i.e. the fields that cannot be gauged
away by some algebraic gauge symmetry and are not expressed in terms of derivatives of some other
fields, are the representatives of $\coh^{q}(\Sigm)$. The independent gauge invariant differential
expressions are the representatives of $\coh^{q+1}(\Sigm)$. The nontrivial (reducible) Bianchi
identities are the representatives of $\coh^{q+j}(\Sigm)$, $j=2,...$ . We sum up the interpretation
of $\coh(\Sigm)$ in the table below\parbox[t][6pt][c]{0pt}{}\par\noindent
\begin{tabular}{|c|p{10.8cm}|}\hline
    cohomology group & \multicolumn{1}{c|}{\rule{0pt}{13pt}interpretation} \\ \hline\hline
    $\coh^{q-i},\ i=1,...,q$ &  \rule{0pt}{13pt}differential gauge parameters at the $i$-th level of reducibility \\ \hline
    $\coh^{q\phantom{+1}}$ & \rule{0pt}{13pt}dynamical fields \\ \hline
    $\coh^{q+1}$ & \rule{0pt}{13pt}independent gauge invariant equations on dynamical fields \\ \hline
    $\coh^{q+i+1},\ i=1,...$ & \rule{0pt}{13pt}Bianchi identities at the $i$-th level \\ \hline
\end{tabular}\par\rule{0pt}{14pt}
The maximal number of derivatives connecting two elements of $\coh^{q'}_{g_1}$ and
$\coh^{q'+1}_{g_2}$ is equal to\footnote{In the (anti)-de Sitter case two covariant derivatives
might appear in the form of a commutator, which is an algebraic expression and not a second order
differential operator. Fortunately, we are able to trace the appearance of such commutator terms.}
$(g_2-g_1+1)$, meaning that if some representatives of $\coh^{q'}_{g_1}$ and $\coh^{q'+1}_{g_2}$
correspond to a parameter and a dynamical field then the gauge transformation law contains
$(g_2-g_1+1)$ derivatives; if $\coh^{q'}_{g_1}$ and $\coh^{q'+1}_{g_2}$ correspond to a dynamical
field and a gauge invariant equation then the latter is up to $(g_2-g_1+1)$-th order in derivative,
etc.

\subsection{Restriction of irreducible modules}\label{SbSRestriction}
When formulated in terms of Young diagrams the restriction rules are the same\footnote{It is worth
stressing that at least one of the weights $(s_1,...,s_\nu)$ of $\mathfrak{so}(2\nu+1)$ or
$\mathfrak{so}(2\nu)$ must be zero in order to get rid of (anti)-selfdual representations both for
the (anti)-de Sitter algebra and its Lorentz subalgebra, which is implied.} both for \sldd{} and
\sodd. Let $\mathfrak{g}$ and $\mathfrak{h}$ denote either \sldd{} and \sld{} or \sodd{} and \sod.

When restricted to the subalgebra $\mathfrak{h}\subset\mathfrak{g}$, irreducible finite-dimensional
representations of $\mathfrak{g}$ decompose into a direct sum of irreducible representations of
$\mathfrak{h}$. We denote this functor as $\res^{\mathfrak{g}}_{\mathfrak{h}}\Xx$, where $\Xx$ is a
Young diagram that determines an irreducible representation of $\mathfrak{g}$. The result of
applying $\res^{\mathfrak{g}}_{\mathfrak{h}}$ to $\Xx=\Y{s_1,...,s_n}$  reads \cite{Barut}
\be\res^{\mathfrak{g}}_{\mathfrak{h}}\Xx=\bigoplus_{k_1,...,k_N}\Xx_{\{k_1,...,k_N\}},\ee where the
multiplicity of each irreducible module $\Xx_{\{k_1,...,k_N\}}$ is one and
\be\label{AppRestrictionFormula}
\Xx_{\{k_1,...,k_N\}}=\begin{cases}\Y{k_1,...,k_n}, & k_1\in[s_2,s_1],...,k_{n-1}\in[s_n,s_{n-1}], k_n\in[0,s_n],\\
\emptyset, & \mbox{otherwise}.\end{cases}\ee So, the result of the restriction is given by various
Young diagrams obtained by removing an arbitrary number of cells from the right of rows of $\Xx$
provided that the length of each shortened row is not less than the length the next row of $\Xx$.

$\res^{\mathfrak{g}}_{\mathfrak{h}}\Xx$ is endowed with a natural structure of a graded vector
space if to each element $\Xx_{\{k_1,...,k_N\}}$ we assign the grade
$g=k_1+k_2+...+k_N-s_2-...-s_n$, so that the grade of the minimal rank element of
$\res^{\mathfrak{g}}_{\mathfrak{h}}\Xx$ is $0$ and the grade of the maximal rank element of
$\res^{\mathfrak{g}}_{\mathfrak{h}}\Xx$ is $s_1$.

It is convenient to define generally reducible $\mathfrak{h}$-modules $\Xx_g$ to be a direct sum of irreducible
$\mathfrak{h}$-modules $\Xx_{\{k_1,...,k_N\}}$ having the same grade $g$ (the same rank)
\be\Xx_g=\bigoplus_{k_1+...+k_n=g}\Xx_{\{k_1,...,k_N\}},\ee so that $\res^{\mathfrak{g}}_{\mathfrak{h}}\Xx=\bigoplus_g\Xx_g$.

Among $\Xx_{\{k_1,...,k_N\}}$ there are elements $\Xx^m$ (those having first $(m-1)$ rows $k_1,...,k_{m-1}$ of maximal length
with the rest of rows $k_m,...,k_n$ having minimal length) that
can be referred to as maximally symmetric\be\label{MaxSymRestriction}\Xx^m=\begin{cases}\Y{s_1,...,\widehat{s_{m}},...,s_n}, & m=1,..,n,\\
\Y{s_1,...,s_n}, & m>n,
\end{cases}\ee i.e. as if the $m$-th weight is thrown away. The grade of maximally symmetric elements $\Xx^k$ is given by \be
g(\Xx^k)=\begin{cases}s_1-s_{k+1}, & 1\leq k\leq n, \\ s_1, & k>n\end{cases}\ee

\subsection{Two simple examples}\label{SbSTwoSimpleCases}
\paragraph{The simplest \Sigm-model for $\boldsymbol \sodd$.} Let $\Ds$ be a rank-$s$ totally symmetric irreducible module of the (anti)-de
Sitter algebra. The action of $\DO$ on an element $W^\Ds\fm{q}$ reads\be\label{SimplestAction} \DO W^{A(s)}\fm{q}=\DL
W^{A(s)}\fm{q}+s\lambda V^A E_MW^{A(s-1)M}\fm{q}-s\lambda E^A V_M W^{A(s-1)M}\fm{q},\ee where we adopt the signs as for the
anti-de Sitter case. The result of the restriction of $W^{A(s)}\fm{q}$ to the Lorentz subalgebra is given by \be
W^{A(s)}\fm{q}\Longleftrightarrow\omega^{a(k)}\fm{q}, \qquad k=0,...,s,\ee If we choose the standard gauge (\ref{UnfldSplitting})
for the compensator $V^A$, the field $\omega^{a(k)}\fm{q}$ can be identified with the traceless part of
$W^{a(k)\bullet(s-k)}\fm{q}$. Note that the contraction of two Lorentz indices of $W^{a(k)\bullet(s-k)}\fm{q}$ with the metric
$\eta_{ab}$ does not vanish \be \eta_{bb}W^{a(k-2)bb\bullet(s-k)}\fm{q}=W^{a(k-2)\bullet(s-k+2)}\fm{q}.\ee We refer to such
expressions with a part of indices of the (anti)-de Sitter algebra restricted to the Lorentz algebra and with the other indices
pointing along $V^A$ as to 'raw'. In terms of the raw fields $W^{a(k)\bullet(s-k)}\fm{q}$ (\ref{SimplestAction}) is rewritten as
\be\DO W^{a(k)\bullet(s-k)}\fm{q}=\DL W^{a(k)\bullet(s-k)}\fm{q}+(s-k)\lambda h_m W^{a(k)m\bullet(s-k-1)}\fm{q}-k\lambda h^a
W^{a(k-1)\bullet(s-k+1)}\fm{q},\nonumber\ee where use is made of $E^a=h^a$, $E^\bullet=0$, $V^a=0$, $V^\bullet=1$. Next, we
 rewrite the raw expressions in terms of the irreducible connections $\omega^{a(k)}\fm{q}$
\begin{align}\DO\omega^{a(k)}\fm{q}&=\DL \omega^{a(k)}\fm{q} +\lambda h_m \omega^{a(k)m}\fm{q}+\lambda f_k\left(h^a\omega^{a(k-1)}\fm{q}-\frac{(k-1)}{d+2k-4}\eta^{aa}h_m\omega^{a(k-2)m}\fm{q}\right)\nonumber\end{align}
where the fields have been rescaled to get rid of some factors and \be\nonumber f_k=\frac{k(s-k-1)(d+s+k-2)}{d+2k-2}.\ee In terms
of Lorentz connections, $\nabla_-$ and $\nabla_+$ of (\ref{ConnectionSplitting}) or the operators $V^AE_M$ and $E^AV_M$ of
(\ref{SimplestAction}) give rise to the two algebraic operators
\begin{align}
    \Sigm\left(\omega^{a(k+1)}\fm{q}\right)&=h_m \omega^{a(k)m}\fm{q},\label{SbSSigmaMinusSO}\\
    \Sigp\left(\omega^{a(k-1)}\fm{q}\right)&=f_k\left(h^a\omega^{a(k-1)}\fm{q}-\frac{(k-1)}{d+2k-4}\eta^{aa}h_m\omega^{a(k-2)m}\fm{q}\right).
\end{align}
It is straightforward to verify that $\Sigm^2=0$ due to $h^ah^b+h^bh^a\equiv0$. For the same reason $\Sigp^2=0$.

\paragraph{The simplest \Sigm-model for $\boldsymbol \sldd$.}
It is useful to consider fields with relaxed trace constraints, for instance, the unfolded
off-shell constraints for symmetric fields of all spins in Minkowski space were found in
\cite{Vasiliev:2005zu} to have the form of zero curvature and covariant constancy conditions with
the fields not subjected to any trace constraints. For this purpose, we take $\Ds$ to be an
irreducible module of \sldd{}. The analogue of the Lorentz algebra is then \sld.

Again, take $\Ds$ be a rank-$s$ totally symmetric tensor representation. The result of the restriction of $W^{A(s)}\fm{q}$ to the
\sld-subalgebra is given by the same number of component fields \be W^{A(s)}\fm{q}\Longleftrightarrow\omega^{a(k)}\fm{q}, \qquad
k=0,...,s.\ee Since no trace constraints are imposed, the 'raw' fields $W^{a(k)\bullet(s-k)}\fm{q}$ are directly identified with
the irreducible \sld-fields $\omega^{a(k)}\fm{q}$ \be \omega^{a(k)}\fm{q}=W^{a(k)\bullet(s-k)}\fm{q}, \qquad k=0,...,s.\ee
$\nabla_-$ and $\nabla_+$ give rise to
\begin{align}
    \Sigm\left(\omega^{a(k+1)}\fm{q}\right)&=h_m \omega^{a(k)m}\fm{q},\label{SbSSigmaMinusSL}\\
    \Sigp\left(\omega^{a(k+1)}\fm{q}\right)&=\tilde{f}_k h^a\omega^{a(k-1)}\fm{q},
\end{align}
where $\tilde{f}_k=(s-k-1)k$. The nilpotency of \Sigpm{} is obvious.

\subsection{Formal definition.}\label{SbGCSigmaMinusDef} Given some Lie algebra $\mathfrak{g}$, its representation $\Ds$
and a commutative subalgebra $\mathfrak{f}\subset\mathfrak{g}$, there is a well-known
definition\footnote{We are grateful to E.Feigin for many valuable discussions on Lie algebra
cohomology and \Sigm{} and for reference \cite{Kumar}.} of (co)homology of the Lie algebra
$\mathfrak{f}$ with coefficients in a $\mathfrak{g}$-module $\Ds$ taken as an
$\mathfrak{f}$-module. \be\nonumber\pl: \Ds \otimes \Lambda^q(\mathfrak{f}) \longrightarrow \Ds
\otimes \Lambda^{q-1}(\mathfrak{f}),\ee \be \label{GCSimplestComplexClassic}\pl(a\otimes
u_1\wedge...\wedge u_q)=\sum_{i=1}^{i=q} (-)^{i+1}u_i(a)\otimes
u_1\wedge...\wedge\hat{u}_i\wedge...\wedge u_q,\quad a\in \Ds, u_i\in\mathfrak{f},\ee where
$u_i(a)$ is the action of $u_i\in\mathfrak{f}\subset\mathfrak{g}$ on a vector $a\in\Ds$.

The above definition leads to the $\Sigm$-cohomology for the case of $\mathfrak{g}=\sldd$, in which
$\mathfrak{f}$ in some base is identified with the subalgebra of matrices having nonvanishing
entries in the first row except for the leftmost entry \be\left(\begin{tabular}{c|c}$0$ & $u$ \\
\hline
\parbox[c][30pt][c]{8pt}{$\ast$} & $\mathfrak{h}$
\end{tabular}\right),\qquad u\in\mathfrak{f}.\ee

Now we turn to orthogonal algebra, $\mathfrak{g}=\sodd$, the Lorentz subalgebra is
$\mathfrak{h}=\sod$, then $\mathfrak{f}$ as a vector subspace is given by antisymmetric matrices
that have zeros everywhere except for the first row and the first column
\be\left(\begin{tabular}{c|c}$0$ & $u$ \\
\hline
\parbox[c][30pt][c]{16pt}{$-u$} & $\mathfrak{h}$
\end{tabular}\right),\qquad u\in\mathfrak{f},\ee
i.e., $\mathfrak{f}$ is nothing but the translation generators. However, $\mathfrak{f}$ is a subspace and
not a subalgebra, so that we cannot use the classical definition. Nevertheless, we will show that
in some cases one still can associate with $\mathfrak{f}$ certain nilpotent operator $\pl$,
$\pl^2=0$, acting on $\Ds$ and hence build a complex.

Suppose we are given a Lie algebra $\mathfrak{g}$, its module $\Ds$ and a subalgebra
$\mathfrak{g}_0\subset\mathfrak{g}$, which is to be identified with the Lorentz subalgebra. There
is a canonical decomposition $\mathfrak{g}=\mathfrak{g}_0\oplus_\alpha\mathfrak{g}_\alpha$ of
$\mathfrak{g}$ as a vector space into irreducible representations $\mathfrak{g}_\alpha$ of the
subalgebra $\mathfrak{g}_0$. Restricting to the subalgebra $\mathfrak{g}_0$, the $\mathfrak{g}$-module
$\Ds$ decomposes into $\mathfrak{g}_0$-modules $\Ds_{k}$\be
\mbox{Res}^{\mathfrak{g}}_{\mathfrak{g}_0}\Ds=\bigoplus_{k}\Ds_{k}.\ee The subalgebra
$\mathfrak{g}_0$ acts diagonally, i.e. $\mathfrak{g}_0(\Ds_k)\subset\Ds_k$. In contrast, the action
of $\mathfrak{g}_\alpha$ takes $\Ds_k$ to some other $\Ds_i$. The morphism $\rho_\alpha:
\mathfrak{g}_\alpha\otimes\Ds\longrightarrow\Ds$ defined as $u\otimes a\longrightarrow u(a)$,
$u\in\mathfrak{g}_\alpha$, $a\in\Ds$, is a homomorphism of two $\mathfrak{g}_0$-modules.

The definition (\ref{GCSimplestComplexClassic}) rests on $\mathfrak{f}$ being a subalgebra. This
may not be the case now for $\mathfrak{g}_\alpha$. Nevertheless we can construct certain commuting
operators, which are beyond the representation of $\mathfrak{g}$ on $\Ds$. To succeed we need the
action of $\mathfrak{g}_\alpha$ on $\Ds$ to be $\mathbb{Z}$-graded.

For classical Lie algebras the notion of rank is well-defined, so let $|\Ds_{k}|$ denote the rank
of $\Ds_{k}$. Let us define another decomposition $\Ds=\bigoplus_g\Ds_g$ of the $\mathfrak{g}$-module
$\Ds$ into generally reducible $\mathfrak{g}_0$-modules such that
\be\Ds_g=\bigoplus_{k:|\Ds_{k}|=g}\Ds_{k}\ee is a direct sum over $\Ds_{k}$ having rank $g$.

The tensor product $\mathfrak{g}_\alpha\otimes\Ds$ of the two $\mathfrak{g}_0$-modules can be
explicitly computed. In view of general properties of tensor products
$\mathfrak{g}_\alpha\otimes\Ds_g$ decomposes into representations with ranks confined in the range
$|g-|\mathfrak{g}_\alpha||$, $g+|\mathfrak{g}_\alpha|$. Therefore, the rank provides us with
natural $\mathbb{Z}$-grading on $\Ds$. Let the operators realizing the action of $\mathfrak{g}$ on
$\Ds$ be denoted as $\vartheta_{\mathfrak{g}}$. Then, there is a decomposition of
$\vartheta_{\mathfrak{g}_\alpha}$ into the parts with definite grade
\be\vartheta_{\mathfrak{g}_\alpha}=\bigoplus_{i\in\mathbb{Z}}\vartheta^i_{\mathfrak{g}_\alpha},\qquad\quad
\vartheta^i_{\mathfrak{g}_\alpha}: \Ds_g\longrightarrow \Ds_{g+i},\ee $\vartheta_{\mathfrak{g}_0}$
has by definition zero grade part only, $\vartheta_{\mathfrak{g}_0}=\vartheta^0_{\mathfrak{g}_0}$.
Obviously, there is a certain $n$ such that $\vartheta^j_{\mathfrak{g}_\alpha}\equiv0$ if $j<-n$ for
any $\alpha$ and we assume that there is a certain $\mathfrak{g}_{\min}$ among
$\mathfrak{g}_{\alpha}$ such that $\mathfrak{g}_{\min}^{-n}\neq0$. If there are several
$\mathfrak{g}_{\alpha}$ such that $\mathfrak{g}_{\alpha}^{-n}\neq0$ then $\mathfrak{g}_{\min}$ is a
direct sum over such $\mathfrak{g}_{\alpha}$. By definition of representation we have
\be[\vartheta_x,\vartheta_y]=\vartheta_{[x,y]},\quad x,y\in\mathfrak{g}\qquad\Longrightarrow\qquad
[\vartheta^{-n}_{x'},\vartheta^{-n}_{y'}]=0,\quad x',y'\in\mathfrak{g}_{\min},\ee i.e.
$\vartheta^{-n}_{\mathfrak{g}_{\min}}$ are commuting operators.

Consequently, given a $\mathbb{Z}$-graded decomposition of action of some algebra on its representation it is possible to single
out commuting operators that belong to the lowest or highest grade. Despite the fact that $\mathfrak{g}_{\min}$ does not form a
subalgebra, the operators $\vartheta^{-n}_{\mathfrak{g}_{\min}}$ do form a commutative subalgebra.

The complex $\Comp(\Ds,\pl)$ is defined in a standard way: \be \pl: \Ds\otimes\Lambda(\mathfrak{g}_{\min}), \qquad\pl:
\Ds_g\otimes\Lambda^q(\mathfrak{g}_{\min})\longrightarrow\Ds_{g-n}\otimes\Lambda^{q-1}(\mathfrak{g}_{\min}),\ee \be \pl(a\otimes
u_1\wedge...\wedge u_q)=\sum_{i=1}^{i=q} (-)^{i+1}\vartheta^{-n}_{u_i}(a)\otimes u_1\wedge...\wedge\hat{u}_i\wedge...\wedge
u_q,\quad a\in \Ds, u_i\in\mathfrak{g}_{\min}.\ee

We collect in the table below some cases that are or may be of interest\parbox[t][6pt][c]{0pt}{}
\par\noindent\begin{tabular}{|x{0.5cm}|x{1.7cm}|x{1.1cm}|x{3cm}|x{6.5cm}|}\hline & $\mathfrak{g}$ & $\mathfrak{g}_0$ & $\mathfrak{g}_\alpha$
& description\tabularnewline\hline 1&
\parbox[c][23pt][c]{0pt}{}\sodd & \sod & \smallpic{\YoungA} & \dSAdS{} fields on-shell \tabularnewline 2& \parbox[c][23pt][c]{0pt}{}\sldd &
\sld & $\parbox[c][5pt][b]{5pt}{\smallpic{\YoungA}}\oplus\parbox[c][5pt][b]{6pt}{$\smallpic{\YoungA}^*$}\oplus\bullet$ & \dSAdS{}
fields off-shell \tabularnewline 3&\parbox[c][23pt][c]{0pt}{}\soddd & \sod &
$2\,\parbox[c][5pt][b]{5pt}{\smallpic{\YoungA}}\oplus\bullet$ & conformal fields on-shell\tabularnewline
4&\parbox[c][23pt][c]{0pt}{}\slddd & \sld &
$2\parbox[c][5pt][b]{5pt}{\smallpic{\YoungA}}\oplus2\parbox[c][5pt][b]{6pt}{$\smallpic{\YoungA}^*$}\oplus4\bullet$ & conformal
fields off-shell \tabularnewline 5&
\parbox[c][23pt][c]{0pt}{}\sld & \sod &
$\smallpic{\YoungB}$ & trace decomposition \tabularnewline 6&\parbox[c][23pt][c]{0pt}{}\sldd & \sod
&
$\parbox[c][5pt][b]{12pt}{\smallpic{\YoungB}}\oplus2\,\parbox[c][5pt][b]{5pt}{\smallpic{\YoungA}}\oplus\bullet$
& unconstrained \dSAdS{} fields\tabularnewline \hline
\end{tabular}

\rule{0pt}{14pt} Note that the signature of $\mathfrak{so}$-algebras does not matter, in what
follows we assume that appropriate real forms are chosen. Item 1 is the case we investigate in the
present paper, it concerns the gauge fields in (anti)-de Sitter space. Item 2: the trace
constraints on fields are fully relaxed so that we have the correct pattern of gauge symmetries but
there can be imposed no field equations. The decomposition $\mathfrak{g}_\alpha$ consists of a
vector, covector and a scalar. In this case, vector (or covector) representation itself forms a
commutative subalgebra, so that $\coh(\Ds,\Sigm)$ coincides with the ordinary Lie algebra
cohomology, the answer can be found in \cite{Kumar}. Item 3: $\mathfrak{g}$ can be taken as the
conformal algebra, this provides a natural framework for conformal fields, which are studied in
\cite{Vasiliev:2009ck}. Note that in this case we again meet the ordinary Lie algebra cohomology.
Item 4: the same case of conformal fields but the description is off-shell. Item 5 corresponds to
the trace decomposition of a tensor with fully relaxed trace constraints in terms of traceless
tensors. Item 6 may be related to the unconstrained approach of
\cite{Francia:2002aa,Francia:2002pt,Bekaert:2002dt,Bekaert:2003az,Francia:2005bu} for the case of
gauge fields in \dSAdS.

We will study the complex $\Comp(\Ds,\Sigm)$ dual to $\Comp(\Ds,\pl)$ for $\mathfrak{g}$ being the (anti)-de Sitter algebra and
$\mathfrak{g}_0$ being its Lorentz subalgebra. The most significant fact for computing \Sigm-cohomology is that on account of
\be[\vartheta_{w},\vartheta^{-n}_{u}]=\vartheta^{-n}_{[w,u]},\qquad[w,u]\in\mathfrak{g}_{\min}, \qquad w\in \mathfrak{g}_0,\quad
u\in\mathfrak{g}_{\min}\ee the differential $\pl$ (or \Sigm) commutes with the action of $\mathfrak{g}_0$ and hence
$\mathfrak{g}_0$ acts on the (co)homology, so that it is convenient to label the representatives of (co)homology by the weights
of irreducible $\mathfrak{g}_0$-modules or by Young diagrams in the case of interest.

\paragraph{\Sigm, specialization to \DSADS.} The \Sigm-complex $\Comp(\Ds,\Sigm)$ is associated with the complex $\Comp(\Ds,\DO)$.
To build $\Comp(\Ds,\Sigm)$ we need an irreducible $\mathfrak{g}$-module $\Ds$ and nondegenerate $\DO$, meaning that it yields
$E^A$ with the maximal rank (or, in the standard gauge, $h^a$), so that very little is needed from $\Comp(\Ds,\DO)$. The
representatives of \Sigm-cohomology are irreducible Lorentz modules whose weights are to be found.

The (anti)-de Sitter algebra as a vector space $\mathfrak{g}$ splits as $\mathfrak{g}=\mathfrak{h}\oplus\mathfrak{p}$, where
$\mathfrak{h}$ is the Lorentz algebra \lorentz{} and $\mathfrak{p}$ is a vector representation of $\mathfrak{h}$, namely, the
translation generators $P_a$ constitute $\mathfrak{p}$. The splitting (\ref{ConnectionDecomposition})
$\Omega=\Omega_L+\nabla_--\nabla_+$ of a $\mathfrak{g}$-connection $\Omega$ implies that $\nabla_--\nabla_+$ corresponds to the
operators $\vartheta_{\mathfrak{p}}$ from the previous subsection that represent the action of the translation generators. So
$\nabla_-$ and $\nabla_+$ are the operators $\vartheta_\mathfrak{p}^{-1}$ and $\vartheta_\mathfrak{p}^{+1}$ such that
$\vartheta_{\mathfrak{p}}=\vartheta_\mathfrak{p}^{-1}+\vartheta_\mathfrak{p}^{+1}$. Both $\nabla_-$ and $\nabla_+$ are algebraic, and
hence algebraic are the induced operators \Sigm{} and \Sigp, we ignore the dependence on the space-time coordinates $x^\mu$.

We define the $q$-cochain $C^q(\Ds)=\bigoplus_g C^q_g(\Ds)$ with \be
C^q_g(\Ds)=\Ds_g\otimes\Lambda^q(\mathfrak{p}).\label{CqgDefinition}\ee Then, $\nabla_\pm$ induce
two nilpotent operators $\Sigpm$ \be\Sigpm: C^q_g(\Ds) \longrightarrow C^{q+1}_{g\pm1}(\Ds).\ee

In tensorial terms, we decompose $\WW^\Ds\fm{q}$ into connections of the Lorentz subalgebra and convert all form indices to fiber
ones with the help of the inverse background vielbein $h^{a\mu}$, $h^{a\mu}h^b_\mu=\eta^{ab}$, so that the Lorentz algebra
$\mathfrak{h}$ starts acting on the former form indices too. $\Ds\otimes\Lambda^q(\mathfrak{p})$ is an $\mathfrak{h}$-module with
the action of $\mathfrak{h}$ on $\Lambda^q(\mathfrak{p})$ induced from that on $\mathfrak{p}$. Moreover, $\mathfrak{h}$ commutes
both with \Sigm{} and \Sigp, which allows us to decompose $\Ds\otimes\Lambda^q(\mathfrak{p})$ into irreducible
$\mathfrak{h}$-modules and parameterize the representative of the \Sigm-cohomology groups by Young diagram of $\mathfrak{h}$.

An element of $C^q_g(\Ds)$ is a degree-$q$ exterior form $\omega^{\Xx}\fm{q}$ with values in a
generally reducible $\mathfrak{h}$-module $\Xx$ belonging to $\Ds_g$. As an $\mathfrak{h}$-module
$\omega^{\Xx}\fm{q}$ can be decomposed into irreducible $\mathfrak{h}$-modules, the decomposition
is equivalent to taking $\mathfrak{h}$-tensor product \be C^q_g(\Ds)\sim\Ds_g\otimes_\mathfrak{h}
\Ya{q}=\bigoplus_{r=0}^{r=q}\bigoplus_{i_r}M^{g,q}_{r,i_r} \Xx^{g,q}_{r,i_r},\ee where the sum is
over irreducible $\mathfrak{h}$-modules $\Xx^{g,q}_{r,i_r}$, with $M^{g,q}_{r,i_r}$ being the
multiplicity of $\Xx^{g,q}_{r,i_r}$. By definition, all irreducible modules in $\Ds_g$ have the
same rank, denote it $|\Ds_g|$. The additional summation index $r$ distinguishes between traces
of different orders, so that the rank $|\Xx^{g,q}_{r,i_r}|$ of $\Xx^{g,q}_{r,i_r}$ is equal to
$|\Ds_g|+q-2r$. In tensorial terms, the trace order is half the number of indices that must be
contracted to get a tensor with the symmetry of $\Xx^{g,q}_{r,i_r}$.

The operator $\Sigm$ preserves two natural gradings - the total rank of the tensor it acts on (the rank of the fiber tensor $g$
plus the form degree $q$) and the trace order. In addition, that $\boldsymbol{\sigma_-}$ preserves the $\mathfrak{h}$-module
structure means that it does not mix different $\mathfrak{h}$ tensors up. Granting this, $\Comp(\Ds, \Sigm)$ decomposes into a
direct sum \be\Comp(\Ds,
\Sigm)=\bigoplus_{q+g}\bigoplus_r\bigoplus_\Xx\Comp(\Ds,\Sigm;\Xx,q+g,r)\label{SigmaComplexDecomposition}\ee of complexes
parameterized by an arbitrary Young diagram \Xx{} from $\Ds_g\otimes\Ya{q}$, the total rank $q+g$ and the trace order $r$.
Restricted on $\Comp(\Ds,\Sigm;\Xx,q+g,r)$, $\Sigm$ is given by a set of linear maps
$\mathbb{R}^{M^{g,q}_{r,i_r}}\otimes\Xx\longrightarrow\mathbb{R}^{M^{g-1,q+1}_{r,i_r}}\otimes\Xx$ acting on the first factor.

Consequently, $\Comp(\Ds,\Sigm)$ is well-defined for any $\mathfrak{g}$-module $\Ds$. The
background vielbein $E^A$ and the compensator field $V^A$ provide a field-theoretical realization
of $\Comp(\Ds,\Sigm)$, at the condition that the vielbein field $E^A$ has the maximal rank. We will
see that $\Comp(\Ds,\Sigm)$ has rich cohomology as distinct from, for example, the de Rham complex.

It is worth stressing that the equation of motion, gauge transformations, Bianchi identities, etc. contain both \Sigm{} and
\Sigp, e.g. $\DL\omega^g+\Sigm(\omega^{g+1})+\Sigp(\omega^{g-1})=0$. Hence, we can chose either \Sigm{} or \Sigp{} to be the
operator \Sigm{} of section \ref{SbSInterpretation} used to interpret the unfolded equations.

We might study the cohomology problem both for \Sigm{} and \Sigp, however, the choice of \Sigm{}, acting from higher rank tensors
of $\Ds_{g+1}$ to lower rank tensors of $\Ds_g$, is more natural since equation
$\DL\omega^g+\Sigm(\omega^{g+1})+\Sigp(\omega^{g-1})=0$ expresses higher rank auxiliary fields $\omega^{g+1}$ in terms of
derivatives of lower rank fields $\omega^g$ if $\Sigm$ has vanishing cohomology at grade $g$, plus possibly lower derivative
terms coming from $\Sigp(\omega^{g-1})$. Therefore, the chain of auxiliary fields starts from dynamical fields having the lowest
possible rank, these fields will be recognized as the field potentials $\phi^\Ss$.

The interpretation of unfolded equations in terms of the \Sigp-cohomology leads to dual
formulations, in which the dynamical fields are tensors of a rank higher than that of $\phi^\Ss$.
An example of such a dual formulation was studied in \cite{Matveev:2004ac}.

\subsection{Distinguished parts of tensor products}\label{SbSHWPMSP}
\paragraph{The highest weight part.}
Let $\Xx=\Y{s^x_1,...,s^x_n}$ and $\Yy=\Y{s^y_1,...,s^y_n}$ be two Young diagrams either of \sld{}
or \sod. Despite the fact that to decompose the tensor product $\Xx\otimes_\slds\Yy$ into a direct
sum of irreducible modules is a complicated problem, we can be sure that at least one irreducible
module is present in $\Xx\otimes\Yy$ whose Young diagram $\Zz=\Y{s^x_1+s^y_1,...,s^x_n+s^y_n}$ is
obtained by row-by-row concatenation. This one is called the highest-weight part, $\hwp(\Xx,\Yy)$.
The highest weight part of $\Xx\otimes\Yy$ is given by a single Young diagram. In the case of
interest the second multiplier is a one-column Young diagram. Let us denote the highest weight part
of $\Xx\otimes\Ya{q}$ as
\be\label{AppSlTensorProductHW}\hwp(\Xx,q)=\begin{cases}\Y{s^x_1+1,...,s^x_q+1,s^x_{q+1},...,s^x_n}, & q<n,\\
\Y{s^x_1+1,...,s^x_n+1,\underbrace{1,...,1}_{q-n}}, & q\geq n.\end{cases}\ee

\paragraph{Remark on $\boldsymbol{\sod}$-tensor products.} Roughly speaking, the difference between the tensor products of \sld{} and \sod{} is that in the
latter case one is able to contract indices with the help of the invariant tensor $\eta_{ab}$ of
$\sod$, i.e. to take traces. Thus, the \sod-tensor product rule for $\Xx\otimes_\sods\Yy$ consists
of taking traces, which removes pairs of cells (one from $\Xx$ and another from $\Yy$), and, then,
adding the rest of the cells of $\Yy$ to a set of diagrams obtained from $\Xx$ at the first stage.
The precise rules can be found in \cite{QuantGrCrystals}, which for the cases of interest are given
in Appendix A.

In general the tensor product of two irreducible \sod-modules decomposes into a direct sum of
irreducible modules whose multiplicities can be greater than one, because the same diagram can be
obtained generally by removing and, then, adding cells from/to different places.

Given some element $\Zz$ of $\Xx\otimes_\sods\Yy$, the number of cells that were removed from $\Xx$
(or $\Yy$) is called the trace order of $\Zz$.

\paragraph{The maximally symmetric part.}
For the case of \sld{} the highest weight part will also be called the maximally symmetric part.
However, for the case of \sod{} the two definitions are different.

For the case of \sod{}, given two irreducible \sod-modules \Xx{} and \Yy, the maximally symmetric
part of $\Xx\otimes\Yy$ with the trace order $r$, $\msp(\Xx,\Yy,r)$, is a sum of the form
$\bigoplus_\alpha\hwp(\Xx^r_\alpha,\Yy^r_\alpha)$, where $\Xx^r_\alpha$ and $\Yy^r_\alpha$ are the
diagrams obtained from $\Xx$ and $\Yy$ by taking all possible traces of order $r$, so that
$\Xx^r_\alpha$ and $\Yy^r_\alpha$ each has $r$ cells less than $\Xx$ and $\Yy$, respectively. The
index $\alpha$ runs over all inequivalent traces of order $r$.

As distinct from the \sld-case, the maximally symmetric part of the \sod-tensor product may contain
many irreducible modules, however, each comes with multiplicity one.

In this paper the second multiplier is always a one column diagram, i.e. $\Yy=\Ya{q}$. Given an
\sod-Young diagram $\Xx=\Yb{(s_1,p_1),...,(s_N,p_N)}$ (the block notation for Young diagrams is
more convenient) and two nonnegative integers $q$, $r$ such that $q\geq r$ the maximally symmetric
part of $\Xx\otimes_\sods\Ya{q}$ with the trace order $r$ is denoted $\msp^\sods(\Xx,q,r)$.  The
process of taking $\msp$ is illustrated on fig.\,\ref{fig:PicGCMSP}. It is evident that different
partitions of $r$ give rise to distinct elements of the \msp. The sum over the traces of all orders
is denoted $\msp(\Xx,q)$, $\msp(\Xx,q)=\sum_r \msp(\Xx,q,r)$.

Thus, the maximally symmetric part for $\Xx\otimes_\sods\Yy$ is obtained by taking all possible traces and, then, adding the rest
of the cells according to the \hwp-rule. In terms of Young diagrams, we see that $\hwp^\slds\equiv\msp^\slds\equiv\hwp^\sods$ and
$\hwp^\sods\subset\msp^\sods$.

\begin{figure}
\begin{align}&\parbox{3.5cm}{{{\bep(200,180){%
\put(0,0){\BRectT{5}{5}{2}{2}{N}}%
\put(0,50){\BRectT{8}{6}{2}{3}{k}}%
\put(0,110){\BRectT{12}{6}{2}{3}{1}}%
}\eep}}}
&&\bigotimes&&\parbox{1.0cm}{\bep(10,180){\put(0,40){\ColumnTDotted{1}{3}{t_1}}\put(0,-10){\ColumnTDotted{1}{2}{t_N}}%
\put(0,10){\ColumnTDotted{1}{3}{t_k}}\put(0,70){\ColumnShadded{10}}}\put(15,120){$q-r$}\put(18,20){$r$}\eep}&&=&&
\parbox{6cm}{{{\bep(160,180){%
\put(0,0){\BRectTM{5}{5}{2}{2}{N}}\put(70,70){\dottedline{9}{0}}\put(165,120){$q-r$}%
\put(0,50){\BRectTM{8}{6}{2}{3}{k}}\put(80,80){\ColumnShadded{3}}\put(70,70){\ColumnShadded{1}}%
\put(0,110){\BRectTM{12}{6}{2}{3}{1}}\put(120,170){\dottedline{4}{0}}%
\put(120,140){\ColumnShadded{3}}\put(110,110){\ColumnShadded{3}}%
\put(160,120){\vector(0,1){50}\vector(0,-1){50}}%
}\eep}}}\nonumber\end{align} \caption{Taking the maximally symmetric part of \sod-tensor product
$\Xx\otimes\Yy$, $\Xx=\Yb{(s_1,p_1),...,(s_N,p_N)}$, $\Yy=\Ya{q}$. Since taking the trace must
result in a Young diagram, different traces correspond to different partitions of $r=t_1+...+t_N$,
such that $t_i\leq p_i$. Then, $t_i$ cells of $\Yy$ are removed from the bottom-right of the $i$-th
block of $\Xx$. Finally, the rest of cells from $\Ya{q}$, i.e. $q-r$ cells, is added to the first
rows, which gives the highest weight part. If $r=0$ then we get the \sld-case, for which
$\hwp(\Xx,q)\equiv\msp(\Xx,q)$.}\label{fig:PicGCMSP}
\end{figure}

\subsection{Restriction and \ensuremath{\hwp,\ \msp}}\label{SbSRestrictionTensorProduct}
Since the Lorentz modules that label the subcomplexes of $\Comp(\Ds,\Sigm)$ come from the tensor products of the restriction of
$\Ds$ by one-column diagrams and the diagrams labelling the \Sigm-cohomology tend to be as symmetric as possible, to write out
the results we need to combine the maximally symmetric part of a tensor product with the restriction functor.

\paragraph{$\boldsymbol{\sldd}$.} Given an \sldd-irreducible module $\Ds=\Y{s_1,...,s_n}$ let $\mspr^\slds(\Ds,q)$ be the element
from $\res^\sldds_\slds\Ds\otimes\Ya{q}$ of the form
\be\mspr^\slds(\Ds,q)=\Y{s_1+1,....,s_q+1,\widehat{s_{q+1}},s_{q+2},...,s_n},\ee i.e., one cell is
added to the right of each row in the range $1,...,q$ and the $(q+1)$-th row is thrown away.
Therefore, one can rewrite the definition of $\mspr$ in terms of $\hwp$ or $\msp$ with the argument
being the maximally symmetric component $\Ds^{q+1}$, defined in (\ref{MaxSymRestriction}),
\be\mspr^\slds(\Ds,q)\equiv\msp^\slds(\Ds^{q+1},q)\equiv\msp^\slds(\Ds^{q+1},\Ya{q})\equiv\hwp^\slds(\Ds^{q+1},q).\ee

\paragraph{$\boldsymbol{\sodd}$.} Let $\Ds=\Y{s_1,...,s_n}$ be a
Young diagram of \sodd{} and such that all weights $(s_1,...,s_n)$ are different. Then,
$\mspr^\sods(\Ds,q,r)$ is defined as
\be\mspr^\sods(\Ds,q,r)=\msp^\sods(\Ds^{q-r+1},q,r)\equiv\msp^\sods(\Ds^{q-r+1},\Ya{q},r),\ee i.e.
$\mspr^\sods(\Ds,q,r)$ is a sum over all diagrams that are obtained from the maximally symmetric
component $\Ds^{q-r+1}$ (see (\ref{MaxSymRestriction}) for the definition) by taking all possible
traces of order $r$ and, then, adding one cell to each of the first $(q-r)$ rows. The weight
$s_{q-r+1}$ to be thrown away is determined by the number of cells that remain to be added after
taking the trace.

If among the weights $(s_1,...,s_n)$ some are equal then the block notation is more convenient, so
we take $\Ds=\Yb{(s_1,p_1),...,(s_N,p_N)}$. The process of taking $\mspr(\Ds,q,r)$ is illustrated
on fig.\,\ref{fig:PicGCMSPR}. Similar to the case where all weights in \Ds{} are different, one
first takes all possible traces and then adds the rest of the cells, i.e. $(q-r)$, according to the
\hwp{} rules. The difference is that certain diagrams must be deleted while taking the highest
weight part. Let some cell be called a filled vacancy if it has been removed (while taking traces)
and then restored (while adding the rest of the cells according to the \hwp-rules). Let $k$ be the
number of the block of \Ds{} to which the $(q-r+1)$-th weight belongs. Then, the diagrams to be
deleted are those diagrams for which there is at least one filled vacancy at the $k$-th block.
\begin{figure}
\begin{align}&
\parbox{7.5cm}{{{\bep(160,180){\put(80,70){\dottedline{4}{0}}\put(90,90){\dottedline{3}{0}}
\put(0,0){\BRectT{5}{5}{2}{2}{N}}\put(90,90){\dottedline{6}{0}}\put(155,120){$q-r$}\put(125,77){${\epsilon'}$}%
\put(0,50){\BRectTA{8}{6}{2}{2}{k}}\put(80,90){\ColumnShadded{2}}\put(125,120){$t_1$}%
\put(0,110){\BRectTM{12}{6}{2}{3}{1}}\put(120,170){\dottedline{3}{0}}%
\put(65,45){\circle*{3}}\put(75,45){\circle*{3}}\put(5,-5){\circle*{3}}\put(15,-5){\circle*{3}}%
\put(25,-5){\circle*{3}}\put(35,-5){\circle*{3}}\put(45,-5){\circle*{3}}%
\put(50,40){\BlockADotted{3}{1}}\put(55,45){\circle*{3}}\put(0,-10){\BlockADotted{5}{1}}%
\put(120,140){\ColumnShadded{3}}\put(110,110){\ColumnShadded{3}}%
\put(150,120){\vector(0,1){50}\vector(0,-1){30}}\put(130,10){\circledDigit{a}}\put(120,80){\vector(0,1){10}\vector(0,-1){10}}%
}\eep}}}&&\parbox{6cm}{{{\bep(160,180){
\put(65,45){\circle*{3}}\put(75,45){\circle*{3}}\put(15,-5){\circle*{3}}%
\put(25,-5){\circle*{3}}\put(35,-5){\circle*{3}}\put(45,-5){\circle*{3}}%
\put(50,40){\BlockADotted{3}{1}}\put(55,45){\circle*{3}}\put(10,-10){\BlockADotted{4}{1}}%
\put(0,0){\BRectT{5}{5}{2}{2}{N}}\put(80,70){\dottedline{7}{0}}\put(155,120){$q'-r'$}%
\put(0,50){\BRectTMA{8}{6}{2}{2}{k}}\put(80,90){\ColumnShadded{2}}\put(125,120){$t_1$}%
\put(0,110){\BRectTM{12}{6}{2}{3}{1}}\put(120,170){\dottedline{3}{0}}%
\put(120,140){\ColumnShadded{3}}\put(110,110){\ColumnShadded{3}}\put(70,70){\ColumnShadded{2}}%
\put(150,120){\vector(0,1){50}\vector(0,-1){50}}\put(125,57){$t_k$}\put(125,77){$\epsilon'$}\put(0,-10){\YoungCcA}\put(130,10){\circledDigit{b}}%
\put(120,80){\vector(0,1){10}\vector(0,-1){10}}\put(80,50){\dottedline{4}{0}}\put(90,90){\dottedline{3}{0}}%
\put(120,60){\vector(0,1){10}\vector(0,-1){10}}%
}\eep}}}\nonumber\end{align} \caption{Illustration for $\Ds=\Yb{(s_1,p_1),...,(s_N,p_N)}$.
\newline (a) Taking $\mspr(\Ds,q,r)$. That the weight $s_{q-r+1}$ from the $k$-th block is thrown
away implies that the blocks $1,...,k-1,k+1,...,N$ remain unchanged and the $k$-th block becomes
shorter by one row. In tensor language, in order to project onto $\Ds^{q-r+1}$ one needs to
contract the compensator $V^A$ with the indices corresponding to the cells marked by $\bullet$ and
then apply Young symmetry and trace projectors. Any trace is determined by a partition
$r=t_1+...+t_N$ such that $t_i\leq p_i$ if $i=1,...,k-1,k+1,...,N$ and $t_k\leq p_k-1$. In taking
the trace $t_i$ cells are removed from the bottom-right of the $i$-th block. The rest of cells,
which are drawn hatched, is added to the first rows. The additional condition for diagrams having
equal rows implies that the cells being added must not overlap with any of the $t_k$ cells that
have been removed from the $k$-th block while taking traces. Therefore, the gap $\epsilon'$ is
always nonnegative.
\newline (b) Taking dual. The same diagram is obtained in another way. The diagram
$\widetilde{\Ds}^{q-r+1}$ is just $\Ds^{q-r+1}$ with one extra cell below the last row. Consider one special
trace of order $r'=r+\epsilon'+1$ of a degree-$q'=q+2\epsilon'+1$ form
$w^{\widetilde{\Ds}^{q-r+1}}\fm{q'}$. First, one takes the trace of $\widetilde{\Ds}^{q-r+1}$ of
order $r+\epsilon'+1$: $r$ cells are removed in the same way as in (a), one cell is removed from the
last row, $\epsilon'$ cells are removed from the $k$-th block in addition to the $t_k$ cells just
removed. Second, the rest of the cells, i.e. $q-r+\epsilon'$ is added to the first rows. Finally,
the same diagram as in (a) is obtained.}\label{fig:PicGCMSPR}
\end{figure}

The sum over the traces of all orders $r=0,1,...,q$ such that $s_{q-r+1}=s_{q+1}$ is denoted
$\mspr(\Ds,q)$. The condition $s_{q-r+1}=s_{q+1}$ implies that all elements of $\mspr(\Ds,q)$ are
from $\Ds_{g}\otimes\Ya{q}$ with $g$ being equal to the grade of $\Ds^{q+1}$. If all weights
$(s_1,...,s_n)$ are different then $\mspr(\Ds,q)=\mspr(\Ds,q,r=0)$ contains a single element.

It is worth stressing that the weight of \sodd{} is $(s_1,...,s_n,0_{n+1},...,0_\nu)$,
$\nu=[(d+1)/2]$, and if $(q-r)>n$ these zero rows should be added to $\Ds$.

\paragraph{duality map.} Now we define the duality map which takes any element of $\mspr(\Ds,q,r)$
to some other element of the complex $\Comp(\Ds,\Sigm)$ that is defined by the same Young diagram.
Given an \sodd-module \Ds, let $\Xx$ be any irreducible \sod-module that is an element of
$\mspr(\Ds,q,r)$ for some $q$ and $r$. By definition, $\Xx$ appears the same time as a trace of
order $r$ in the decomposition of $C^q_{g,r}(\Ds)$ (\ref{CqgDefinition}) into irreducible
\sod-modules, where $g=g(\Ds^{q-r+1})$. Let $\epsilon'$ be defined for $\Xx$ according to
fig.\,\ref{fig:PicGCMSPR}a. Provided that $g$ is not the maximal possible value of the grade, the
dual to $\Xx$ is an \sod-module $\widetilde{\Xx}$ that is defined by the same Young diagram $\Xx$
and is the element of $C^{q'}_{g+1,r'}(\Ds)$, where $q'=q+2\epsilon'+1$, $r'=r+\epsilon'+1$. See
fig.\,\ref{fig:PicGCMSPR}b for the illustration. \be\begin{tabular}{ccccccc}
 & & $C^q_{g,r}(\Ds)$ & & $C^{q'}_{g+1,r'}(\Ds)$ & & \\
 & & $\cup$ & &  $\cup$  & & \\
 $\omega^{\Ds_g}\fm{q}$ & $\xleftarrow{\quad\displaystyle\pi\quad}$ & ${\Xx}$ & $\xrightarrow{\qquad\qquad}$ &
 $\widetilde{\Xx}$ & $\xrightarrow{\quad\displaystyle\widetilde{\pi}\quad}$
 & $\omega^{\Ds_{g+1}}\fm{q}$ \\
 & & $\cap$ & &  & & \\
 & & $\mspr(\Ds,q,r)$ & &  & &
\end{tabular}\ee
Indeed, the diagram $\widetilde{\Xx}=\Xx$ belongs to $\widetilde{\Ds}^{q-r+1}\otimes\Ya{q'}$, where
$\widetilde{\Ds}^{q-r+1}\in\Ds_{g+1}$, namely
\be\widetilde{\Ds}^{q-r+1}=\Y{s_1,...,s_{q-r},\widehat{s_{q-r+1}},s_{q-r+2},...,s_n,1}.\ee The
trace of order $r+{\epsilon'}+1$ is to be taken as follows: the trace of order $r$ is taken as for
$\Xx$, in doing so extra $\epsilon'$ cells are removed from the block to which the $(q-r+1)$-th
weight belongs, then the only cell in the last row of $\widetilde{\Ds}^{q-r+1}$ is removed. The
rest of $q'-r'=q-r+{\epsilon'}$ cells is added according to the \hwp-rules, which results in the
same diagram $\Xx$.

It is technically very difficult to define the duality map in terms of irreducible Lorentz tensor
because to do so we need to use the explicit form of Young and trace projectors in order to embed
an irreducible tensor with the symmetry of $\Xx$ into the Lorentz connection $\omega^{\Ds_g}\fm{q}$
with the help of certain projector $\pi$ and into $\omega^{\Ds_{g+1}}\fm{q}$ with the help of
projector $\widetilde{\pi}$. Fortunately, to make contact between the formulation in terms of
generalized Yang-Mills fields of the (anti)-de Sitter algebra and metric-like fields of section
\ref{SFieldsConnections} the explicit form of $\pi$ and $\widetilde{\pi}$ is not needed.

\subsection{Sigma-minus cohomology, the result}\label{SbSGCSigmaResult}
Let us first state the main result on the \Sigm-cohomology for the case of \sldd{}, and then for the case of interest \sodd{}.
The proof is left to the Appendix B since it is rather technical.

\begin{Theorem} Let $\Ds=\Y{s_1,...,s_n}$ be a Young diagram defining an irreducible \sldd-module and
$\Comp(\Ds,\Sigm)$ be the associated \Sigm-complex.
Then, \be\coh^q(\Ds,\Sigm)=\begin{cases}\mspr(\Ds,q), & q=0,...,d-1,\\
\emptyset, & q\geq d, \end{cases}\ee where the grade of a single element of $\coh^q(\Ds,\Sigm)$ is $g(\Ds^{q+1})$.
\end{Theorem}
For \sldd{} \Sigm{} has a plain algebraic meaning inasmuch as $V_0$ can be identified with the
commutative subalgebra $\mathfrak{p}$ of \sldd{} that is a covector representation of the \sld{}
subalgebra. Then, the definition of \Sigm coincides with the Lie cohomology of $\mathfrak{p}$ with
values in the \sldd-module \Ds, the solution can be found, for example, in \cite{Kumar}.

\begin{Theorem} Let $\Ds=\Y{s_1,...,s_n}$ be a Young diagram defining an irreducible \sodd{}-module (the signature is irrelevant and
\sodd{} can be viewed as the (anti)-de Sitter algebra) and $\Comp(\Ds,\Sigm)$ be the associated
\Sigm-complex, then \be\coh^q(\Ds,\Sigm)=\coh^q(\Ds,\Sigm)^{reg}\oplus\coh^q(\Ds,\Sigm)^{irreg},
\ee where $\coh^q(\Ds,\Sigm)^{reg}$ is the regular part of the cohomology
\be\coh^q(\Ds,\Sigm)^{reg}=\sum_{k=0}^{k=q}\mspr(\Ds,q,k),\ee and the irregular part
$\coh(\Ds,\Sigm)^{irreg}$ is given by the elementwise dualization of the regular part,
\be\coh(\Ds,\Sigm)^{irreg}=\{\widetilde{\omega}: \omega\in
\coh(\Ds,\Sigm)^{reg}\}=\widetilde{\coh}(\Ds,\Sigm)^{reg},\ee i.e. the representatives of
$\coh(\Ds,\Sigm)^{irreg}$ are obtained by applying the duality map to a representative of each
cohomology class of $\coh^q(\Ds,\Sigm)^{reg}$ - different classes of $\coh^q(\Ds,\Sigm)^{reg}$ are
mapped to different classes in $\coh(\Ds,\Sigm)^{irreg}$.

Note that the grade of $\widetilde{\omega}$ is greater by one than that of $\omega$, with the form degree and trace order
depending on the number of equal weights in $\omega$ and on its degree and trace order. Therefore, different representatives of
$\coh^q(\Ds,\Sigm)^{reg}$ having the same degree, grade and trace order can give rise to classes of $\coh(\Ds,\Sigm)^{irreg}$
with different degrees and trace orders but necessarily having the same grade. It is worth noting also that the duality map
applied to a representative at the highest grade gives nothing.
\end{Theorem}

The latter theorem encompasses all the special cases addressed in the literature: in
\cite{Shaynkman:2001ip} the precise field theoretical meaning was given to $\coh(\Sigm)$, and the
example of $\Comp(\Y{k\rightarrow\infty},\Sigm)$ was investigated in detail; $\coh^q(\Sigm)$ at
lower degrees $q=0,1,2$ for the complex $\Comp(\Y{s-1,s-1},\Sigm)$ related to massless
spin-$s\geq2$ field was found in \cite{Bekaert:2005vh}, previously the field-theoretical
interpretation of this result was known as the Central on-mass-shell theorem \cite{Vasiliev:1986td,
Fradkin:1987ks}; for the purpose of constructing a Lagrangian the cohomology groups corresponding
to the dynamical field and to the Weyl tensor for a field $(\Ss,q,t=1)$, where $q$ is equal to the
length of the shortest column in $\Ss$, were found in \cite{Alkalaev:2003qv, Alkalaev:2005kw,
Alkalaev:2006rw}; $\coh^q(\Y{s-1,s-t},\Sigm)$, $q=0,1$ corresponding to a partially-massless
spin-$s$ field $(\Y{s},q=1,t)$ were important for \cite{Skvortsov:2006at}; in
\cite{Boulanger:2008up, Boulanger:2008kw} $\coh^q(\Sigm)$ at lower degrees were found for the following fields
$(\Y{2,1},1,1)$, $(\Y{3,1},1,1)$ and $(\Ss,1,1)$ with $\Ss=\Y{s_1,s_2,...,s_n}$ or $\Ss=\Y{s_1,s_1,s_2,...,s_n}$ such that $s_1-s_2\geq4$.

\begin{Corollary} For $\Ds$ such that all of the weights $s_i$ are different, the result turns to a very simple form
because of \be\mspr(\Ds,q,r)=\msp(\Ds^{q-r+1},q,r)\ee and hence
\be\coh^q(\Ds,\Sigm)^{reg}=\sum_{k=0}^{k=q}\msp(\Ds^{q-k+1},q,k).\ee The duality map applied to any
representative of a nontrivial cohomology class with some $q,g,r$ except for those at the maximal
grade produces a representative of the cohomology class with $q+1,g+1,r+1$ labelled by the same
Young diagram.
\end{Corollary}

\subsection{Examples}\label{SbSExamplesSigmaCohomology}
\paragraph{$\boldsymbol{\Ds=\Y{s-1,s-1}}$.} The main theorem applied to ${\Ds=\Y{s-1,s-1}}$ gives the following list of
\Sigm-cohomologies at lower degrees\parbox[t][6pt][c]{0pt}{}\par\noindent
\begin{tabular}{|x{1.1cm}|x{4cm}|x{4cm}|x{4cm}|}
  \hline
   $q\backslash g$ & $0$ & $1$ & $s-1$\tabularnewline \hline
  0 & \emptypar{20pt}$\AYoungM{5}{s-1}$ & $\emptyset$ & $\emptyset$  \tabularnewline
  1 & \emptypar{20pt}$\AYoungM{6}{s}\oplus\AYoungM{4}{s-2}$ &  $\emptyset$  & $\emptyset$   \tabularnewline
  2 & \emptypar{30pt}$\emptyset$  & $\AYoungM{6}{s}\oplus\AYoungM{4}{s-2}$ & $\BYoungM{6}{6}{s}{}$ \tabularnewline
  3 & \emptypar{30pt}$\emptyset$ & $\AYoungM{5}{s-1}$ & $\CYoungM{6}{6}{1}{s}{}{}\oplus\BYoungM{6}{5}{s}{s-1}$  \tabularnewline \hline
\end{tabular}\par\rule{0pt}{14pt}Note that $\epsilon'=1$ (see fig.\,\ref{fig:PicGCMSPR} for the definition of $\epsilon'$) for the only representative of $\coh^{q=0}_{g=0,r=0}$, hence the duality map
takes it to the class at degree $q+2\epsilon'+1=3$. Analogously, $\epsilon'=0$ for both in
$\coh^{q=1}_{g=0}$, hence the duality map takes them to the class at degree two. There are no duals
for those at the maximal grade $s-1$.

Consider a gauge theory with the gauge field given by a one-form $W^\Ds\fm{1}$, which, as is
well-known \cite{Vasiliev:2001wa}, describes a massless spin-$s$ field, i.e. $(\Y{s},1,1)$. The
interpretation of the \Sigm-cohomology is as follows: $\coh^0$ corresponds to a traceless
rank-$(s-1)$ gauge parameter, $\xi^{a(s-1)}$. The dynamical field $\coh^1$ is represented by two
traceless tensors of ranks $s$ and $s-2$, which can be combined into a doubly traceless Fronsdal
field $\phi^{a(s)}$. The gauge transformation law is of first order,
$\delta\phi^{a(s)}=D^a\xi^{a(s-1)}$. In $\coh^2$ we see the second order equations, which are in
one-to-one correspondence with the dynamical fields, suggesting the system admits a Lagrangian
\cite{Lopatin:1987hz, Vasiliev:2001wa}. The Weyl tensor $C^{a(s),b(s)}$ is also present in
$\coh^2$, which is the order $s$ derivative of fields. In $\coh^3$ there are Bianchi identities
both for Fronsdal equations, corresponding to the gauge symmetry with $\xi^{a(s-1)}$ and for the
Weyl tensor, implying that it is constructed out of $\phi^{a(s)}$ rather than being an independent
object.

\paragraph{$\boldsymbol{\Ds=\Y{s_1,s_2}}$.} In this case the table of \Sigm-cohomology at lower degrees reads\parbox[t][6pt][c]{0pt}{} \par\noindent
\begin{tabular}{|x{0.7cm}|x{1.9cm}|x{1.9cm}|x{1.9cm}|x{1.9cm}|x{4cm}|}
  \hline
   $q\backslash g$ & $0$ & $1$ & $s_1-s_2$ & $s_1-s_2+1$ & $s_1$ \tabularnewline \hline
  0 & \emptypar{20pt}$\AYoungM{4}{s_2}$ & $\emptyset$ & $\emptyset$ & $\emptyset$ & $\emptyset$  \tabularnewline
  1 & \emptypar{20pt}$\AYoungM{3}{s_2-1}$ & $\AYoungM{4}{s_2}$ &   $\AYoungM{6}{s_1+1}$ &  $\emptyset$  & $\emptyset$   \tabularnewline
  2 & \emptypar{30pt}$\emptyset$  & $\AYoungM{3}{s_2-1}$ & $\AYoungM{5}{s_1}$  & $\AYoungM{6}{s_1+1}$ & $\BYoungM{6}{5}{s_1+1}{s_2+1}$\tabularnewline
  3 & \emptypar{30pt}$\emptyset$ & $\emptyset$& $\emptyset$ & $\AYoungM{5}{s_1}$ & $\CYoungM{6}{5}{1}{s_1+1}{s_2+1}{}\oplus\BYoungM{6}{4}{s_1+1}{s_2}$  \tabularnewline \hline
\end{tabular}\par
\rule{0pt}{14pt}Consider a gauge theory with gauge field $W^\Ds\fm{1}$, which describes a
partially-massless spin-$(s_1+1)$ field of depth $t=s_1-s_2+1$ \cite{Skvortsov:2006at}. Indeed,
there is a gauge parameter $\xi^{a(s_2)}$ in $\coh^0$; the dynamical fields are $\phi^{a(s_2-1)}$,
$\phi^{a(s_2)}$ and the primary field $\phi^{a(s_1+1)}$ with the highest rank. The appearance of
fields with lower ranks, which cannot be generally associated with the traces of a single field, is
because partially-massless fields lie between massless and massive. For a Lagrangian description of
a massive spin-$(s_1+1)$ field in addition to a traceless field $\phi^{a(s_1+1)}$ one needs
supplementary traceless fields of ranks $s_1-1$, $s_1-2$, ... $1$, $0$, which vanish on-mass-shell
\cite{Fierz:1939ix, Singh:1974qz}. For partially-massless fields this chain becomes shorter because
of disappearance of fields with ranks $s_2-2$,...,$0$. However, not all of the supplementary fields
can now be excluded, these are the fields $\phi^{a(s_2-1)}$ and $\phi^{a(s_2)}$. The gauge
transformation law has schematically the form $\delta \phi^{a(s_1+1)}=D^a...D^a\xi^{a(s_2)}+...$\,.
In $\coh^2$ we see the wave equation for $\phi^{a(s_1+1)}$, Weyl tensor and two more constraints on
supplementary fields.

That there is no Bianchi identity in $\coh^3$ for the gauge symmetry with $\xi^{a(s_2)}$ is due to
the fact that \Sigm{} is an operator that is responsible for expressing fields of higher rank in
terms of derivatives of lower rank fields and hence cannot track out the Bianchi identities of the
form $D^b...D^b G_{b(s_1-s_2+1)a(s_2)}+...\equiv0$, where $G^{a(s_1+1)}=\square
\phi^{a(s_1+1)}+...$ is the equation on $\phi^{a(s_1+1)}$. In this case Bianchi identities
correspond to the reversed situation when lower rank fields are expressed in terms of divergences
of higher rank fields.

\paragraph{$\Ds$ is a $\boldsymbol{(s-1)\times(q+1)}$ block diagram, $\boldsymbol{\Ds=\Yb{(s-1,q+1)}}$.}
Consider now a gauge theory with the field $W^\Ds\fm{q}$, which describes a massless field with
spin $\Ss=\Yb{(s,q)}$. According to \cite{Metsaev:1995re, Brink:2000ag} it is only for these fields
that the number of degrees of freedom in Minkowski space is equal to that in \dSAdS{}. The fields
with $\Ss=\Yb{(s,q)}$ are the true massless fields in this sense. With
$\Xx=\Yb{(s-1,q)}=\Ds^1=...=\Ds^{q+1}$ the \Sigm-cohomology reads
\parbox[t][6pt][c]{0pt}{}
\par\noindent
\begin{tabular}{|x{1.1cm}|x{4cm}|x{4cm}|x{4cm}|}\hline
  $q\backslash g$ & $0$ & $1$ & $s-1$\tabularnewline \hline
  $0$ & \emptypar{20pt}\msp(\Xx,0)=\Xx & $\emptyset$ & $\emptyset$  \tabularnewline
  .. & ... & ... & ....\tabularnewline
  $q-1$ & \emptypar{20pt}\parbox{2.1cm}{\msp(\Xx,q-1)} &  $\emptyset$  & $\emptyset$   \tabularnewline
  $q$ & \emptypar{20pt}\parbox{2.1cm}{\msp(\Xx,q)} &  $\emptyset$  & $\emptyset$   \tabularnewline
  $q+1$ & \emptypar{20pt}$\emptyset$ &  \parbox{2.1cm}{\msp(\Xx,q)}  & \mspr(\Ds,q+1)=\Ds   \tabularnewline
  $q+2$ & \emptypar{20pt}$\emptyset$ &  \parbox{2.1cm}{\msp(\Xx,q-1)} & \parbox{2.3cm}{\mspr(\Ds,q+2)}  \tabularnewline
  .. & ... & ... & ....\tabularnewline
  $2q+1$ & \emptypar{20pt}$\emptyset$ & \parbox{2.1cm}{\msp(\Xx,0)}  &  \parbox{2.3cm}{\mspr(\Ds,2q+1)}  \tabularnewline
  $2q+2$ & \emptypar{20pt}$\emptyset$ & $\emptyset$  &  \parbox{2.3cm}{\mspr(\Ds,2q+2)}  \tabularnewline\hline
\end{tabular}\par\rule{0pt}{14pt}
The gauge parameter at the deepest level of reducibility given by $\coh^0$ is just a traceless
tensor with the symmetry of $\Xx$. Note that $\mspr(\Ds,r)=\msp(\Xx,r)$ if $r\leq q$. The gauge
parameter in $\coh^r$, $r=1,...,q-1$ along with the primary component
\be\Yb{(s,r),(s-1,q-r)}=\hwp(\Xx,r)\subset\msp(\Xx,r)\ee contain certain traces that are needed for
the gauge symmetry to be realized off-shell. The explicit expression for $\xi^\Ds\fm{r}$ reads
\be\xi^{a(s),...,b(s),c(s-1),...,u(s-1)}=\xi^{a(s-1),...,b(s-1),c(s-1),...,u(s-1),\bullet(s-1)|\overbrace{\scriptstyle
a...b}^r}\ee Similarly for the dynamical field in $\coh^{q}$
\be\phi^{a(s),...,u(s)}=\xi^{a(s-1),...,u(s-1),\bullet(s-1)|\overbrace{\scriptstyle a...u}^q}\ee We
see that there is a one-to-one correspondence between the second order equations in $\coh^{q+1}$ and the dynamical
fields in $\coh^{q}$. There is also a generalized Weyl tensor in $\coh^{q+1}$, which is an
irreducible tensor of the Lorentz algebra with the symmetry $\Ds$. By virtue of the definition of
$\mspr$ for diagrams with equal rows, $\mspr(\Ds,q+1)$ contains only one irreducible component,
which has the symmetry of $\Ds$ itself. For higher degrees $q+2$,... $\mspr(\Ds)$ contains also
certain traces. It is easy to see the duality of the form $\coh^{q-k}_{g=0}\sim\coh^{q+k+1}_{g=1}$,
$k=0,...,q$, implying that there is a one-to-one correspondence between equations of motion and
dynamical fields, gauge symmetries at the level-$k$ and the order-$k$ Bianchi identities.

\section{Interpretation of results: Gauge fields vs. Gauge connections}\label{SInterpretation} According to
\cite{Skvortsov:2009zu}, a gauge field defined by a triple $(\Ss,q,t)$ can be described by the gauge connection $W^\Ds\fm{q}$ of
the (anti)-de Sitter algebra $\mathfrak{g}$, where \Ds
\begin{align}\label{GaugeFieldConnection}&\Ss=\parbox{3.4cm}{{\bep(80,80)\unitlength=0.38mm%
\put(0,0){\RectCRowUp{5}{4}{3}{$s_{q+1}$}{$...$}{$s_n$}}%
\put(0,30){\RectCRowUp{9}{8}{7}{$...$}{$s_{q-1}$}{$s_q$}}%
\put(0,60){\RectARowUp{10}{$s_1$}}\put(32,-3){$\msv$}\eep}}&&
(\Ss,q,t)&&\Longleftrightarrow&&(\Ds,q),&& \Ds=\parbox{3.6cm}{\boldpic{\bep(80,80)\unitlength=0.4mm%
\put(0,0){\RectCRowUp{5}{4}{3}{$s_{q+1}$}{$...$}{$s_n$}}%
\put(0,30){\RectCRowUp{9}{8}{6}{$...$}{$s_q-1$}{$s_q-t$}}
\put(0,60){\RectARowUp{10}{$s_1-1$}}\put(32,-3){$\mathfrak{g}$}\eep}}\end{align} or with the indices written explicitly
\begin{align}&\phi^{a(s_1),...,v(s_n)}&&\Longleftrightarrow&&
W^{A(s_1-1),...,B(s_q-1),C(s_q-t),D(s_{q+1}),...,F(s_p)}_{\mu_1...\mu_q}dx^{\mu_1}\wedge...\wedge dx^{\mu_q}\end{align} A field
$(\Ss,q,t)$ can be described by the connection $W^\Ds\fm{q}$ of $\mathfrak{g}$ in the sense that there is the inclusion of exact
sequences, discussed in detail below,
\begin{align}& \qquad... && && \qquad... && &&\qquad... && && \qquad... && && ... \nonumber\\
&\stackB{$\mathfrak{g}$-module}{\Verma{E_0}{\Ss_0}}&&\hookrightarrow&&\stackB{$\phi^{\Ss_0}$}{on-shell}
&&\hookrightarrow&&\stackB{$\bar{\phi}^{\Ss_0}$}{off-shell}&&\hookrightarrow&&\stackB{$e^{\Ll_0}\fm{q}$}{frame-like}&&\hookrightarrow&&
W^\Ds\fm{q}\nonumber\\ & \qquad\uparrow && && \qquad\uparrow && &&\qquad\uparrow && && \qquad\uparrow && && \uparrow \nonumber\\
&\stackB{$\mathfrak{g}$-module}{\Verma{E_1}{\Ss_1}}&&\hookrightarrow&&\stackB{$\xi^{\Ss_1}$}{on-shell}
&&\hookrightarrow&&\stackB{$\bar{\xi}^{\Ss_1}$}{off-shell}&&\hookrightarrow&&\stackB{$\xi^{\Ll_1}\fm{q-1}$}{frame-like}
&&\hookrightarrow&&\xi^\Ds\fm{q-1}\nonumber\\ & \qquad\uparrow && && \qquad\uparrow && &&\qquad\uparrow && && \qquad\uparrow &&
&& \uparrow \nonumber\\ & \qquad... && && \qquad... && &&\qquad... && && \qquad... && && ..., \nonumber
\end{align}
which is exact in vertical arrows, with horizontal arrows denoting the inclusion maps. The leftmost
vertical arrows are the arrows from (\ref{ExactSequence}); the rightmost vertical arrows are
realized as the action of $\DO$ in the complex $\Comp(\Ds,\DO)$; the vertical arrows next to them
are realized as the $\DO$ in the complex $\Comp(\Ds,\DO)$ with $\Ds$ being treated as an
\lorentz-module.

In order to obtain an off-shell formulation for a $(\Ss,q,t)$ field one has to get rid of
differential constraints (\ref{FullSystemB}) on the dynamical field $\phi^{\Ss_0}$ and gauge
parameters $\xi^{\Ss_i}$, which implies that the field content needs to be extended. The extended
fields $\bar{\phi}^{\Ss_0}$ and gauge parameters $\bar{\xi}^{\Ss_i}$ have the same symmetry type
but are no longer irreducible Lorentz tensors, satisfying certain trace constraints that are weaker
than the full tracelessness. We refer to such fields and gauge parameters with relaxed trace
constraints as to {\it extensions}. The fields $\phi^{\Ss_0}$, $\xi^{\Ss_i}$ are the highest weight
parts\footnote{see section \ref{SbSHWPMSP} for the definition.} of the decomposition of
$\bar{\phi}^{\Ss_0}$ and $\bar{\xi}^{\Ss_i}$ into irreducible Lorentz tensors.

The problem with the vertical arrows from the second and third columns, which are realized on
irreducible metric-like fields and on their extensions, respectively, is in that the explicit use
of Young symmetrizers is needed. Let us now define the horizontal arrows that denote the inclusion
maps.

$\Verma{E_i}{\Ss_i}$ is realized on the solutions of (\ref{FullSystemA})-(\ref{FullSystemB}) imposed on the irreducible tensor
field of the Lorentz algebra having the symmetry of $\Ss_i$, which for $i=0$ and for $i=1,...,q$ corresponds to imposing the
equations of motion on the field $\phi^{\Ss_0}$ and to imposing the gauge fixing conditions on the gauge parameters
$\xi^{\Ss_i}$, respectively.

The extended fields are embedded as a maximally symmetric parts\footnotemark[\value{footnote}] into
the connections of the Lorentz algebra. The dynamical field $\bar{\phi}^{\Ss_0}$ and the gauge
parameters $\bar{\xi}^{\Ss_i}$, $i=1,...,q$ are embedded into the generalized frame field
$e^{\Ll_0}\fm{q}$ and $\xi^{\Ll_i}\fm{q-i}$, respectively. The modules $\Ll_i$, $i=0,...,q$ are
certain irreducible Lorentz-modules coming from the restriction of $\Ds$, namely,
$\Ll_i=\Ds^{q-i+1}$.

Written in terms of tensor fields $\bar{\phi}^{\Ss_0}$, $\bar{\xi}^{\Ss_i}$ of the Lorentz algebra,
all expressions, for example, the gauge transformation law and the equations of motion, are
extremely complicated due to the presence of Young symmetrizers and trace projectors. In contrast,
when reformulated in terms of the connection $W^\Ds\fm{q}$ of $\mathfrak{g}$, the theory of any
gauge field $(\Ss,q,t)$ has a very simple form.

The main result of \cite{Skvortsov:2009zu} is that certain components of $W^\Ds\fm{q}$,
$\xi^\Ds\fm{q-1}$, ..., $\xi^\Ds\fm{0}$ were identified with $\phi^\Ss$, $\xi^{\Ss_1}$, ...,
$\xi^{\Ss_q}$ and it was proved that the correct mass-like terms determined by $E_i$ and $\Ss_i$
arise provided that certain equations in terms of $R^\Ds\fm{q+1}$ are imposed on $W^\Ds\fm{q}$ and
certain gauge fixing conditions in terms of $W^\Ds\fm{q}$, $\xi^\Ds\fm{q-1}$, ..., $\xi^\Ds\fm{1}$
are imposed on $\xi^{\Ss_1}$, ..., $\xi^{\Ss_q}$.

Joining together the interpretation of $\coh(\Sigm)$ presented in section \ref{SbSInterpretation} and the theorem on the
structure of $\coh(\Ds,\Sigm)$ yields the following.

The highest grade representatives of $\coh^q$, given by\footnote{see  section
\ref{SbSRestrictionTensorProduct} for the definition.}   $\mspr(\Ds,q)$ with the grade
$g=g(\Ds^{q+1})$, are to be interpreted as the primary dynamical fields. We see that
$\Ss_0=\hwp(\Ds^{q+1},q)$ belongs to $\mspr(\Ds,q)$ with the rest of elements of $\mspr(\Ds,q)$
having smaller rank. Thus, a representative of $\mspr(\Ds,q)$ is given by a tensor field
$\bar{\phi}^{\Ss_0}$ having the symmetry of $\Ss_0$, but $\bar{\phi}^{\Ss_0}$ is not generally
irreducible, containing certain traces. The on-shell field $\phi^{\Ss_0}$ is embedded into
$\bar{\phi}^{\Ss_0}$ as the highest weight part, $\phi^{\Ss_0}=\hwp(\Ds^{q+1},q)$. Actually, it is
easy to find among the fields $\omega^{\Xx}\fm{q}$ coming from the restriction of $W^\Ds\fm{q}$ the
one that contains $\bar{\phi}^{\Ss_0}$. It is the generalized frame field $e^{\Ll_0}\fm{q}$ with
$\Ll_0=\Ds^{q+1}$.

Likewise for the level-$i$ gauge parameter $\xi^{\Ss_i}$. The extension $\bar{\xi}^{\Ss_i}$ is the
representative of $\coh^{q-i}$ at the highest grade, $\coh^{q-i}=\mspr(\Ds^{q-i+1},q-i)$. The
on-shell gauge parameter $\xi^{\Ss_i}$ enters $\bar{\xi}^{\Ss_i}$ as the highest weight part,
$\xi^{\Ss_i}=\hwp(\Ds^{q-i+1},q-i)$.

The number of derivatives connecting the highest grade fields of $\coh^{q-i}$ and $\coh^{q-i-1}$ is
equal to $g(\Ds^{q-i+1})-g(\Ds^{q-i})+1$. Substituting the explicit form of $\Ds$ in terms of
$(\Ss,q,t)$ gives exactly the difference of the lowest energies $E_{i-1}-E_i$, which also counts
the number of derivatives for a field-theoretical realization.

Note that for massless unitary fields, i.e. those having $t=1$ and $q$ equal to the height of the first block of $\Ss$, $\coh^{q-i}_{g>0}=\emptyset$,
$i=0,...,q$ and thus the primary dynamical field $\bar{\phi}^{\Ss_0}$ appears at the lowest grade,
so do its gauge parameters at all levels. For nonunitary massless fields and for partially-massless
fields in addition to the primary dynamical field $\bar{\phi}^{\Ss_0}$ certain other dynamical
fields having smaller rank can appear at lower grade.

Note also that the highest rank representatives of some $\coh^{q'}_{g'}$ correspond to the on-shell
situation, in which all lower rank representatives of $\coh^{q'}_{g'}$ are zero by virtue of gauge
fixing conditions. So to make contact with the on-shell description in terms of metric-like fields,
presented in section \ref{SFieldsConnections}, it is sufficient to interpret the highest rank
representatives only.

As for the field equations the situation is more complicated. Since in a general case of
mixed-symmetry field, i.e. the one having $\Ss=\Y{s_1,s_2,...}$ with $s_1\neq s_2>0$, all of the
first order constraints (\ref{FullSystemB}) cannot be achieved via imposing gauge conditions on a
single gauge parameter $\bar{\xi}^{\Ss_1}$, the full system of equations of motion consists both of
second and first order equations. Thus we cannot expect the number of equations to be equal to the
number of component fields in $\bar{\phi}^{\Ss_0}$. The representatives of $\coh^{q+1}_{g}$ with
$g$ equal to the grade of the primary dynamical field, i.e. to $g(\Ds^{q+1})$ correspond to certain
first order gauge invariant equations for $\bar{\phi}^{\Ss_0}$.

For $t=1$ we see that the highest rank representatives of $\coh^{q+1}_{g}$ have the symmetry of all
the gauge parameters\footnote{Recall that a massless spin-$\Ss$ field in Minkowski space has a
number of gauge symmetries with the parameters whose Young diagrams are obtained by removing one
cell from $\Ss$ in all possible ways.} for a massless spin-$\Ss$ field in Minkowski space, except
for $\xi^{\Ss_1}$. The number of representatives of $\coh^{q+1}_{g}$ with the highest
rank equals the number of first order constraints (\ref{FullSystemB}) minus one. One constraint
of (\ref{FullSystemB}) can be imposed as a gauge condition for $\xi^{\Ss_1}$. In the Minkowski case
the rest of the constraints (\ref{FullSystemB}) can be imposed with the help of other gauge
parameters, one parameter - one constraint.

For $t>1$, i.e. for partially-massless fields, the gauge symmetry with $\xi^{\Ss_1}$ is so weak that none of the constraints
(\ref{FullSystemB}) can be imposed with the help of $\xi^{\Ss_1}$. Thus, the number of representatives of $\coh^{q+1}_{g}$
with the highest rank equals the number of first order constraints in (\ref{FullSystemB}).

Similar statements can be made about the correspondence of the the highest rank representatives of $\coh^{q+i}_{g}$ and level-$i$
gauge parameters of a massless spin-$\Ss$ field in Minkowski space. This correspondence is not accidental since the first order
constraints (\ref{FullSystemB}) are the same for gauge fields in Minkowski and \dSAdS, with $D_m$ being the covariant derivative
in the space of interest. The difference is that for massless fields in Minkowski space these results can be achieved via gauge
fixing and for gauge fields in \dSAdS{} most of the constraints (all for $t>1$) are to be imposed as independent equations.

As for second order field equations that are the representatives of $\coh^{q+1}_{g+1}$, we see that at least there is a
representative in $\coh^{q+1}_{g+1}$ that has the symmetry of $\Ss_0$. It is for this representative that the mass-like term was calculated in \cite{Skvortsov:2009zu} and was shown to coincide with the group-theoretical result (\ref{WEMassFormula}). There is no one-to-one correspondence $\coh^{q}_{g}\leftrightarrow\coh^{q+1}_{g+1}$ between the second order equations and primary fields
$\bar{\phi}^{\Ss_0}$ since by virtue of Bianchi identities a number of the second order equations corresponding to the traces of
$\bar{\phi}^{\Ss_0}$ can be obtained as the derivative of certain first order constraints from $\coh^{q+1}_{g}$.

The only highest grade representative of $\coh^{q+1}$ is given by $\mspr(\Ds,q+1,0)=\msp(\Ds^{q+2},q+1)=\hwp(\Ds^{q+2},q+1)$, it
has the symmetry $\Ss_{-1}$ of a Weyl tensor for a field $(\Ss,q,t)$.

Let us consider certain higher degree cohomology groups for $t=1$. The representatives of $\coh^{q+2}_{g+1}$ correspond to the
Bianchi identities. As is expected, there is a representative of $\coh^{q+2}_{g+1}$ having the symmetry of $\xi^{\Ss_1}$.
Actually, there are also the representatives having the symmetry of all gauge parameters of a massless spin-$\Ss$ field in
Minkowski space. This suggests the enhancement of the gauge symmetry in the flat limit $\lambda^2\rightarrow0$
\cite{Brink:2000ag}.

\paragraph{Physical degrees of freedom.} The \Sigm-cohomology can be used to directly count the number of physical degrees of freedom,
as it was demonstrated in the case of massless fields in Minkowski space in \cite{Skvortsov:2008vs}. However instructive it might
be, there is no need to count degrees of freedom explicitly. It is sufficient to look at $\coh^{q+1}(\Ds,\Sigm)$.

Firstly, suppose that not all of the equations in $\coh^{q+1}(\Ds,\Sigm)$ are imposed. It implies
that, in addition to the Weyl tensor and its descendants coupled to the gauge module, some other
components of the field curvature $R^\Ds\fm{q+1}$ are nonzero on-shell, these can be parameterized
by new fields, which are analogous to Weyl tensor. We can analyze the Bianchi identities and solve
them with some other fields, and so on. As a result, an infinite-dimensional module grows at each
place where some equation was not imposed, which gives rise to new degrees of freedom, thus making
the system reducible.

Contrariwise, if the equation $R^\Ds\fm{q+1}=[\mbox{Weyl tensor}]$ were describing more physical degrees of freedom than the
number of states in the corresponding irreducible representation $\Irrep{\Ss}{E_0}$, there should be a possibility to further
impose certain gauge invariant differential equations that would make the system irreducible. This contradicts the statement that
all gauge invariant independent equations are classified through \Sigm-cohomology.

Consequently, once all components of the field curvature except for the Weyl tensor and its
descendants are set to zero, the system automatically describes the correct number of physical
degrees of freedom.

\paragraph{Remarks on the Weyl module.} Recall that the generalized Weyl tensor is by definition the lowest order gauge-invariant combination of derivatives of the
dynamical field $\phi^{\Ss}$ that is allowed to be nonzero on-mass-shell. The Weyl tensor is a representative of
$\coh^{q+1}(\Ds,\Sigm)$ at the highest nontrivial grade.

It is difficult to write down explicit expressions for the Weyl module inasmuch as we are faced
with Young symmetrizers since the Weyl tensor and its descendants are tensors of the Lorentz
algebra. The problem is to adjust coefficients in front of \Sigm{} and \Sigp{} acting on the fields
from the Weyl module. In somewhat different setup it was done in \cite{Boulanger:2008up,
Boulanger:2008kw}.

The spectrum of fields of the Weyl module is in the results of \cite{Alkalaev:2003qv, Boulanger:2008up, Boulanger:2008kw,
Skvortsov:2009zu}. To determine this spectrum one can use the following heuristic consideration: given the symmetry types $\Ss_0$ and $\Ss_1$ of a dynamical field and its gauge parameter, respectively, one marks the extra cells of $\Ss_0$ as compared to
$\Ss_1$. The marked cells correspond to derivatives in the gauge transformation law. Then one adds cells to $\Ss_1$, emulating
various derivatives of the dynamical field, until one of the new cells is found in the same column with a marked cell. The latter situation correspond to implicit antisymmetrization of two derivatives, which is identically zero in Minkowski space or gives a
tensor of a lower rank in (anti)-de Sitter space. Therefore, a diagram with a new cell being in the same column with a marked
cell corresponds to certain gauge invariant expression. The diagram with the smallest number of added cells is the Weyl tensor,
all the others are its descendants. See fig. \ref{fig:PicGCWeylModule} for the illustration.

\begin{figure}
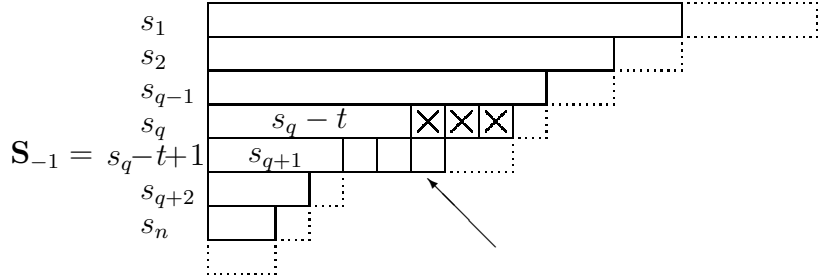

\caption{The spectrum of the Weyl module is shown. The cells corresponding to derivatives are marked. The descendants of the Weyl
tensor are obtained by adding cells in the places outlined by a dotted line. The arrow shows the place where two covariant
derivatives, one from the gauge transformations and another one from the expression of the Weyl tensor in terms of gauge
potential $\phi^{\Ss}$, happens to be in the same column.}
\begin{align}
\Ss_{-1}=\parbox{9.6cm}{{\bep(240,90)(-40,0)\unitlength=0.45mm%
\put(-20,42){$s_q$}\put(-20,52){$s_{q-1}$}\put(-20,62){$s_2$}\put(-20,72){$s_1$}\put(-30,32){$s_q\!\!-\!t\!\!+\!1$}%
\put(0,10){\RectBRowUp{3}{2}{}{}}\put(-20,22){$s_{q+2}$}\put(-20,12){$s_n$}%
\put(0,30){\RectCRowUp{10}{6}{4}{}{$s_q-t$}{$s_{q+1}$}}\put(40,30){\YoungC}\put(60,40){\YoungCcC}%
\put(60,40){\Rect{3}{1}}\put(40,30){\Rect{3}{1}}\put(85,8){\vector(-1,1){20}}%
\put(140,70){\RectDotted{4}{1}}\put(120,60){\RectDotted{2}{1}}\put(100,50){\RectDotted{2}{1}}\put(90,40){\RectDotted{1}{1}}%
\put(30,20){\RectDotted{1}{1}}\put(70,30){\RectDotted{2}{1}}\put(20,10){\RectDotted{1}{1}}\put(0,0){\RectDotted{2}{1}}%
\put(0,60){\RectBRowUp{14}{12}{}{}}\eep}}\nonumber\end{align} \label{fig:PicGCWeylModule}
\end{figure}

Note that a Weyl tensor alone does not determine $(\Ss,q,t)$, as it was the case for massless fields in Minkowski space, but the
Weyl module does, of course.

We see that the spectrum of the Weyl module looks almost as the one coming from a restriction of a tensor of the (anti)-de Sitter
algebra whose symmetry is given by a Young diagram with the first row tending to infinity. Therefore, the method for calculating
the \Sigm-cohomology developed in Appendix B can be applied to the Weyl module too. Consequently, we can find a rather simple and
complete answer for the structure of $\coh(\Sigm)$ of the Weyl module in terms of $\coh(\Sigm)$ of the gauge module. Namely,
there is a one-to-one correspondence $\coh^{k}_{i}(\mbox{Weyl},\Sigm)\leftrightarrow\coh^{q+k+1}_{g'+i}(\mbox{gauge},\Sigm)$,
$i=0,1,...$ where $g'$ is the grade of the Weyl tensor in $\coh^{q+1}(\mbox{gauge},\Sigm)$, i.e. $g'=g(\Ds^{q+2})$, between the
(reducible) Bianchi identities for the Weyl tensor in the gauge module and those in the Weyl module. This result confirms that
the two modules are glued properly.

\section{The simplest mixed-symmetry field}\setcounter{equation}{0}\label{SExampleTwoRow}
To illustrate we consider a massless unitary field of spin $\Y{s_1,s_2}$, i.e. $\Ss=\Y{s_1,s_2}$ and $q=t=1$. The exact sequence
(\ref{ExactSequence}) defining the irreducible representation $\Irrep{E_0}{\Y{s_1,s_2}}$ with $E_0=d+s_1-3$ given by
(\ref{AllLevelsE}) reads
\be\ComplexC{\Verma{E_0+1}{\Y{s_1-1,s_2}}}{\Verma{E_0}{\Y{s_1,s_2}}}{\Irrep{E_0}{\Y{s_1,s_2}}}\nonumber.\ee

\paragraph{On-shell metric-like formulation, \cite{Metsaev:1995re, Metsaev:1997nj}.} The field potential $\phi^{a(s_1),b(s_2)}$ is an irreducible Lorentz tensor field having the symmetry
of $\Ss$ and satisfies (\ref{FullSystemA})-(\ref{FullSystemB})
\begin{align}
(\Box+m^2)\phi^{a(s_1),b(s_2)}&=0,\label{TwoRowWave}\\
\DL_c \phi^{a(s_1-1)c,b(s_2)}=\DL_c\phi^{a(s_1),b(s_2-1)c}&=0
\end{align}
where the mass-like parameter is determined by $(\Ss,1,1)$ according to (\ref{WEMassFormula})\be
m^2=\lambda^2\left((s_1-2)(d+s_1-3)-s_1-s_2\right)\ee The equations are invariant under the gauge transformations $\delta
\phi^{a(s_1),b(s_2)}=\DL^a\xi^{a(s_1-1),b(s_2)}$, where the gauge parameter is an irreducible Lorentz tensor having the symmetry
of $\Ss_1=\Y{s_1-1,s_2}$ and is subjected to (\ref{FullSystemA})-(\ref{FullSystemB}) equations with \be
{m_\xi}^2=\lambda^2\left((s_1-1)(d+s_1-2)-s_1-s_2+1\right).\ee

\paragraph{Off-shell metric-like formulation.} The extended field content is given by the field potential $\phi^{a(s_1),b(s_2)}$
having the symmetry of $\Ss$ and satisfying \be\label{TwoRowFieldTracelessness}\eta_{cc}\phi^{a(s_1),b(s_2-2)cc}\equiv0, \qquad
\eta_{cc}\eta_{dd}\phi^{a(s_1-4)ccdd,b(s_2)}\equiv0,\ee and thus not being irreducible. The gauge parameter
$\xi^{a(s_1-1),b(s_2)}$ needs not be extended in this rather simple case, thus being irreducible
\be\label{TwoRowGaugeParamTracelessness}\eta_{cc}\xi_1^{a(s_1-3)cc,b(s_2)}\equiv\eta_{cc}\xi_1^{a(s_1-2)c,cb(s_2-1)}\equiv\eta_{cc}\xi_1^{a(s_1-1),b(s_2-2)cc}\equiv0.
\ee The algebraic constraints (\ref{TwoRowFieldTracelessness}) imposed on the field $\phi^{a(s_1),b(s_2)}$ implies that it
consists of three irreducible Lorentz tensors, two of them corresponding to nontrivial traces,
\be\label{TwoRowFieldDecomposition}\phi^{a(s_1),b(s_2)}\longleftrightarrow{\minorpic{\RectBRow{6}{4}{$\scriptstyle
s_1$}{$\scriptstyle s_2$}}}\oplus{\minorpic{\RectBRow{5}{4}{$\scriptstyle s_1-2$}{$\scriptstyle
s_2$}}}\oplus{\minorpic{\RectBRow{5}{4}{$\scriptstyle s_1-1$}{$\scriptstyle s_2-1$}}}.\ee The gauge parameter is a single
irreducible Lorentz tensor,
\be\label{TwoRowGaugeParamDecomposition}\xi_1^{a(s_1-1),b(s_2)}\longleftrightarrow{\minorpic{\RectBRow{6}{4}{$\scriptstyle
s_1-1$}{$\scriptstyle s_2$}}}.\ee The gauge transformations have the same form \be\delta
\phi^{a(s_1),b(s_2)}=D^a\xi^{a(s_1-1),b(s_2)}\label{TwoRowGaugeLaw}.\ee Despite the fact that the gauge parameter
$\xi^{a(s_1-1),b(s_2)}$ is no longer required to have vanishing divergences (\ref{FullSystemB}), one can verify that the
algebraic constraints (\ref{TwoRowFieldTracelessness}) and (\ref{TwoRowGaugeParamTracelessness}) are consistent with
(\ref{TwoRowGaugeLaw}).

The field equations consist of the two independent equations
\begin{align}
&D_nD^n\phi^{a(s_1),b(s_2)}-D^aD_n\phi^{a(s_1-1)n,b(s_2)}+\frac12D^aD^a\phi^{a(s_1-2)n\phantom{n},b(s_2)}_{\phantom{a(s_1-2)n}n}+\nonumber\\
&\qquad+2\lambda^2\eta^{aa}\phi^{a(s_1-2)n\phantom{n},b(s_2)}_{\phantom{a(s_1-2)n}n}+
2\lambda^2\eta^{ab}\phi^{a(s_1-1)n,\phantom{n}b(s_2-1)}_{\phantom{a(s_1-1)n,}n}+m^2\phi^{a(s_1),b(s_2)}=0,\label{TwoRowSecondOrder}\\
&D^a \phi^{a(s_1-1)n,b(s_2-1)}_{\phantom{a(s_1-1)n,b(s_2-1)}n}-D_n\phi^{a(s_1),b(s_2-1)n}=0,\label{TwoRowFirstOrder}
\end{align}
the one that reduces to the wave equation (\ref{TwoRowWave}) after imposing certain gauge, and the other that excludes low spin
states with the spins $\Y{s_1,s_2-i}$, $i=1,...,s_2$ inasmuch as $\DL_c\phi^{a(s_1),b(s_2-1)c}=0$ cannot be imposed as a gauge
fixing condition, and is, in fact, an independent constraint, whose gauge-invariant implementation is given by
(\ref{TwoRowFirstOrder}).

To conclude, it can be shown that (\ref{TwoRowSecondOrder}), (\ref{TwoRowFirstOrder}) together with the gauge transformations
(\ref{TwoRowGaugeLaw}) imply the correct number of physical degrees of freedom, which corresponds to $\Irrep{E_0}{\Y{s_1,s_2}}$.

\paragraph{Unfolded formulation, \DSADS-covariant.} According to the statement of \cite{Alkalaev:2003qv}, the gauge module
for a gauge field $(\Y{s_1,s_2},1,1)$ is given by a single connection of the (anti)-de Sitter
algebra $W^{A(s_1-1),B(s_1-1),C(s_2)}\fm{1}$ that takes values in the irreducible representation
with the symmetry of $\Ds=\Y{s_1-1,s_1-1,s_2}$. The gauge transformations and the field curvature
read
\begin{align}\label{TwoRowAdSGauge}
\delta W^{A(s_1-1),B(s_1-1),C(s_2)}\fm{1}&=\DO \xi^{A(s_1-1),B(s_1-1),C(s_2)}\fm{0}, \\
R^{A(s_1-1),B(s_1-1),C(s_2)}\fm{2}&=\DO W^{A(s_1-1),B(s_1-1),C(s_2)}\fm{1}.\label{TwoRowAdSStrength}
\end{align}
The field curvature is manifestly gauge invariant $\delta R^{A(s_1-1),B(s_1-1),C(s_2)}\fm{2}=0$.

The correct field equations are imposed by setting all components of the field curvature to zero
except for the Weyl tensor and its descendants. The Weyl tensor is an irreducible Lorentz tensor
$C^{a(s_1),b(s_1)}$ having the symmetry of $\Y{s_1,s_1}$. Among its descendants
$C^{a(s_1+i),b(s_1),c(j)}$, $i=0,1,2,...$, $j=0,...,s_2$ those with $i=0$ couple with the gauge
module. The set $C^{a(s_1),b(s_1),c(j)}$, $j=0,...,s_2$ can be embedded into the irreducible tensor
$C^{A(s_1),B(s_1),C(s_2)}\fm{0}$ of the (anti)-de Sitter algebra subjected to \be
V_M\left(C^{A(s_1),B(s_1-1)M,C(s_2)}\fm{0}-\frac{s_2}{s_1-s_2+1}C^{A(s_1),B(s_1-1)C,C(s_2-1)M}\fm{0}\right)\equiv0,\ee
which removes the components of the $\res(\Y{s_1,s_1,s_2})$ having the symmetry of
$\Y{s_1,s_1-i,j}$ with $i=1,...,s_1-s_2$. Therefore, the field equations have the form \be
R^{A(s_1-1),B(s_1-1),C(s_2)}\fm{2}=E_ME_NC^{A(s_1-1)M,B(s_1-1)N,C(s_2)}\fm{0},\label{TwoRowAdSEquations}\ee
and we do not consider the constraints on $C^{A(s_1),B(s_1),C(s_2)}\fm{0}$ following from the
Bianchi identity $\DO R^{A(s_1-1),B(s_1-1),C(s_2)}\fm{2}\equiv0$.

\paragraph{\Sigm-map.} In the table below we draw the diagrams corresponding to the elements of $\coh^{q}_{g}(\Ds,\Sigm)$ for low
$q$, which are relevant for the field $\phi^{a(s_1),b(s_2)}$.\\
\begin{tabular}{|c|c|c|c|@{\hspace{2pt}}c@{\hspace{2pt}}|c|}
  \hline
   $q\backslash g$ & 0 & 1 & $s_1\!-\!s_2\!-\!1$ & $\,s_1\!-\!s_2$ & $s_1\!-\!1$  \\ \hline
  0 & \emptypic{15}$\BYoung{6}{4}{s_1-1}{s_2}$ & $\emptyset$ & $\emptyset$ & $\emptyset$ & \\ \hline
  1 & \emptypic{20}$\substack{\parbox{42pt}{\BYoung{7}{4}{s_1}{s_2}}{\displaystyle\oplus}\,\parbox{30pt}{\BYoung{5}{4}{s_1-2}{s_2}}
  \\ {\displaystyle\oplus}\,\parbox{36pt}{\BYoung{6}{4}{s_1-1}{s_2-1}}}\parbox[t][12pt][c]{0pt}{}$ &  & $\emptyset$ & $\emptyset$ & $\emptyset$  \\ \hline
  2& \emptypic{15}$\BYoung{7}{4}{s_1}{s_2-1}\oplus\BYoung{5}{4}{s_1-2}{s_2-1}$ & $\BYoung{7}{4}{s_1}{s_2}\oplus\BYoung{5}{4}{s_1-2}{s_2}$ & \emptypic{15}$\BYoung{6}{6}{s_1}{}$ & $\emptyset$ & $\emptyset$  \\ \hline
  3 & $\emptyset$ & \emptypic{20}$\substack{\parbox{44pt}{\BYoung{7}{4}{s_1}{s_2-1}}{\displaystyle\oplus}\,\parbox{32pt}{\BYoung{5}{4}{s_1-2}{s_2-1}}
  \\ {\displaystyle\oplus}\,\parbox{36pt}{\BYoung{6}{4}{s_1-1}{s_2}}}\parbox[t][12pt][c]{0pt}{}$ & \emptypic{15}$\BYoung{6}{5}{s_1}{s_1-1}$ & \emptypic{15}$\BYoung{6}{6}{s_1}{}$ & \emptypic{20}$\CYoung{6}{6}{4}{s_1}{s_1}{s_1+1}$  \\ \hline
  4 & $\emptyset$ & \emptypic{15}\BYoung{6}{4}{s_1-1}{s_2-1} &
  \emptypic{15}$\BYoung{5}{5}{s_1-1}{s_1-1}$ & \emptypic{15}$\BYoung{6}{5}{s_1}{s_1-1}$ &
  \emptypic{30}$\substack{\parbox{38pt}{\CYYoung{6}{6}{4}{s_1}{s_1}{s_2+1}{\YoungA}}{\displaystyle\oplus}\,\parbox{38pt}{\CYoung{6}{6}{4}{s_1}{s_1}{s_2}}\\
  {\displaystyle\oplus}\,\parbox{36pt}{\CYoung{6}{5}{4}{s_1}{s_1-1}{s_2+1}}}\parbox[t][22pt][c]{0pt}{}$  \\ \hline
\end{tabular}

\paragraph{Unfolded formulation, Lorentz-covariant.} The Lorentz-covariant frame-like formulation is constructed by reducing the representation
of the (anti)-de Sitter algebra $\mathfrak{g}$ down to the representations of the Lorentz algebra
\be
\res^{\mathfrak{g}}_{\lorentz}\left(\parbox{2.3cm}{\boldpic{\RectCRow{6}{6}{4}{$s_1-1$}{$s_1-1$}{$s_2$}}}\right)
=\bigoplus_{j=0}^{j=s_1-s_2-1}
\bigoplus_{i=0}^{i=s_2}\parbox{2.5cm}{\RectCRow{6}{4}{2}{$s_1-1$}{$s_2+j$}{i}},\ee hence, the gauge
field $W^\Ds\fm{1}$ decomposes into the following set of \lorentz-connections \be
W^{A(s_1-1),B(s_1-1),C(s_2)}\fm{1}\longleftrightarrow \omega^{a(s_1-1),b(s_1-i),c(j)}\fm{1},\qquad
i\in[1,s_1-s_2], \quad j\in[0,s_2].\ee The gauge parameter and the field curvature decompose in a
similar way.

The physical field $\phi^{a(s_1),b(s_2)}$, which is a representative of $\coh^{q=1}_{g=0}(\Sigm)$, is embedded into the
generalized frame field \be\label{TwoRowFrame}
e^{a(s_1-1),b(s_2)}\fm{1}=W^{a(s_1-1),b(s_2)\bullet(s_1-s_2-1),\bullet(s_2)}\fm{1}\ee as the \msp{}, which is given by
\be\phi^{a(s_1),b(s_2)}=e^{a(s_1-1),b(s_2)|a},\qquad\quad e^{a(s_1-1),b(s_2)|c}=e^{a(s_1-1),b(s_2)}_\mu h^{\mu c}.\ee The
decomposition of $e^{a(s_1-1),b(s_2)|c}$ into irreducible \lorentz-tensors reads
\begin{align}
{\minorpic{\parbox{44pt}{\RectBRow{6}{4}{$\scriptstyle s_1-1$}{$\scriptstyle
s_2$}}}}\otimes{\minorpic{\YoungA}}&=\left[\rule{0pt}{14pt}\parbox{51pt}{{\minorpic{\RectBRow{7}{4}{$\scriptstyle
s_1$}{$\scriptstyle s_2$}}}}\oplus\parbox{38pt}{\minorpic{\RectBRow{5}{4}{$\scriptstyle s_1-2$}{$\scriptstyle
s_2$}}}\oplus\parbox{40pt}{\minorpic{\RectBRow{5}{3}{$\scriptstyle s_1-1$}{$\scriptstyle
s_2-1$}}}\right]\oplus\left[\rule{0pt}{18pt}\parbox{44pt}{\minorpic{\RectCYoung{6}{4}{{$\scriptstyle s_1-1$}}{{$\scriptstyle
s_2$}}{\YoungA}}}\oplus\parbox{44pt}{\minorpic{\RectBRow{6}{5}{$\scriptstyle s_1$}{$\scriptstyle s_2+1$}}}\right].
\end{align}
The terms in the first brackets correspond to $\coh^{q=1}_{g=0}(\Sigm)$, the field components in the second brackets are
\Sigm-exact inasmuch as they can be gauged away by the gauge parameters that have the same symmetry type.

The gauge parameter $\xi^{a(s_1-1),b(s_2)}$, which is a representative of $\coh^{q=0}_{g=0}(\Sigm)$, is defined as
\be\xi^{a(s_1-1),b(s_2)}=\xi^{a(s_1-1),b(s_2)}\fm{0}=\xi^{a(s_1-1),b(s_2)\bullet(s_1-s_2-1),\bullet(s_2)}\fm{0}.\ee

\paragraph{From frame-like to metric-like, field equations.} Having identified the representatives of
$\coh^{q=1}_{g}(\Sigm)$ and $\coh^{q=0}_{g}(\Sigm)$ corresponding to the dynamical field $\phi^{a(s_1),b(s_2)}$ and to the gauge
parameter $\xi^{a(s_1-1),b(s_2)}$, let us now turn to $\coh^{q=2}_{g}(\Sigm)$ whose representatives give the field equations.

The 'raw' field curvature at the lowest grade, i.e. the field curvature for the generalized frame
field $e^{a(s_1-1),b(s_2)}\fm{1}$, and the two 'raw' field curvatures at grade-one for the
auxiliary fields $\omega^{a(s_1-1),b(s_2+1)}\fm{1}$ and $\omega^{a(s_1-1),b(s_2),c}\fm{1}$ read
\begin{align}
&\label{TwoRowLowestStrength}R^{a(s_1-1),b(s_2)|\Sa\Sa}=D^{\Sa}e^{a(s_1-1),b(s_2)|\Sa}+(s_1-s_2-1)\omega^{a(s_1-1),b(s_2)\Sa|\Sa}+s_2\omega^{a(s_1-1),b(s_2),\Sa|\Sa}\\
&R^{a(s_1-1),b(s_2+1)|\Sa\Sa}=D^{\Sa}\omega^{a(s_1-1),b(s_2+1)|\Sa}+(s_1-s_2-2)W^{a(s_1-1),b(s_2+1)\bullet(s_1-s_2-3)\Sa,\bullet(s_2)|\Sa}+\nonumber\\
&\qquad\qquad+s_2W^{a(s_1-1),b(s_2+1)\bullet(s_1-s_2-2),\bullet(s_2-1)\Sa|\Sa}-\eta^{a\Sa}W^{a(s_1-2)\bullet,b(s_2+1)\bullet(s_1-s_2-2),\bullet(s_2)|\Sa}+\nonumber\\
&\qquad\qquad\qquad\qquad-\eta^{b\Sa}e^{a(s_1-1),b(s_2)|\Sa}\\
&R^{a(s_1-1),b(s_2);c|\Sa\Sa}=D^{\Sa}\omega^{a(s_1-1),b(s_2),c|\Sa}+(s_1-s_2-1)W^{a(s_1-1),b(s_2)\bullet(s_1-s_2-2)\Sa,c\bullet(s_2-1)|\Sa}+\nonumber\\
&\qquad+(s_2-1)W^{a(s_1-1),b(s_2)\bullet(s_1-s_2-1),c\bullet(s_2-2)\Sa|\Sa}-\eta^{a\Sa}W^{a(s_1-2)\bullet,b(s_2)\bullet(s_1-s_2-1),c\bullet(s_2-1)|\Sa}+\nonumber\\
&\qquad\qquad-\eta^{b\Sa}W^{a(s_1-1),b(s_2-1)\bullet(s_1-s_2),c\bullet(s_2-1)|\Sa}-\eta^{c\Sa}e^{a(s_1-1),b(s_2)|\Sa},
\end{align}
where we have substituted (\ref{TwoRowFrame}) and
\begin{align}
\omega^{a(s_1-1),b(s_2+1)|m}&=W^{a(s_1-1),b(s_2+1)\bullet(s_1-s_2-2),\bullet(s_2)|m},\\
\omega^{a(s_1-1),b(s_2);c|m}&=W^{a(s_1-1),b(s_2)\bullet(s_1-s_2-1),c\bullet(s_2-1)|m}.
\end{align}
The 'raw' field $\omega^{a(s_1-1),b(s_2);c|m}$ have partial Young symmetry properties,
\be(s_2+1)\omega^{a(s_1-1),b(s_2);b|m}=-(s_1-s_2-1)\omega^{a(s_1-1),b(s_2+1)|m}.\ee When expressed in terms of
$\phi^{a(s_1),b(s_2)}$, the representative of $\coh^{q=2}_{g=0}$ \be R^{a(s_1-1),b(s_2-2)m|a}_{\phantom{a(s_1-1),b(s_2-2)m|a}m}=
D^a e^{a(s_1-1),b(s_2-1)n|}_{\phantom{a(s_1-1),b(s_2-1)n|}n}-D_ne^{a(s_1-1),b(s_2-1)n|a}\ee give the same expression as
(\ref{TwoRowFirstOrder}). The simplest way to take the representative of $\coh^{q=2}_{g=1}$ and derive (\ref{TwoRowSecondOrder})
is to compute
\begin{align}
&(s_1-s_2-t)R^{a(s_1-1),b(s_2)\bullet(s_1-s_2-t-1)n,\bullet(s_2)|a}_{\phantom{a(s_1-1),b(s_2)\bullet(s_1-s_2-t-1)n,\bullet(s_2)|a}n}+
s_2R^{a(s_1-1),b(s_2)\bullet(s_1-s_2-t),\bullet(s_2-1)n|a}_{\phantom{a(s_1-1),b(s_2)\bullet(s_1-s_2-t),\bullet(s_2-1)n|a}n}\nonumber=0,
\end{align}
where the terms with the derivative of $\omega^{a(s_1-1),b(s_2+1)|m}$ and $\omega^{a(s_1-1),b(s_2),c|m}$ can be expressed from
\begin{align}
&R^{a(s_1-1),b(s_2)\bullet(s_1-s_2-1),\bullet(s_2)|an}=D^ae^{a(s_1-1),b(s_2)|n}-D_ne^{a(s_1-1),b(s_2)|a}+\nonumber\\
&\qquad\qquad\qquad\qquad-(s_1-s_2-1)\omega^{a(s_1-1),b(s_2)n|a}-s_2\omega^{a(s_1-1),b(s_2);n|a}=0,
\end{align} that is obtained from (\ref{TwoRowLowestStrength}) by symmetrizing one index $\Sa$ with $a(s_1-1)$.

Consequently, formulated in terms of a single connection $W^\Ds\fm{q}$ (\ref{TwoRowAdSGauge}), (\ref{TwoRowAdSStrength}),
(\ref{TwoRowAdSEquations}) the theory has a very simple form and the representatives of the \Sigm-cohomology give all relevant
quantities. Technical complications arise when passing to a metric-like formulation, in which the origin of the trace constraints
(\ref{TwoRowFieldTracelessness}-\ref{TwoRowGaugeParamTracelessness}) is not self-evident and the field equations are more
involved.

\section{Conclusions}\label{SConclusions}
In this paper we have presented the results on the \Sigm-cohomology for the algebraic complex
$\Comp(\Ds,\Sigm)$ associated with the differential complex $\Comp(\Ds,\DO)$ of gauge connections
generated by an arbitrary irreducible finite-dimensional representation $\Ds$ of the (anti)-de
Sitter algebra. The complex $\Comp(\Ds,\Sigm)$ arises if we would like to reinterpret the fields of
the (anti)-de Sitter algebra in terms of the Lorentz one.

As distinct from the rich complex $\Comp(\Ds,\Sigm)$, by virtue of the Poincare lemma
$\Comp(\Ds,\DO)$ is locally exact in degrees greater than zero and provides no interesting
information.

The results on the \Sigm-cohomology are important for constructing Lagrangians inasmuch as the
Lagrangian equations must set to zero the representatives of $\coh(\Sigm)$ that are associated with
the equations of motion, as it is established by direct method, for example, for massless symmetric
spin-$s$ fields in (anti)-de Sitter space in \cite{Vasiliev:1980as, Lopatin:1987hz}, for massless
arbitrary-spin fields in Minkowski space in \cite{Skvortsov:2008sh} and for massless two-column
fields in (anti)-de Sitter space in \cite{Alkalaev:2003hc}.

Restriction to the \Sigm-cohomology in the sector of fields and gauge parameters yields the minimal
formulation for a given gauge field, so that all algebraic gauges are imposed and all auxiliary
fields are expressed in terms of dynamical ones. The differential form structure is lost in
$\coh(\Sigm)$ in that the representatives of $\coh(\Sigm)$ are certain Lorentz tensors, which may
be embedded into the forms of the degree dictated by $\coh(\Sigm)$, however, the constraints
involving both the form and the fiber indices arise (these are formulated in terms of the
background frame and its inverse). Therefore, the minimal formulation operates with a collection of
metric-like fields, which may look strange, e.g. having complicated trace constrains.

That the \Sigm-cohomology of the gauge module in the sector of the Weyl tensor together with its
Bianchi identities is perfectly glued to the \Sigm-cohomology of the Weyl module suggests that the
spectrum of fields of the Weyl module is correct. Thus, the next problem is to find the higher-spin
algebras \cite{Konshtein:1988yg, Vasiliev:2004cm} for mixed-symmetry fields and construct the
corresponding nonlinear equations, which are believed to exist \cite{Metsaev:1993mj,
Metsaev:1993ap, Metsaev:2005ar, Boulanger:2004rx} and are likely to be constructed within the
unfolded approach.

\section*{Acknowledgements}
The author is grateful to M.A.Vasiliev, K.B.Alkalaev, O.V.Shaynkman, N.Boulanger, P.Sundell and especially to E.Feigin for many
helpful discussions on \Sigm. The author wishes to thank M.A.Vasiliev for reading the manuscript and giving many valuable
comments. The work was supported in part by grants RFBR No. 08-02-00963, LSS-1615.2008.2, INTAS No. 05-7928, UNK LPI, by the Landau
scholarship and by the scholarship of the Dynasty foundation.

\appendix
\renewcommand{\theequation}{\Alph{section}.\arabic{equation}}
\section*{Appendix A: Tensor products}\label{AppTensorProducts}
\setcounter{equation}{0}\setcounter{section}{1} Let $\Xx$ and $\Yy$ be two irreducible representations of some Lie algebra
$\mathfrak{g}$, then we can take the tensor product of $\Xx\otimes_{\mathfrak{g}}\Yy$ and decompose it into the direct sum
\be\Xx\otimes_{\mathfrak{g}}\Yy=\bigoplus_{\Zz} C^{\Zz}_{\Xx,\Yy}\Zz\ee over irreducible modules $\Zz$ with multiplicities given
by the Littlewood-Richardson coefficients $C^{\Zz}_{\Xx,\Yy}$. For the reader's convenience we present below the tensor product
rules for $\mathfrak{g}$ being $\sld$ or $\sod$ and for $\Yy$ being a one column Young diagram.

\paragraph{$\boldsymbol{\sld}$-tensor product.} Let $\Xx=\Yb{(s_i,p_i)}$ and $\Yy=\Ya{q}$ be irreducible representations of \sld, then the decomposition of the tensor product
$\Xx\otimes\Ya{q}$ is of the form
\be\label{AppTensorProductGL}\Xx\otimes_\sld\Ya{q}=\bigoplus_{\alpha_1+...+\alpha_{N+1}=q}\Xx^{\{\alpha_j\}},\ee
where the multiplicity of each irreducible representation
$\Xx_{\{\alpha_j\}}$ is $1$ and the sum is over all Young diagrams
$\Xx^{\{\alpha_j\}}$
\be\label{AppSlTensorProduct}\Xx^{\{\alpha_1,...,\alpha_{N+1}\}}=\parbox{4.5cm}{\unitlength=0.35mm\SlNProduct}\qquad:\alpha_i\leq
p_i,\quad\mbox{for}\quad i=1,...,N,\ee with
$\alpha_1+...+\alpha_{N+1}=q$. The diagrams of total height greater
than $d$ correspond to identically zero tensors and must be
discarded. The first column must be removed from the diagrams of
height $d$, so that the resulted diagram has height $(d-1)$ at most.

Since $\Ya{q}$ corresponds to a representation on antisymmetric tensors, two cells of $\Ya{q}$ must not appear in the same row in
$\Xx$, which determines the shape of $\Xx^{\{\alpha_j\}}$.

\paragraph{$\boldsymbol{\sod}$-tensor product.} To decompose the tensor product $\Xx\otimes_\sod\Ya{q}$ of two \sod{} representations
$\Xx=\Yb{(s_i,p_i)}$ and $\Yy=\Ya{q}$ is a more complicated problem because of ability to take
traces with the help of the invariant tensor $\eta_{ab}$. The decomposition of
$\Xx\otimes_\sod\Ya{q}$ has the form
\be\label{AppSoProduct}\Xx\otimes_\sod\Ya{q}=\bigoplus_{\{\alpha_j,\beta_i\}}
N_{\{\alpha_j,\beta_i\}}\Yy^{\{\alpha_j,\beta_i\}},\ee where the sum is over all Young diagrams
$\Yy^{\{\alpha_j,\beta_i\}}$
\be\label{AppSoTensorProduct}\Yy^{\{\alpha_j,\beta_i\}}=\parbox{4.5cm}{\unitlength=0.40mm\TracedTensorProduct}\qquad:
\alpha_i+\beta_i\leq p_i,\quad\mbox{for}\quad i=1,...,N,\ee provided there is a nonnegative integer
$\rho$ such that \be\label{AppRhoDefinition}
q=\sum_{i=1}^{i=N}\left(\alpha_i+\beta_i\right)+\alpha_{N+1}+2\rho,\ee The multiplicity
$N_{\{\alpha_j,\beta_i\}}$ of $\Yy^{\{\alpha_j,\beta_i\}}$ is given by the number of integer
partitions \be\label{AppPartitions}
N_{\{\alpha_j,\beta_i\}}=\partition{\epsilon_1,...,\epsilon_N|\rho},\quad
\epsilon_i=p_i-\alpha_i-\beta_i,\ee of $\rho$ into the sum of $N$ integers $k_1+...+k_N=\rho$ such
that $0\leq k_i\leq\epsilon_i$. The trace order $r$ for $\Yy_{\{\alpha_j,\beta_i\}}$ is \be
r=\sum_{i=1}^{i=N}\beta_i+\rho.\ee

The meaning of the above is as follows. Before adding the cells of
$\Ya{q}$ to $\Xx$, one can take traces, i.e. to remove pairs of
cells, one from $\Yy$ and another one from $\Xx$. Since $\Yy$
corresponds to antisymmetric tensor representations, two cells
cannot be removed from the same row of $\Xx$. Therefore, each trace
of order $r$ corresponds to some integer partition of $r$ into the
sum $t_1+...+t_N$ provided that $t_i\leq p_i$. A subcolumn of height
$t_i$ is removed from the bottom-right of the $i$-th block. Then,
the rest of the cells from $\Yy$, i.e. $(q-r)$, can be added to what
$\Xx$ has turned into after taking traces. Recall that the $i$-th
block consists now of two subblocks $\Yb{s_i,p_i-t_i}$ and
$\Yb{s_i-1,t_i}$. There are two types of places to which cells can
now be added: $\alpha_i$ cells are added to the top-right of the
subblock $\Yb{s_i,p_i-t_i}$; $(t_i-\beta_i)$ cells are added to the
top-right of the subblock $\Yb{s_i-1,t_i}$. The latter leads to the
possibility to get the same diagrams in many different ways, i.e.
results in the multiplicity greater than one. Different partitions
of $\rho$ into $\sum_i(t_i-\beta_i)$ provided the trace order
$r=\sum_i t_i$ and all $\beta_i$ are fixed results in identical
Young diagrams. So $\rho$ is the number of cells that were first
removed and then restored.

When the height of some diagram $\Yy^{\{\alpha_j,\beta_i\}}$ is greater than $[d/2]$, the antisymmetric invariant tensor
$\epsilon_{a_1...a_d}$ has to be used to transform it to a diagram with height less than $[d/2]$ or to impose (anti)selfduality
conditions when the height is $[d/2]$ for $d$ even. We implicitly assume that the rules described above are applicable to all
tensors products considered in the paper.

\section*{Appendix B: \Sigm-cohomology}
\setcounter{equation}{0}\setcounter{section}{2}
\subsection*{The case of $\mathfrak{sl}(d+1)$} Below we compute the \Sigm{}-cohomology for \sldd{}
and \sodd{}, starting with the case of \sldd. As it has been already
mentioned, the \Sigm{}-cohomology in the case of \sldd{} is closely
related to the ordinary Lie algebra cohomology. This is not so in
the case of \sodd, for which a different method should be developed.
We first apply the new method to the case of \sldd{}, so that one
can check the results. The method is to embed $\Comp(\Ds,\Sigm)$
into the tensor product of much more simple complexes associated to
one-row Young diagrams, $\Ds=\Y{s}$, then the cohomology of
$\Comp(\Ds,\Sigm)$ with $\Ds$ of general shape can be obtained with
the help of certain projectors, whose kernels we are able to find.

Set $\mathfrak{g}=\sldd$ and $\mathfrak{h}=\sld$. Let $V$ be a
fundamental (vector) representation of $\mathfrak{g}$. For any
nonzero (compensator) vector\footnote{To be strict, a vector from
the dual space is also needed, we skip obvious details since we can
talk about tensors of \sodd{} modulo traces rather than of
\sldd-tensors.} $v\in V$, we have the decomposition $V=V_0\oplus
V_1$, where $V_1$ is a one-dimensional subspace spanned by the
vector $v$ and $V_0$ is a vector representation of $\mathfrak{h}$.
We define a complex \be\label{SldComplex}\Sigm:\quad
T^m(V)\otimes\Lambda^q(V_0)\longrightarrow
T^m(V)\otimes\Lambda^{q+1}(V_0),\ee where $T^m(V)$ is the $m$-th
tensor power of $V$, $\Lambda^q(V_0)$ is the $q$-th exterior power
of $V_0$ and \Sigm{} is a nilpotent operator
\begin{align}
\Sigm:&(X_1\otimes...\otimes X_m)\otimes z_1\wedge...\wedge z_q\longrightarrow\nonumber\\
&\longrightarrow\sum_{i=1}^{i=m} (X_1\otimes...\widehat{\otimes
X_{i-1}}\otimes v\otimes X_{i+1}\otimes...\otimes X_m)\otimes
\rho_v(X_i)\wedge z_1\wedge...\wedge z_q,\end{align} where
$X_1,...,X_m\in V$, $z_1,...,z_q\in V_0$ and $\rho_v(X)$ is a
projector onto $V_0$, i.e. $\rho_v(v)=0$, $\rho_v(x)=x$ for $x\in
V_0$.

In order to single out from $T^m(V)$ an irreducible
$\mathfrak{g}$-module $\Ds$ with rank $m=\mbox{rank}(\Ds)$, the
Young symmetrizer $\pi_\Ds$ is needed. A Young symmetrizer is a
weighted sum over all permutations of $m=\mbox{rank}(\Ds)$ factors
\be \pi_\Ds[X_1\otimes...\otimes X_m]=\sum_{\{\sigma\}}f(\sigma)
X_{\sigma_1}\otimes...\otimes X_{\sigma_m},\ee where the weight
function $f(\sigma)$ is determined by $\Ds$, e.g. for $\Ds=\Y{m}$
$f(\sigma)=(m!)^{-1}$. It is not hard to see that the Young
symmetrizer $\pi_\Ds$ commutes with the action of $\Sigm$.
Therefore, given any irreducible $\mathfrak{g}$-module $\Ds$ the
nilpotent operator \be\Sigm:\quad
\Ds\otimes\Lambda^q(V_0)\longrightarrow
\Ds\otimes\Lambda^{q+1}(V_0),\ee is well-defined, so is the
corresponding complex on $\Ds\otimes\Lambda(V_0)$, which we denote
$\Comp(\Ds,\Sigm)$. The definition just given is intermediate in a
sense that it does not deal with explicit indices as in section
\ref{SbSTwoSimpleCases}, but seems to depend on the choice of the
compensator as compared to the invariant definition of section
\ref{SbGCSigmaMinusDef}.

On account of the embedding $\Ds\hookrightarrow T^m(V)$ with $m=\mbox{rank}(\Ds)$, any element of $\mathfrak{h}$-module $\Xx$
from $\res^\mathfrak{g}_\mathfrak{h}\Ds$ can be written as a sum over elements of the form \be
\pi_\Ds(\overbrace{v\otimes...\otimes v}^{m-k}\otimes \pi_\Xx( x_1\otimes...\otimes x_k)),\ee where $x_1,...,x_k\in V_0$ and
$\Xx\in\res^\mathfrak{g}_\mathfrak{h}\Ds$.

As it has been already mentioned, $\Ds\otimes\Lambda(V_0)$ can be considered as an $\mathfrak{h}$-module, with the action of
$\mathfrak{h}$ on $\Lambda(V_0)$ induced from that on $V_0$. Due to $\mathfrak{h}v=0$, \Sigm commutes with the action of
$\mathfrak{h}$ and, hence, both the elements of the complex and the representatives of \Sigm-cohomology can be considered as
$\mathfrak{h}$-modules so that one can deal with irreducible $\mathfrak{h}$-modules only.

\paragraph{de Rham complex.} Consider the de Rham complex $\Rham$ on the polynomials in $d$ commuting variables $y^a$ with
the action of the de Rham differential $\pl$ defined by
\be\label{GCRhamActionA}\pl(\omega(y^a|\theta^b))=\theta^c\frac{\pl}{\pl
y^c}\omega(y^a|\theta^b),\ee where Grassmann variables $\theta^a$
are the analogs of $dx^a$. Rewritten in components, the action of
$\pl$ on the component of degree $k$ and $q$ in $y^a$ and
$\theta^b$, respectively, reads
\be\label{GCRhamActionB}\pl(\omega^{a(k)|\mu[q]})=k\omega^{a(k-1)\mu|\mu[q]},\ee
where the antisymmetrization over the form indices $\mu$ is implied.
As is well known, the cohomology of the de Rham complex is
concentrated in the constant polynomials in $y^a$ and $\theta^b$,
i.e. \be\begin{tabular}{|x{1.1cm}|x{2cm}|x{2cm}|}
  \hline
   $q\backslash g$ & $0$ & $>0$ \tabularnewline \hline
  0 & \emptypar{20pt}$\bullet$ & $\emptyset$   \tabularnewline
  $>0$ & \emptypar{20pt}$\emptyset$ &  $\emptyset$    \tabularnewline\hline
\end{tabular}\ee
Decomposing $\omega^{a(k)|\mu[q]}$ into irreducible
$\mathfrak{h}$-modules, we have
\be\label{AppRhamDecomposition}\omega^{a(k)|\mu[q]}\sim\parbox{60pt}{\bep(60,50)\put(0,40){\RectT{6}{1}{\TextTop{k}}}\put(0,0){\RectT{1}{4}{\TextCenter{q}}}\eep}\oplus
\parbox{70pt}{\bep(70,30)\put(10,30){\RectT{6}{1}{\TextTop{k}}}\put(0,0){\RectT{1}{4}{\TextCenter{q}}}\eep}.\ee
Evidently, the second, the less antisymmetric component of
(\ref{AppRhamDecomposition}) is exact inasmuch as the component with
the same symmetry type is in $\omega^{a(k+1)|\mu[q-1]}$, the two
components forming a so-called contractible pair. Therefore, the
total space of $\Rham$ is decomposed into a direct sum of
contractible pairs plus constants $\omega^{|}$ that represent the
only nontrivial cohomology class.

\paragraph{de Rham complex $\Rham^s$ with constraints or $\Comp(\Y{s},\Sigm)$.} Consider now the de Rham complex $\Rham^s$
on polynomials in $y^a$ having degree not greater than $s$. Obviously, it can be realized as the space of degree $s$ polynomials
in $d+1$ variables $y^a$ and $y^\bullet$, $\pl$ is defined by the same formula. Therefore, $\Rham^s$ is the simplest example of
the $\Sigm$-complex $\Comp(\Ds,\Sigm)$ with $\Ds=\Y{s}$, and $\omega^{a(k)|\mu[q]}$ is identified with the projection
$W^{a(k)\bullet(s-k)|\mu[q]}$ of a single form $W^{A(s)|\mu[q]}$ valued in \sldd-module $\Y{s}$, c.f. (\ref{SbSSigmaMinusSL}).

It is easy to find the cohomology of $\Rham^s$: since it is the restriction of the de Rham complex,
in addition to the de Rham cohomology we will have new cohomology classes with representatives
coming from those contractible pairs at grade-$s$ that get broken over restriction - these former
exact forms represent now nontrivial cohomology classes since the $s+1$ grade becomes trivial.
Thus, \be\label{GCRhamSCoh}\begin{tabular}{|x{1.1cm}|x{3cm}|x{3cm}|}
  \hline
   $q\backslash g$ & $0$ & $s$ \tabularnewline \hline
  0 & \emptypar{20pt}$\aO^0=\bullet$ & $\emptyset$   \tabularnewline
  1 & \emptypar{20pt}$\emptyset$ &  $\aB^1=\parbox{53pt}{\AYoungM{6}{s+1}}$    \tabularnewline
  2 & \emptypar{30pt}$\emptyset$  & $\aB^2=\parbox{53pt}{\BYoungM{6}{1}{s+1}{}}$  \tabularnewline \hline
\end{tabular}\ee
where the notation $\aO^0$, $\aB^q$ was introduced to label the
cohomology classes.

\paragraph{$\Rham^{s_1,s_2}$ complex.} Of use for us will be also the complex $\Rham^{s_1,s_2}$, $s_2>0$,
obtained from $\Rham^{s_1}$ by restricting further polynomials in
$y^a$ to have degree not less than $s_2$, or more formally
\be\ComplexC{\Rham^{s_2}}{\Rham^{s_1,s_2}}{\Rham^{s_1}}.\ee The need
for $\Rham^{s_1,s_2}$ is due to Young symmetrizers, which confine
the rows of Young diagrams coming from
$\res^\mathfrak{g}_\mathfrak{h}\Ds$ to be between two integers, the
smallest of which being generally greater than zero.

Again, the cohomology of $\Rham^{s_1,s_2}$ is easy to find - it is sufficient to find contractible pairs in $\Rham^{s_1}$ at
grade $s_2$ that get broken, yielding new cohomology \be\label{GCRhamSSCoh}\begin{tabular}{|x{1.1cm}|x{3cm}|x{3cm}|}
  \hline
   $q\backslash g$ & $s_2$ & $s_1$ \tabularnewline\hline
  0 & \emptypic{15}$\aA^0=\parbox{34pt}{\AYoungM{4}{s_2}}$ & $\emptyset$   \tabularnewline
  1 & \parbox[c][25pt][c]{0pt}{}$\aA^1=\parbox{34pt}{\BYoungM{4}{1}{s_2}{}}$ &  $\aB^1=\parbox{53pt}{\AYoungM{6}{s_1+1}}$    \tabularnewline
  2 & \parbox[c][40pt][c]{0pt}{}$\aA^2=\parbox{34pt}{\CYoungM{4}{1}{1}{s_2}{}{}}$  & $\aB^2=\parbox{53pt}{\BYoungM{6}{1}{s_1+1}{}}$  \tabularnewline \hline
\end{tabular}\ee
It is worth mentioning that the representatives at higher degrees are obtained by adding cells to the bottom-left of $\aA^0$ and
$\aB^1$.

\paragraph{$\Rham(\Ds,\pl)$ complex.} Given any diagram $\Ds=\Y{s_1,...,s_n}$ we define the complex
$\Rham(\Ds,\pl)$ as a tensor product of $\Rham^{s_i}$, \be\Rham(\Ds,\pl)=\Rham^{s_1}\otimes \Rham^{s_2}\otimes...\otimes
\Rham^{s_{n-1}}\otimes \Rham^{s_n},\ee with the action of the total differential $\pl$ defined through the action of $\pl$ on
each multiplier $\omega^i\fm{q_i}\in\Rham^{s_i}$ as
\begin{align}\nonumber\pl&\left(\omega^1\fm{q_1}\otimes\omega^2\fm{q_2}\otimes...\otimes\omega^n\fm{q_n}\right)=
\pl(\omega^1\fm{q_1})\otimes\omega^2\fm{q_2}\otimes...\otimes\omega^n\fm{q_n}+(-)^{q_1}\omega^1\fm{q_1}\otimes\pl(\omega^2\fm{q_2})\otimes...\otimes\omega^n\fm{q_n}+...\end{align}

A simple fact from the spectral sequences theory tells us that the cohomology $\coh^q(\Ds,\pl)$ of
$\Rham(\Ds,\pl)$ is just the tensor product of cohomology groups at each factor. However, the
complex $\Rham(\Ds,\pl)$ is still far from $\Comp(\Ds,\Sigm)$ inasmuch as (1) no Young conditions
are imposed; (2) each of the factors possesses its own copy $\Lambda(V_0)$, i.e. the elements of
the complex are differential multi-forms\footnote{Multi-form is an element of the direct product of
several copies of the exterior algebra. With application to higher-spin theories multi-forms were
studied in \cite{Medeiros:2002ge, Medeiros:2003dc, Bekaert:2002dt, Bekaert:2006ix}.} rather than
just forms. The complex in question $\Comp(\Ds,\Sigm)$ can be extracted from $\Rham(\Ds,\pl)$ by
applying two projectors $\pi_\Ds$ and $\piLambda$, where $\piLambda$ singles out the most
antisymmetric part of a multiform with no effect on coefficients, i.e.
\be\piLambda:\Lambda^{q_1}\otimes...\otimes\Lambda^{q_n}\longrightarrow\Lambda^{q_1+q_2+...+q_n}.\ee
It is not hard to see that the projectors $\pi_\Ds$ and $\piLambda$ commute both with $\pl$ and with each
other, the latter is evident since $\pi_\Ds$ affects only the coefficients of multi-forms while
$\piLambda$ affects only multi-forms. However, in getting the cohomology $\coh(\Ds,\Sigm)$ of
$\Comp(\Ds,\Sigm)$ by applying $\pi_\Ds$ and $\piLambda$ to the cohomology $\coh(\Ds,\pl)$ of
$\Rham(\Ds,\pl)$ we may meet two obstructions: (1) certain cohomology classes can fall into the
kernel of $\pi_\alpha$, $\alpha=\Ds,\boldsymbol{\Lambda}$; (2) there can be contractible pairs
$E=\pl(F)$ such that $F$ does not belong to $\ker(\pi_\alpha)$ but $E\in\ker(\pi_\alpha)$ and thus
$F$ becomes a representative of a nontrivial cohomology class for $\Comp(\Ds,\Sigm)$. It turns out
that it is rather simple to find the kernel of $\pi_\alpha$ and we can also track the appearance of
new cohomology through (2).

\paragraph{The properties of $\boldsymbol{\piDs}$.}

To begin with, let us note that given a Young diagram, say $\Xx$, and an irreducible tensor with
the symmetry of \Xx, say $C^\Xx$, written in symmetric basis, it is not necessary for the number of
indices of some sort over which the total symmetrization in $C^\Xx$ is performed to equal the
length of the corresponding row in \Xx. We refer to the irreducible $\mathfrak{h}$-tensors with the
number of indices of each sort being equal to the length of the corresponding row as to the tensors
with canonical arrangement of indices and noncanonical otherwise. For instance, given $\Xx=\Y{k,m}$
then $C^{a(k),b(m)}$ is a canonical arrangement and $C^{a(k-i)b(i),b(m)}$ with $i>0$ is not. Also
note that an $\mathfrak{h}$-tensor obtained by contracting a number of compensators with an
irreducible $\mathfrak{g}$-tensor is not generally irreducible. For the example of a
$\mathfrak{g}$-tensor with the symmetry of $\Ds=\Y{s_1,s_2}$ we have \be\label{GCTwoRowProjectorsA}
W^{a(k)\bullet(s_1-k),b(m)\bullet(s_2-k)}=\sum_{j=0}^{k+j\leq s_1}\alpha^{k,m}_j\theta(s_1-k-j)
C^{a(k)b(j),b(m-j)},\ee where $\theta(k\geq0)=1$, $\theta(k<0)=0$ and it is natural to set
$\alpha^{k,m}_{j=0}=1$, so that
\be\pi_{\Xx}\left(W^{a(k)\bullet(s_1-k),b(m)\bullet(s_2-m)}\right)=C^{a(k),b(m)}\label{AppTwoRowDecomposition}\ee
for some $\Xx=\Y{k,m}$ provided that $\Xx\in\res^\mathfrak{g}_\mathfrak{h}\Ds$. Extraction from
(\ref{GCTwoRowProjectorsA}) of irreducible components with the symmetry different from $\Xx$ we may
call noncanonical projection. Symmetrization of all 'a'-indices with one index 'b' in
(\ref{AppTwoRowDecomposition}) does not yield zero except for the term with $\alpha^{k,m}_{j=0}$,
rather it gives a recurrent equation for $\alpha^{k,m}_j$. The solution is
$\alpha^{k,m}_j=\frac{(-)^j(s_1-k)!(k-m+j)!}{(s_1-k-j)!(k-m+2j)!}$, which does not degenerate in
the range of definition. Since each irreducible module in $\res^\mathfrak{g}_\mathfrak{h}\Ds$
appears once, there is no confusion with noncanonical projections.

Let $C^{a(k),b(m)}$ be an irreducible tensor with the symmetry of $\Xx=\Y{k,m}$. It can be embedded
into the elements $\omega^{a(k+i)|b(m-i)}$ of $\Rham^{s_1,s_2}$ in a canonical way if $i=0$ and
noncanonically if $i>0$. Since each element of $\res^\mathfrak{g}_\mathfrak{h}\Ds$ comes with
multiplicity one\footnote{This is true for the cases considered in the paper with ($\mathfrak{g}$,
$\mathfrak{h}$) being (\sldd, \sld) or (\sodd, \sod).}, the projector $\piDs$ maps all components
having the symmetry of $\Xx$ to the same element of $\res^\mathfrak{g}_\mathfrak{h}\Ds$, possibly
modulo an overall factor, if $\Xx\in\res^\mathfrak{g}_\mathfrak{h}\Ds$ and to zero otherwise.
Therefore, it is sufficient to deal with irreducible tensors with canonical arrangements of
indices.

The kernel of $\pi_\Ds$ is easy to find: if for some canonical $F$ we have $\pl(F)\neq0$ and $\pl(F)\in\ker(\piDs)$, then it
implies that in $\pl(F)$ the number of indices of some sort is less than the length of the corresponding row in $\Ds$, i.e.
$\pl(F)$ does not belong to $\res^\mathfrak{g}_\mathfrak{h}\Ds$. This being said, new cohomology appears when passing from
$\Rham^{s_1,...,s_n}$ to $\Comp(\Ds,\Sigm)=\piDs\piLambda\left[\Rham^{s_1,...,s_n}\right]$, the new cohomologies are given by
those new in $\Rham^{s_i,s_{i+1}}$ as compared to $\Rham^{s_i}$.

\paragraph{tensor product cohomology.} Consider the projection $\pi_\Ds\piLambda\left[\coh(\Ds,\pl)\right]$
of the cohomology $\coh(\Ds,\pl)$, which is given by taking the tensor product
$\coh(\Rham^{s_1,s_2})\otimes \coh(\Rham^{s_2,s_3})\otimes...$ and applying then
$\pi_\Ds\piLambda$. Consider the tensor product
$\omega=\omega^1\otimes...\otimes\omega^{n-1}\otimes\omega^n$ of the representatives $\omega^i$ of
nontrivial cohomology classes of $\Rham^{s_{i},s_{i+1}}$. It turns out that if at least one of the
representatives $\omega^1$, ... , $\omega^{n-1}$ corresponds to a cohomology class that is
characterized by a Young diagram with more than one row then $\pi_\Ds\piLambda\left[\omega\right]$
is a representative of the trivial cohomology class in $\Comp(\Ds,\Sigm)$. Recalling the notation
of (\ref{GCRhamSCoh}) and (\ref{GCRhamSSCoh}), we can put it differently by saying that left
multipliers of the form $\aA^{q>0}\otimes$ and $\aB^{q>1}\otimes$ yield trivial cohomology in
$\Comp(\Ds,\Sigm)$. Indeed, let us write down all four options for the tensor product of cohomology
where the first multiplier is represented by a Young diagram with more than one row (two for
simplicity)
\begin{align*}
&\aA^{q>0}\otimes\aA&&\sim&&\parbox{50pt}{\RectBYoung{5}{$s_2$}{\YoungCcA}}\otimes\parbox{40pt}{\RectARow{3}{$s_3$}}&&\xrightarrow{\pi_\Ds\piLambda}&&
\parbox{50pt}{\bep(50,20)\put(0,0){\RectBYoung{5}{$s_2$}{\YoungCcA}}\put(10,0){\RectARow{3}{$s_3$}}\eep},\\
&\aA^{q\geq0}\otimes\aB&&\sim&&\parbox{50pt}{\RectBYoung{5}{$s_2$}{\YoungCcA}}\otimes\parbox{60pt}{\RectAYoung{5}{$s_2$}{\YoungCcA}}&&\xrightarrow{\pi_\Ds\piLambda}&&\emptyset,\\
&\aB^{q>1}\otimes\aA&&\sim&&\parbox{70pt}{\RectAYoung{6}{$s_1$}{\YoungCcAA}}\otimes\parbox{40pt}{\RectARow{3}{$s_3$}}&&\xrightarrow{\pi_\Ds\piLambda}&&
\parbox{80pt}{\bep(80,20)\put(0,0){\RectAYoung{7}{$s_1$}{\YoungCcAA}}\put(10,0){\RectARow{3}{$s_3$}}\eep},\\
&\aB^{q>1}\otimes\aB&&\sim&&\parbox{80pt}{\RectAYoung{7}{$s_1$}{\YoungCcAA}}\otimes\parbox{60pt}{\RectAYoung{5}{$s_2$}{\YoungCcA}}&&\xrightarrow{\pi_\Ds\piLambda}&&
\parbox{80pt}{\bep(80,20)\put(0,0){\RectAYoung{7}{$s_1$}{\YoungCcAA}}\put(10,0){\RectAYoung{5}{$s_2$}{\YoungCcA}}\eep}\sim0,\\
\end{align*}
where checked cells correspond to the form indices in tensor language, e.g.
{\RectBYoung{5}{$s_2$}{\YoungCcA}} correspond to a representative of the form $C^{a(s_2),\mu}$,
which is closed inasmuch as tensors with less than $s_2$ indices 'a' belong to the kernel of
$\pi_\Ds$. So, $\aA^{q\geq0}\otimes\aB$ is mapped to zero since the first row in a Young diagram
cannot be shorter than the second one; $\aB^{q>1}\otimes\aB$ is also mapped to zero because two
form indices of different sorts appear in the same group of symmetric indices, which gives zero
after applying $\piLambda$; both $\aA^{q>0}\otimes\aA$ and $\aB^{q>1}\otimes\aB$ are mapped to
exact forms in $\Comp(\Ds,\Sigm)$ inasmuch as one form index in the second group of indices can now
result from applying $\pi_\Ds\piLambda\pl$ to $C^{a(s_2),b(s_3+1)}$ and $C^{a(s_1)\mu,b(s_2+1)}$,
respectively - roughly speaking on account of definite Young symmetry and the fact that forms of
different sorts become identical via $\piLambda$, certain arrangements of indices in a tensor can
now be obtained through $\pi_\Ds\piLambda\pl$, which is impossible in $\Rham(\Ds,\pl)$. Note that
the disappearance of classes represented by a tensor product of Young diagrams with more than one
row concerns left multipliers in the tensor product and has no effect on the last multiplier
$\Rham^{s_n}$ in $\Rham^{s_1,...,s_n}$, which is always the rightmost one.

\paragraph{The properties of $\boldsymbol{\piLambda}$.} Since a component with some definite Young symmetry can enter
more than one element of $\Rham^{s_1,s_2}$ even for $q=0$, one can adjust coefficients if front of
them to get a closed form after applying $\piLambda$, e.g.
\be\label{GCExampleLambda}\omega=\sum_{i=0}^{i=n}\frac{(-)^i}{(n-i)!i!}C^{a(n-i)b(i)},\qquad
\piLambda\left[\pl(\omega)\right]=0,\ee where it has been taken into account that the action $\pl$
on some multi-form $B^{a(k)|b(m)}\fm{q_1,q_2}$ reads as
\be\label{GCSigmaMinusActionExample}\pl\left(B^{a(k)|b(m)}_{\mu[q_1]|\nu[q_2]}\right)=
kB^{a(k-1)\mu|b(m)}_{\mu[q_1]|\nu[q_2]}+mB^{a(k)|b(m-1)\nu}_{\mu[q_1]|\nu[q_2]},\ee where indices
$\mu$ and $\nu$ correspond to two different sorts of forms. The projector $\piLambda$ roughly
speaking replaces all form indices of different sorts with indices of just one sort, say $\mu$,
with further antisymmetrization. It is important that $\pl$ consists of the two parts in this
example, with one decreasing the number of $a$'s and another one decreasing the number of $b$'s.
Therefore, having some $B^{a(k)|b(m)}\fm{q_1,q_2}$ possessing only one $\mathfrak{h}$-irreducible
component and trying two solve the closeness condition we can meet two situations: (1)
$\piLambda\left[\pl\left(B^{a(k)|b(m)}\fm{q_1,q_2}\right)\right]\neq0$ and to compensate we can
introduce $B^{a(k-1)|b(m+1)}\fm{q_1,q_2}$ and $B^{a(k+1)|b(m-1)}\fm{q_1,q_2}$ -
(\ref{GCExampleLambda}) is an example of this type. However, the process runs out with
$B^{a(0)|b(m+k)}\fm{q_1,q_2}$ and $B^{a(k+m)|b(0)}\fm{q_1,q_2}$. Taking into account the properties
of $\piDs$ and the range of tensor ranks of $\Rham^{s_1,s_2}$ we see that $k+m=s_2$, i.e. the
tensor has the lowest possible grade and hence must be totally symmetric,
$B^{a(k)|b(m)}\fm{q_1,q_2}\equiv B^{a(k)b(m)}\fm{q_1,q_2}$ thus representing no new cohomology
class; (2) $\piLambda\left[\pl\left(B^{a(k)|b(m)}\fm{q_1,q_2}\right)\right]=0$ implies that
$B^{a(k)|b(m)}\fm{q_1,q_2}$ is of the form (with canonical arrangement of indices)
\begin{align*}
&B^{a(k)|b(m)}\fm{q_1,q_2}&&\sim&&\parbox{70pt}{\bep(70,30)\put(0,0){\RectAYoung{6}{$k$}{\YoungCcAAA}}\put(10,10){\RectARow{5}{$m$}}\eep}&&\xrightarrow{\displaystyle\quad\pl\quad}&&
\parbox{70pt}{\bep(70,30)\put(0,0){\RectAYoung{5}{$k-1$}{\YoungCcBAA}}\put(10,10){\RectARow{5}{$m$}}\eep}\oplus
\parbox{70pt}{\bep(70,30)\put(0,0){\RectAYoung{6}{$k$}{\YoungCcAAA}}\put(10,10){\RectAYoung{4}{$m-1$}{\YoungCcA}}\eep},
\end{align*} where checked cells correspond to form indices of different sorts such that $\pl$ gives zero only after applying
$\piLambda$, e.g. $\pl\left(C^{a(k)\nu,b(m)\mu}\right)\neq0$ and $\piLambda\left[\pl\left(C^{a(k)\nu,b(m)\mu}\right)\right]=0$.
However, all these candidates for new cohomology either are exact analogously to the case of $\aA^{q>0}\otimes\aA$ and
$\aB^{q>1}\otimes\aB$, or are equivalent to the old classes of the form $\aA\otimes\aA$, $\aB\otimes\aA$ or $\aB\otimes\aB$.

\paragraph{To sum up,}\hspace{-0.4cm} nontrivial cohomology classes are generated by various tensor products of the most symmetric
representatives of $\coh(\Rham^{s_i,s_{i+1}})$, $i<n$, and $\coh(\Rham^{s_n})$. Since any row in a
Young diagram cannot be longer than the previous one, products of the form $\aA\otimes\aB$ are
forbidden. Consequently, \be\coh^q(\Ds,\Sigm)=
\begin{cases}
\piDs\piLambda\left[\overbrace{\aB^1\otimes...\otimes\aB^1}^q\otimes\overbrace{\aA^0\otimes...\otimes\aA^0}^{n-1-q}\otimes\,\aO^0\right], & q<n\\
\piDs\piLambda\left[\overbrace{\aB^1\otimes...\otimes\aB^1}^{n-1}\otimes\,\aB^{q-n+1}\right],& q\geq n,
\end{cases}\ee
which coincides both with the well-known result of Lie algebra cohomology theory, see e.g.
\cite{Kumar}, and with the first theorem in section \ref{SbSGCSigmaResult}. It is easy to remove
the brackets - the diagram of $\mathfrak{h}$ is obtained by concatenation of diagrams of each
multiplier.

\subsection*{The case of $\mathfrak{so}(d+1)$}\label{SbSsod}
In the case of \sodd{}, $\mathfrak{g}=\sodd$ and $\mathfrak{h}=\sod$, we follow along the same path, representing
$\Comp(\Ds,\Sigm)$ as a projection of a tensor product of complexes associated with one-row \sodd-diagrams. It allows us to easly
find candidates for cohomology since the projectors $\piDs$, $\piLambda$ and one new projector $\piCross$ either do not lead to
new cohomology ($\piLambda$, $\piCross$) or it is simple to track the appearance of new cohomology ($\piDs$), neither is it
difficult to find their kernels. At the final step we compute the Euler characteristic to find the dimension of
$\ker(\piDs\piLambda\piCross)$.

\paragraph{harmonic de Rham complex.} Consider the de Rham complex $\hRham$ on harmonic polynomial in $d$ variables
$y^a$, i.e. $\frac{\pl^2}{\pl y^c\pl y_c}\omega(y^a|\theta^b)\equiv0$. The action of the differential $\pl$ is given by the same
formula (\ref{GCRhamActionA}), or in components by (\ref{GCRhamActionB}). The harmonicity condition in terms of components
$\omega^{a(k)|\mu[q]}$ is equivalent to the vanishing trace condition for indices $a$. The cohomology of the harmonic de Rham
complex is well-known, for example in the framework of the unfolded approach it was found in \cite{Shaynkman:2000ts},
\be\begin{tabular}{|x{1.1cm}|x{2cm}|x{2cm}|x{4cm}|}
  \hline
   $q\backslash g$ & $0$ & $1$ & $>1$\tabularnewline \hline
  0 & \emptypar{20pt}$\bullet$ & $\emptyset$ & $\emptyset$  \tabularnewline
  1 & \emptypar{20pt}$\emptyset$ &  $\bullet$  & $\emptyset$     \tabularnewline
  $>1$ & \emptypar{30pt}$\emptyset$  & $\emptyset$ & $\emptyset$ \tabularnewline\hline
\end{tabular}\ee There is one new cohomology class as compared to the de Rham complex $\Rham$, which is due to the
tracelessness condition.

\paragraph{harmonic de Rham complex with constraints.} Analogously, we define $\hRham^s$ to be the harmonic de Rham complex on polynomials
with degree not greater than $s$. As it was the case for \sldd{}, the complex $\hRham^s$ turns out
to be the simplest example of the $\Sigm$-complex $\Comp(\Y{s},\Sigm)$. The component
$\omega^{a(k)|\mu[q]}$ is identified with the traceless part of the projection
$W^{a(k)\bullet(s-k)|\mu[q]}$ of a single form $W^{A(s)|\mu[q]}$ valued in \sodd-module $\Y{s}$,
c.f. (\ref{SbSSigmaMinusSO}). The cohomology of $\hRham^s$ is given by that of $\hRham$ plus some
new cohomology classes because of breaking certain contractible pairs at grade $s$ due to the
degree constraint \be\label{AppCohTableHarmA}\begin{tabular}{|x{1.1cm}|x{2cm}|x{2cm}|x{6cm}|}
  \hline
   $q\backslash g$ & $0$ & $1$ & $s$\tabularnewline \hline
  0 & \emptypar{20pt}$\aO^0=\bullet$ & $\emptyset$ & $\emptyset$  \tabularnewline
  1 & \emptypar{20pt}$\emptyset$ &  $\aO^1=\bullet$  & $\aB^1=\AYoungM{6}{s+1}$     \tabularnewline
  2 & \emptypar{30pt}$\emptyset$  & $\emptyset$ & $\aB^2=\parbox{53pt}{\BYoungM{6}{1}{s+1}{}},\,\,\aC^2=\AYoungM{5}{s}$ \tabularnewline
  3 & \emptypar{30pt}$\emptyset$  & $\emptyset$ & $\aB^3=\parbox{53pt}{\CYoungM{6}{1}{1}{s+1}{}{}},\,\,\aC^3=\BYoungM{5}{1}{s}{}$ \tabularnewline\hline
\end{tabular}\ee
Note that the representatives having the symmetry of one-row Young diagrams occur at degree up to two.

\paragraph{$\hRham^{s_1,s_2}$ complex.} We also need the complex $\hRham^{s_1,s_2}$ that is $\Rham^{s_1}$
on polynomials whose degree is not less than $s_2$, or more formally
\be\ComplexC{\hRham^{s_2}}{\hRham^{s_1,s_2}}{\hRham^{s_1}}.\ee The cohomology of $\hRham^{s_1,s_2}$
reads as \be\label{AppCohTableHarmB}\begin{tabular}{|x{1.1cm}|x{5cm}|x{6cm}|}
  \hline
   $q\backslash g$ & $s_2$ & $s_1$\tabularnewline \hline
  0 & \emptypar{20pt}$\aA^0=\AYoungM{4}{s_2}$ & $\emptyset$   \tabularnewline
  1 & \emptypar{20pt}$\aA^1=\parbox{37pt}{\BYoungM{4}{1}{s_2}{}},\,\,\aD^1=\AYoungM{3}{s_2-1}$ &  $\aB^1=\AYoungM{6}{s_1+1}$     \tabularnewline
  2 & \parbox[c][40pt][c]{0pt}{}$\aA^2=\parbox{37pt}{\CYoungM{4}{1}{1}{s_2}{}{}},\,\,\aD^2=\parbox{30pt}{\BYoungM{3}{1}{s_2-1}{}}$   & $\aB^2=\parbox{53pt}{\BYoungM{6}{1}{s_1+1}{}},\,\,\aC^2=\AYoungM{5}{s_1}$ \tabularnewline\hline
\end{tabular}\ee

\paragraph{$\hRham(\Ds,\pl)$ complex.} Given a Young diagram $\Ds=\Y{s_1,...,s_n}$ the complex
$\hRham(\Ds,\pl)$ is \be\hRham(\Ds,\pl)=\hRham^{s_1}\otimes \hRham^{s_2}\otimes...\otimes
\hRham^{s_{n-1}}\otimes \hRham^{s_n},\ee where the total differential $\pl$ is defined as in the
\sldd-case. With regard to $\hRham(\Ds,\pl)$, its cohomology is just the tensor product of
multipliers' cohomology.

\paragraph{projectors $\boldsymbol{\piDs}$ and $\boldsymbol{\piLambda}$.} Concerning $\piDs$ and $\piLambda$, the following statements are still true:
(1) it is sufficient to consider only irreducible $\mathfrak{h}$-tensors with canonical
arrangements of indices; (2) the Young symmetry conditions via $\piDs$ induce the appearance of the
new cohomology out of $\hRham^{s_i}$ at grade $s_{i+1}$ such that the whole cohomology is
equivalent to that of $\hRham^{s_i,s_{i+1}}$; (3) multiplying from the left by a cohomology
represented by a Young diagram with more than one row makes the corresponding tensor product
cohomology class trivial, so that $\aA^{q>0}\otimes...$, $\aD^{q>1}\otimes...$,
$\aB^{q>1}\otimes...$, $\aC^{q>2}\otimes...$ result in trivial cohomology after imposing the
projectors; (4) the products of the form $\aA\otimes\aB$, $\aD\otimes\aB$, $\aA\otimes\aC$ and
$\aD\otimes\aC$ are mapped to zero by $\piDs$ if $s_1>s_2$, if $s_1=s_2$, then from the
above-listed only $\aA\otimes\aC$ is allowed. In general, the presence of equal rows in $\Ds$
complicates the answer greatly.

\paragraph{projector $\boldsymbol{\piCross}$.} We need one more projector $\piCross$ that removes cross-traces since each
$\omega(y^a_i|\theta^b_i)$ of $\hRham^{s_i}$ is harmonic, however,
$\omega(y^a_i,y^a_{j}|\theta^b_i,\theta^b_j)$ of $\hRham^{s_i}\otimes\,\hRham^{s_{j}}$ is not
harmonic in $y^a_iy^b_j$ if $i\neq j$, $\frac{\pl^2}{\pl y^c_i\pl
y_{cj}}\omega(y^a_i,y^a_{j}|\theta^b_i,\theta^b_j)\neq0$. In terms of components it implies that
the cross-traces - the traces contracting indices from different groups - do not vanish, e.g. \be
C^{a(k)|b(m)}=C^{a(k)|b(m)}_0+\left(\eta^{ab}C^{a(k-1)|b(m-1)}_1+...\right)+\left(\eta^{ab}\eta^{ab}...\right)+...\,\,\,.\ee
To arrive at $\Comp(\Ds,\Sigm)$, cross-traces must be factored out, this is what $\piCross$ does.
The action (\ref{GCSigmaMinusActionExample}) of $\pl$ consists of replacing one index from each
group with the form index, hence $\pl$ cannot increase the cross-trace order. Therefore, if some
$\omega$ is not a cross-trace itself then $\pl(\omega)$ also is not. Hence it is impossible to have
$\pl\omega\neq0$ and $\piCross\left[\pl\omega\right]=0$ if $\omega$ does not belong to
$\ker(\piCross)$. Consequently, $\piCross$ does not give rise to new cohomology.

\paragraph{result if $\boldsymbol{s_i\neq s_{i+1}}$.} If all weights in $\Ds$ are different the answer is very simple:
the representatives of cohomology classes are of two types. Recalling the notation of
(\ref{AppCohTableHarmA}) and (\ref{AppCohTableHarmB}), the ones of the first type have the form
\be\label{GCSOSigmaMinusDiffA}\coh^q_{r,g}(\Ds,\Sigm)=\piTot\left[\overbrace{\aB^1(\aC^2)\otimes...\otimes\aB^1(\aC^2)}^{q-r}\otimes\aA^0(\aD^1)...\otimes\aA^0(\aD^1)\otimes\aO^0\right],\quad
q-r<n\ee where $\aB^1(\aC^2)$ implies that either $\aB^1$ or $\aC^2$ can appear, analogously for
$\aA^0(\aD^1)$ and $\piTot=\piDs\piLambda\piCross$. The degree $q$, grade $g$ and trace order $r$
are \begin{align} &q=\#\aB^1+2\#\aC^2+\#\aD^1,&& g=s_1-s_{q-r+1},&& r=\#\aC^2+\#\aD^1\end{align} If
$q-r\geq n$ then \be\label{GCSOSigmaMinusDiffC}\coh^q_{r,g=s_1}(\Ds,\Sigm)=
\piTot\left[\overbrace{\aB^1(\aC^2)\otimes...\otimes\aB^1(\aC^2)\otimes\aB^{q'}(\aC^{q'+1})}^n\right],\ee
where $q'=q-r-n+1$. The representatives of the second type have the form
\be\label{GCSOSigmaMinusDiffB}\coh^{q+1}_{r+1,g+1}(\Ds,\Sigm)=\piTot\left[\overbrace{\aB^1(\aC^2)\otimes...\otimes\aB^1(\aC^2)}^{q-r}\otimes\aA^0(\aD^1)...\otimes\aA^0(\aD^1)\otimes\aO^1\right],\ee
where $\aO^0$ is replaced with $\aO^1$, which shifts by one the degree, grade and trace order. The
representative (\ref{GCSOSigmaMinusDiffB}) is obtained from (\ref{GCSOSigmaMinusDiffA}) by the
duality map.

All above-stated can be reformulated in terms of $\mspr(\Ds,q,r)$ as in the second theorem of
section \ref{SbSGCSigmaResult}.

\paragraph{some weights in $\Ds$ are equal.} In the case where some rows in $\Ds$ are equal the block notation is more convenient, so let
$\Ds=\Yb{(s_1,p_1),...,(s_N,p_N)}$. At first sight we have a degeneracy so that the multiplicity of
some diagram in $\coh^q_{g,r}$ can differ from $0$ and $1$. Indeed, if $s_i=s_{i+1}$ for some $i$
then $\aA^0$ equals $\aC^2$ as diagrams and hence if there is more than one group of equal rows in
$\Ds$ then different partitions $q=q_1+q_2+...$ corresponding to
\be...\otimes\underbrace{...\otimes\aA^0\otimes...\otimes\aA^0\otimes\overbrace{\aC^2\otimes...\otimes\aC^2}^{q_1}}_{p_i}\otimes...
...\otimes\underbrace{...\otimes\aA^0\otimes...\otimes\aA^0\otimes\overbrace{\aC^2\otimes...\otimes\aC^2}^{q_2}}_{p_j}.......,\ee
result in representatives of different cohomology classes with identical Young diagrams. It seems to
be not enough to specify $q$, $g$, $r$ and a Young diagram in order to distinguish between
different cohomology classes. This will be proved not to be the case.

Because of the decomposition (\ref{SigmaComplexDecomposition}) for $\Comp(\Ds,\Sigm)$, the
cohomology of $\hRham(\Ds,\pl)$ plus those new induced by Young conditions tells us not only which
subcomplexes $\Comp(\Ds,\Sigm;\Xx,q+g,r)$, \be0\longrightarrow
V\fm{0}\xrightarrow{\,\,\,\,\,\Sigm\,\,}...\xrightarrow{\,\,\,\,\,\Sigm\,\,}
V\fm{q-1}\xrightarrow{\,\,\,\,\,\Sigm\,\,} V\fm{q}\xrightarrow{\,\,\,\,\,\Sigm\,\,}
V\fm{q+1}\xrightarrow{\,\,\,\,\,\Sigm\,\,}...\ee can have nonvanishing cohomology but also
determines the degree $q$, grade $g$ and by definition the trace order $r$, where cohomology can be
nontrivial. We see that for each $\Comp(\Ds,\Sigm;\Xx,q+g,r)$ that may have nontrivial cohomology
there is a unique place in terms of $q$ and $g$ where it can happen. Due to the degeneracy, the
multiplicity of $\Xx$ in $\coh^q_{g,r}$ can be greater than one. Making use of the fact that Euler
characteristic of $\Comp(\Ds,\Sigm;\Xx,q+g,r)$ can be computed either as $\chi=\sum_q(-)^q
dim(\coh^q)$ or $\chi'=\sum_q(-)^q dim(V\fm{q})$, we can determine the multiplicity since all but
one summands in $\chi$ are equal to zero - in this case the Euler characteristic determines the
dimension of cohomology of $\coh(\Ds,\Sigm;\Xx,q+g,r)$ modulo sign factor, of course. It presents
no difficulty to compute $\chi'$, we just count the multiplicity of $\Xx$ appearing as the trace of
order $r$ in $\Ds_g\otimes \Ya{q}$ for different $q$ and $g$ while keeping $q+g$ fixed.

\paragraph{$\boldsymbol{\sod}$-tensor product of restricted representations.} Let us address the question of which Young diagrams can appear
in the tensor product $\Ds_{\{k_1,...,k_N\}}\otimes \Ya{q}$ with
$\Ds_{\{k_1,...,k_N\}}\in\res^{\sodd}_{\sod}\Ds$ and $\Ya{q}$ being a one-column diagram of height
$q$. It is useful to define $\Delta_i=s_i-s_{i+1}$. Applying tensor product rules, given in Appendix A, one gets
\be\Ds_{\{k_1,...,k_N\}}\otimes_\son\Ya{q}=\bigoplus_{\{\alpha_j,\beta_i,\gamma_i\}}N^{\{\alpha_j,\beta_i\}}_{\{k_1,...,k_N\}}\Ds_{\{k_1,...,k_N\}}^{\{\alpha_j,\beta_i,\gamma_i\}},\ee
\be\label{AppSoTensorProductABG}\Ds_{\{k_1,...,k_N\}}^{\{\alpha_j,\beta_i,\gamma_i\}}=\parbox{4.5cm}{\unitlength=0.40mm\TracedTensorProductABG}\qquad:
\substack{\displaystyle \alpha_i+\beta_i\leq p_i-1,\quad\mbox{for}\quad i=1...N,\\
\displaystyle \gamma_i\in[-1,0,...,\Delta_i+1]}\ee The explicit formula for the multiplicity
$N^{\{\alpha_j,\beta_i\}}_{\{k_1,...,k_N\}}$ can be written in terms of integer partitions. Important is the very 'geometry' of
$\Ds_{\{k_1,...,k_N\}}^{\{\alpha_j,\beta_i,\gamma_i\}}$, e.g. $\gamma_i\in[-1,\Delta_i+1]$ if $i<N$ and
$\gamma_i\in[0,\Delta_i+1]$ if $i=N$, $\gamma_i=\Delta_i+1$ implies $\alpha_i=p_i-1$, etc.

We choose some $\Xx=\Ds_{\{k_1,...,k_N\}}^{\{\alpha_j,\beta_i,\gamma_i\}}$ and do not fix any $\Ds_{\{k_1,...,k_N\}}$, counting
contributions of all from $\Ds_g$ that lead to $\Xx$ in the tensor product.

The multiplicity $N^{\{\alpha_j,\beta_i\}}_{\{k_1,...,k_N\}}$ can be greater than one inasmuch as
there are in general many ways to remove cells (take traces) and then add them back in order to get
the same diagram $\Xx$. The source of cells is $\Ya{q}$, of course. Let us refer to a cell in
$\Xx$, that can be obtained by removing a number of cells, including this one, and then adding the
same number of cells back as to a vacancy. More than one cell may be needed because it can be that
to remove some cell, according to the tensor product rules, one has first to remove a number of
adjacent cells. At least two cells are required to fill a vacancy, one to take trace and one to
restore the original cell. The vacancies correspond to $\epsilon_i$ and to the rightmost cell in
$\gamma_i$ provided $\gamma_i\in[0,\Delta_i]$. Note that there is only one way to get the cells
corresponding to $\alpha_i$ and $\gamma_k=\Delta_k+1$ (or $\beta_i$ and $\gamma_k=-1$), these cells
have to be added (or removed), not to mention the cells in the 'interior' of $\Xx$ that are not
affected by the tensor product rules. It is convenient to single the constant parts $Q$, $R$ and
$G$ out of $q$, $r$ and $g$ that do not vary when passing from one $\Ds_{\{k_1,...,k_N\}}$ to
another,
\begin{align*}
q&=Q+q', & Q&=\sum_{j=1}^{N+1} \alpha_j + \sum_{i=1}^N \beta_i+N_{-1}+N_{\Delta+1}, & q'&= N_{\gamma>k}+N_{\gamma<k}+2\rho,\\
g&=G+g', & G&=\sum_{i=1}^{i=N}\left(\gamma_i+\delta_{\gamma_i,-1}-\delta_{\gamma_i,\Delta_i+1}\right), & g'&=N_{\gamma<k}-N_{\gamma>k}\\
r&=R+r', & R&=\sum_{i=1}^N \beta_i+N_{-1}, & r'&=N_{\gamma>k}+\rho
\end{align*}
\begin{align*}
N_{-1}&=\#\{i:\gamma_i=-1\},&N_{\Delta+1}&=\#\{i:\gamma_i=\Delta_i+1\},\\
N_{\gamma<k}&=\#\{i:\gamma_i<k_i\,\gamma_i\neq-1,\Delta_i+1\},&N_{\gamma>k}&=\#\{i:\gamma_i>k_i,\gamma_i\neq-1,\Delta_i+1\},
\end{align*} and $\rho$, $N_{\gamma<k}$, $N_{\gamma>k}$ are the parts that depend on a particular $\Ds_{\{k_1,...,k_N\}}$.
For example, $N_{\gamma<k}$ is equal to the number of those $k_i$ in $\Ds_{\{k_1,...,k_N\}}$ that are greater than $\gamma_i$.
Note that by virtue of the tensor product rules $\gamma_i=k_i,k_i\pm1$. It is $\rho$ that governs the multiplicity, $\rho$ is
equal to the number of vacancies to be filled. Therefore, the multiplicity of $\Xx$ in $\Ds_{\{k_1,...,k_N\}}\otimes\Ya{q}$ is
given by the number of partitions of $\rho$ among the vacancies given by $\epsilon_i$ and those $\gamma_i$ that equal $k_i$
(modulo certain subtleties to be considered below).

\paragraph{Euler characteristic.} Let us proceed to the computation of the Euler characteristic. It makes no difference to compute it only for
$\Xx$ of special form dictated by $\hRham(\Ds,\pl)$ or in the general case, so we do it for
arbitrary $\Xx$. Firstly, we construct for $\Xx$ the generating function $F(z,t)$ such that the
coefficient of $z^{g'}t^{q'}$ is equal to the number of ways to get $\Xx$ in
$\Ds_{\{k_1,...,k_N\}}\otimes\Ya{Q+q'}$ with $\sum_i k_i=G+g'$. Thus, $z$ counts the excess over
the base level $G$, and $t$ counts the number $q'$ of cells needed to get $\Xx$.

It is important for the computations to be simple that the whole diagram $\Xx$ can be cut into
pieces such that the generating function can be first constructed for each of the pieces and then
the total $F(z,t)$ is just the product of generating functions over the pieces. In the table 1
below we collect all different types of such pieces together with generating functions, where
$f_\epsilon(t)=(1-t^{2\epsilon+2})/(1-t^2)$. It is easy to see that the generating function for the
Euler characteristics of $\Comp(\Ds,\Sigm;\Xx,q+g,r)$ is just $F(-t,t)$, where the degree of $t$ is
equal to $2r'$ and the coefficient of $t^{2r'}$ up to a sign equals the Euler characteristic of
$\Comp(\Ds,\Sigm;\Xx,q+g,R+r')$. Note that $F(-t,t)$ can be a polynomial rather than a monomial
because the same diagram $\Xx$ can appear in the tensor product at different values of $r$.

At the Table 1 below, we collected all possible types of pieces, into which the diagram $\Xx$ is
decomposed. Looking at the Table, we see that $\Xx$ such that at least one of $\gamma_i$ does not
take an extremal value $\{-1,0,\Delta_i,\Delta_i+1\}$ results in vanishing Euler characteristic,
which exactly correspond to the fact the cohomology of $\hRham^{s_i,s_{i-1}}$ is concentrated in
the lowest and highest grade. Surprising is that $\chi=0$ if the subsequence
$\gamma_{i-1}=0,\gamma_i=\Delta_i$ occurs in $\Xx$ and hence the piece no. 2 can occur only once in
$\Xx$.

\begin{table}\label{table:TensorProductPieces}
\caption[]{Independent pieces constituting the diagram
$\Ds_{\{k_1,...,k_N\}}^{\{\alpha_j,\beta_i,\gamma_i\}}$, together with generating functions and
Euler
characteristics\parbox[t][6pt][c]{0pt}{}}\noindent\begin{tabular}{|x{0.5cm}|x{3cm}|x{2.5cm}|x{2cm}|x{5.5cm}|}
    \hline
    \rule{0pt}{14pt} & illustration & $F(z,t)$ & $F(-t,t)$ & description \tabularnewline
    \hline
    1 &\parbox{40pt}{\SpecialCaseA} & $1+zt$ &  $1-t^2$ &
    \parbox{5.5cm}{\rule{0pt}{16pt}The last block. There are two ways to get $\gamma_N=0$: (1)
    take a diagram with $k_N=0$ and do nothing; (2) take a diagram with $k_N=1$ (z), and then take a trace (t)\parbox[t][6pt][c]{0pt}{}} \tabularnewline \hline
    2 & \parbox[c][60pt][c]{60pt}{\SpecialCaseG} & $f_{\epsilon_i}(t)$ &  $f_{\epsilon_i}(t)$ &
    \parbox{5.5cm}{\rule{0pt}{16pt}A group of $\epsilon_i$ 'isolated' vacancies. One can remove $k$ cells and then add them back with any
    $k\in[0,\epsilon_i]$, which yields $1+t^2+...+t^{2\epsilon_i}$\parbox[t][6pt][c]{0pt}{}} \tabularnewline \hline
    3 & \SpecialCaseC & $f_{\epsilon_i+1}(t)+ztf_{\epsilon_i}(t)$ & $1$ & \parbox{5.5cm}{\rule{0pt}{16pt}A group of $\epsilon_i$
    vacancies that are not 'isolated', being linked to $\gamma_i=0$. If $k_i=0$ then $\epsilon_i$ effectively increases to $\epsilon_i+1$;
    if $k_i=1$, i.e. $g'=+1$, then this one extra cell must be removed\parbox[t][6pt][c]{0pt}{}}\tabularnewline \hline
    4 & \SpecialCaseF & $f_{\epsilon_i+1}(t)+z^{-1}tf_{\epsilon_i}(t)$ & $t^{2(\epsilon_i+1)}$ &
    \parbox{5.5cm}{\rule{0pt}{16pt}A group of $\epsilon_i$ vacancies is linked to $\gamma_i=\Delta_i$. If $k_i=\Delta_i$ we have
    $f_{\epsilon_i+1}$; in the second case of $k_i=\Delta_i-1$, i.e. $g'=-1$, one extra cell must be added\parbox[t][6pt][c]{0pt}{}}\tabularnewline \hline
    5 & \SpecialCaseB & $f_{\epsilon_i+2}(t)+t^2f_{\epsilon_i}(t)+f_{\epsilon_i+1}(t)(z+z^{-1})t$ & $0$ &
    \parbox{5.5cm}{\rule{0pt}{16pt}This case include parts of the previous two cases; if $k_{i-1}=0$ and $k_i=\Delta_i$ then
    the 'effective' $\epsilon_i$ is equal to $\epsilon_i+2$\parbox[t][6pt][c]{0pt}{}}\tabularnewline \hline
    6 & \SpecialCaseD & $1+t^2+t(z+z^{-1})$ & $0$ & \parbox{5.5cm}{\rule{0pt}{16pt} In this cases $\gamma_i$ does not
    take extreme values. If $k_i=\gamma_i$ then we can either do nothing or remove one cell and then add it back;
    if $k_i=\gamma_i\pm1$ then one cell must be added (removed)\parbox[t][6pt][c]{0pt}{}} \tabularnewline \hline
\end{tabular}
\end{table}
Given $\Xx$, depending on whether the piece no. 2 is present or not, the function $F(-t,t)$ can have one of the two forms
\begin{align}
I:& &&F(-t,t)=\prod_{i:\gamma_i=\Delta_i}t^{2(\epsilon_i+1)},\\
II:&&&F(-t,t)=\prod_{i:\gamma_i=\Delta_i}t^{2(\epsilon_i+1)}(1-t^{2\epsilon'+2}),
\end{align} where $\epsilon'$ is the $\epsilon_i$ that correspond to the piece no. 2 from the table.

\par\noindent{\it I.} In the first case we have all $\gamma_i$ taking one of the maximal values $\{\Delta_i,\Delta_i+1\}$, i.e.
$g=s_1$. The diagram $\Xx$ consists of blocks of the form
\be\label{GCSOSigmaMinusDiffD}\aB\aC_j\sim\pi_{\Yb{s_j,p_j}}\piLambda\piCross
\left[\underbrace{\overbrace{\aB^1\otimes...\otimes\aB^1}^{\alpha_j}\otimes\overbrace{\aC^2\otimes...\otimes\aC^2}^{\epsilon_j+1}}_{p_j}\right].\ee
That $r'=\sum_i(\epsilon_i+1)$ takes the maximal value implies that no $\aA^0$ can occur, only $\aC^2$ can. The first case
correspond to $q-r> n$ and to the maximal grade so that none of the representatives has a dual pair.

\par\noindent{\it II.} In the second case $F(-t,t)$ consists of two monomials, each corresponding to a cohomology class with the
same $\Xx$ but with different $r'$, the difference is $\epsilon'+1$. The second class is obtained via the duality map. Here we
refer to fig.\,\,\ref{fig:PicGCWeylModule} illustrating both classes. The representatives are given by diagrams of the form
\be\label{GCSOSigmaMinusDiffCohA}\piTot\left[\aB\aC_1\otimes...\otimes\aB\aC_{k-1}\otimes\aB\aA\aD_k\otimes\aA\aD_{k+1}\otimes...\otimes\aA\aD_N^0\right],\ee
where
\begin{align}\nonumber\aB\aA\aD_k&\sim\pi_{\Yb{s_k,p_k}}\piLambda\piCross
\left[\underbrace{\overbrace{\aB^1\otimes...\otimes\aB^1}^{\alpha_k}\otimes\overbrace{\aA^0\otimes...\otimes\aA^0}^{\epsilon_k}\otimes
\overbrace{\aD^1\otimes...\otimes\aD^1}^{\beta_k}\otimes\aA^0}_{p_k}\right],\\
\label{GCSOSigmaMinusDiffF}\aA\aD_j&\sim\pi_{\Yb{s_j,p_j}}\piLambda\piCross
\left[\underbrace{\overbrace{\aA^0\otimes...\otimes\aA^0}^{\epsilon_j}\otimes\overbrace{\aD^1\otimes...\otimes\aD^1}^{\beta_j}
\otimes\aA^0}_{p_j}\right],\end{align} and the last block $\aA\aD_N^0$ ends with $\aO^0$ instead of $\aA^0$. The integer $k$
corresponds to the block that has the form of the piece no. 2, i.e. $\epsilon_k=\epsilon'$. Thus, all $\gamma_j$ in the range
$j=1,...,k-1$ take one of the maximal values $\{\Delta_j,\Delta_j+1\}$, the rest of $\gamma_i$ with $i=k,...,N$ take one of the
minimal values $\{-1,0\}$.

The representative of the second cohomology class, which is dual to the first, has the form
\be\label{GCSOSigmaMinusDiffCohB}\piTot\left[\aB\aC_1\otimes...\otimes\aB\aC_{k-1}\otimes\aB\aC\aD_k\otimes\aA\aD_{k+1}\otimes...\otimes\aA\aD_N^1\right],\ee
где \be\label{GCSOSigmaMinusDiffG}\aB\aC\aD_k\sim\pi_{\Yb{s_k,p_k}}\piLambda\piCross
\left[\underbrace{\overbrace{\aB^1\otimes...\otimes\aB^1}^{\alpha_k}\otimes\overbrace{\aC^2\otimes...\otimes\aC^2}^{\epsilon_k}\otimes
\overbrace{\aD^1\otimes...\otimes\aD^1}^{\beta_k}}_{p_k}\right],\ee and the last block $\aA\aD_N^1$ ends with $\aO^1$ instead of
$\aA^0$. The degree $q$, grade $g$ and trace order $r$ are shifted by $2\epsilon'+1$, $1$ and $\epsilon'+1$, respectively.

In conclusion, let us note that despite the possibility of great degeneracy mentioned at the beginning, the projectors somehow
remove degeneracy so that the multiplicity of any $\Xx$ in $\coh^q_{g,r}$ is either zero or one. To be strict, only the shape of
(\ref{GCSOSigmaMinusDiffCohA}) and (\ref{GCSOSigmaMinusDiffCohB}) is relevant, the same diagram can be obtained in many different
ways generally (we can replace some $\aC^2$ with $\aA^0$ in blocks $1,...,k$, and then make the same number of inverse
replacements in blocks $k+1,...,N$). What has been proved is that the multiplicity of $\Xx$ determined by
(\ref{GCSOSigmaMinusDiffCohA}) and (\ref{GCSOSigmaMinusDiffCohB}) is equal to one. Nevertheless, it can be shown that the
suggested representatives (\ref{GCSOSigmaMinusDiffCohA}) and (\ref{GCSOSigmaMinusDiffCohB}) are indeed not exact. It is not hard
to see that the answer just obtained coincides with that in terms of $\mspr(\Ds,q,r)$ given in the second theorem of
Section \ref{SbSGCSigmaResult}.

\providecommand{\href}[2]{#2}\begingroup\raggedright\endgroup

\end{document}